\documentclass[10pt,twocolumn,letterpaper]{article}

\usepackage{cvpr}              % To produce the CAMERA-READY version
\usepackage{subcaption}
\definecolor{cvprblue}{rgb}{0.21,0.49,0.74}
\usepackage[pagebackref,breaklinks,colorlinks,allcolors=magenta]{hyperref}
\usepackage{makecell}
\def\paperID{*****} % *** Enter the Paper ID here
\def\confName{CVPR}
\def\confYear{2025}

\title{NTIRE 2025 Challenge on Short-form UGC Video Quality Assessment and Enhancement: Methods and Results}

\author{Xin Li\textsuperscript{$\dagger$,$*$} \quad Kun Yuan\textsuperscript{$\dagger$,$*$} \quad Bingchen Li\textsuperscript{$\dagger$} \quad Fengbin Guan\textsuperscript{$*$} \quad  Yizhen Shao\textsuperscript{$\dagger$} \quad Zihao Yu\textsuperscript{$*$} \\ Xijun Wang\textsuperscript{$\dagger$}  \quad Yiting Lu\textsuperscript{$*$} \quad Wei Luo\textsuperscript{$*$} \quad Suhang Yao\textsuperscript{$\dagger$} \quad Ming Sun\textsuperscript{$\dagger$,$*$} \quad Chao Zhou\textsuperscript{$\dagger$,$*$} \\  Zhibo Chen\textsuperscript{$\dagger$,$*$} \quad Radu Timofte\textsuperscript{$\dagger$,$*$}  \quad Yabin Zhang \quad Ao-Xiang Zhang \quad Tianwu Zhi \quad Jianzhao Liu \\ Yang Li \quad Jingwen Xu \quad Yiting Liao \quad Yushen Zuo \quad Mingyang Wu \quad Renjie Li \quad Shengyun Zhong \\ Zhengzhong Tu \quad Yufan Liu \quad Xiangguang Chen \quad Zuowei Cao \quad Minhao Tang  \quad Shan Liu \\ Kexin Zhang \quad Jingfen
Xie \quad Yan Wang \quad Kai Chen \quad Shijie Zhao \quad Yunchen Zhang \\ 
Xiangkai Xu \quad Hong Gao \quad Ji Shi \quad Yiming Bao \quad Xiugang Dong \quad Xiangsheng Zhou \quad Yaofeng Tu \\ Ying Liang \quad Yiwen
Wang \quad Xinning Chai \quad Yuxuan Zhang \quad Zhengxue Cheng \quad
Yingsheng Qin \\ Yucai Yang \quad Rong Xie \quad Li Song \quad Wei Sun \quad Kang Fu \quad Linhan Cao \quad Dandan Zhu \\ Kaiwei Zhang \quad Yucheng Zhu \quad Zicheng Zhang \quad Menghan Hu \quad Xiongkuo Min \quad Guangtao Zhai \\  Zhi Jin \quad Jiawei Wu \quad Wei Wang \quad Wenjian Zhang \quad Yuhai Lan \quad Gaoxiong Yi \quad Hengyuan Na \\ Wang Luo \quad Di Wu \quad MingYin Bai \quad Jiawang Du \quad Zilong Lu \quad Zhenyu Jiang \quad Hui Zeng \\ Ziguan Cui \quad Zongliang Gan \quad Guijin Tang  \quad Xinglin Xie \quad Kehuan Song \quad Xiaoqiang Lu\\ Licheng Jiao\quad Fang Liu\quad Xu Liu\quad Puhua Chen \quad Ha Thu Nguyen \quad Katrien De Moor \\ Seyed Ali Amirshahi \quad Mohamed-Chaker Larabi \quad Qi Tang \quad Linfeng He\quad Zhiyong Gao\\ Zixuan Gao\quad Guohua Zhang\quad Zhiye Huang \quad Yi Deng\quad Qingmiao Jiang \quad Lu Chen\\ Yi Yang\quad Xi Liao\quad Nourine Mohammed Nadir \quad Yuxuan Jiang\quad Qiang Zhu\quad Siyue Teng\\ Fan Zhang\quad Shuyuan Zhu\quad  Bing Zeng\quad David Bull \quad Meiqin Liu\quad Chao Yao\quad Yao Zhao
 }

\begin{document}
\maketitle
\renewcommand{\thefootnote}{}
\footnotetext{$^{*}$X. Li(\textcolor{magenta}{xin.li@ustc.edu.cn}), K. Yuan (\textcolor{magenta}{yuankun03@kuaishou.com}), F. Guan, Z. Yu, Y. Lu, W. Luo, M. Sun, C. Zhou, Z. Chen, and R. Timofte are the challenge organizers of Track 1 - VQA.}
\footnotetext{$^{\dag}$X. Li, K. Yuan, B. Li, Y. Shao, X. Wang, S. Yao, M. Sun, C. Zhou, Z. Chen, and R. Timofte are the challenge organizers of Track 2 - KwaiSR.}

\footnotetext{The other authors are participants of the NTIRE 2025 Short-form UGC Video Quality Assessment and Enhancement Challenge.}
\footnotetext{The NTIRE2025 website:~\url{https://cvlai.net/ntire/2025/}}
\footnotetext{The KVQ database:~\url{https://lixinustc.github.io/projects/KVQ/}}
\footnotetext{The KwaiSR database:~\url{https://github.com/lixinustc/KVQE-Challenge-CVPR-NTIRE2025}}

\begin{abstract}
This paper presents a review for the NTIRE 2025 Challenge on Short-form UGC Video Quality Assessment and Enhancement. The challenge comprises two tracks: (i) Efficient Video Quality Assessment (KVQ), and (ii) Diffusion-based Image Super-Resolution (KwaiSR). Track 1 aims to advance the development of lightweight and efficient video quality assessment (VQA) models, with an emphasis on eliminating reliance on model ensembles, redundant weights, and other computationally expensive components in the previous IQA/VQA competitions. Track 2 introduces a new short-form UGC dataset tailored for single-image super-resolution, \ie, the KwaiSR dataset. It consists of 1,800 synthetically generated S-UGC image pairs and 1,900 real-world S-UGC images, which are split into training, validation, and test sets using a ratio of 8:1:1. The primary objective of the challenge is to drive research that benefits the user experience of short-form UGC platforms such as Kwai and TikTok. This challenge attracted 266 participants and received 18 valid final submissions with corresponding fact sheets, significantly contributing to the progress of short-form UGC VQA and image super-resolution. The project is publicly available at ~\url{https://github.com/lixinustc/KVQE-Challenge-CVPR-NTIRE2025}.
\end{abstract}
    
\section{Introduction}
\label{sec:intro}
Recently, short-form user-generated content (S-UGC) platforms such as Kwai and TikTok have emerged as mainstream streaming platforms for information sharing and dissemination. Unlike traditional UGC or professionally generated content (PGC), short-form UGC videos own advantages, including mobile-friendly browsing mode, high user engagement, and abundant content creation~\cite{li2024ntireKVQ-Challengereport,KSVQE,ntire2025shortugc_data}. However, since the content creation of S-UGC does not require professional acquisition devices (\textit{e.g.,} the mobile phone) and experienced users, the source S-UGC videos might inevitability suffer from sub-optimal subjective quality. Meanwhile, the complicated video processing techniques, like preprocessing, transcoding, and enhancement in the S-UGC platform will further lead to unexpected quality changes. It is essential to develop powerful S-UGC video quality assessment and enhancement methods to promote the development of Short-form UGC platforms. 

There are lots of studies that have been devoted to video/image quality assessment (VQA)\cite{SimpleVQA,Fast-vqa,RQ-VQA,ICME2021UGC-VQA,DBLP:conf/mm/XieYQWSZZ24,zhang2023blindLIQE,tu2021ugc,yu2024sfiqa,zhao2023quality,DBLP:conf/mm/LiuWYSTZWL23,DBLP:journals/corr/abs-2503-10259} and image super-resolution\cite{swinir,swinv2-transformer,HAT,EDSR,wang2021realesrgan,li2024sed,yang2022aim,li2021close,DBLP:conf/eccv/QuYZXHSZ24} 
. Based on the availability of reference information, UGC VQA methods can be categorized into full-reference~\cite{sun2021deep, LPIPS}, no-reference~\cite{sun2022deep,tu2021ugc,internvqa,yu2024video}, and reduced-reference~\cite{soundararajan2012video,ma2012reduced} approaches. With the rapid advancement of large language models (LLMs) and large multimodal models (LMMs), recent works~\cite{wu2023qalign,lu2025qadapt,wu2024qinstruct} take the first step to explore their reasoning and understanding capabilities to enhance the interactivity and explainability of VQA framework. In parallel, the evolution of model architectures has significantly improved the performance of image super-resolution. Existing approaches can be roughly classified into four categories: CNN-based~\cite{SR3_EDSR-cnn,SR2——VDSR-cnn,li2023learningDIL}, Transformer-based~\cite{sr_transformer1,sr2_transformer2,sr2_transformer3,sr3_swinfir-transformer,HST}, Mamba-based~\cite{guo2024mambair,ren2024mambacsr}, and MLP-based methods~\cite{tu2022maxim,li2024ucip}. Among them, Transformer-based models excel at capturing global contextual dependencies, while Mamba-based methods offer linear-time sequence processing and possess learned state space dynamics, enabling competitive or superior performance with significantly lower computational overhead. However, S-UGC image super-resolution and efficient S-UGC VQA have been underexplored.

To advance the development of short-form UGC (User-Generated Content) platforms, we organized the NTIRE 2025 Challenge on Short-form UGC Video Quality Assessment and Enhancement, in collaboration with the NTIRE 2025 Workshops. This challenge aims to establish a practical and comprehensive benchmark for evaluating and enhancing the quality of short-form UGC content. We welcome the collaborative efforts of all participants to push the boundaries of short-form video quality. The challenge consists of two tracks: (i) Efficient Short-form UGC Video Quality Assessment. This track introduces an innovative evaluation methodology that combines coarse-grained quality scoring with fine-grained rankings for difficult samples. The VQA models must adhere to a computational limit of 120 GFLOPs; (ii) Diffusion-based Image Super-Resolution for Short-form UGC Images in the Wild. This track focuses on enhancing the subjective quality of S-UGC images, which utilizes a combination of user studies and no-reference quality metrics to better reflect subjective quality.

%% cross-referencing NTIRE 2025 associated challenges

This challenge is one of the NTIRE 2025~\footnote{\url{https://www.cvlai.net/ntire/2025/}} Workshop associated challenges on: ambient lighting normalization~\cite{ntire2025ambient}, reflection removal in the wild~\cite{ntire2025reflection}, shadow removal~\cite{ntire2025shadow}, event-based image deblurring~\cite{ntire2025event}, image denoising~\cite{ntire2025denoising}, XGC quality assessment~\cite{ntire2025xgc}, UGC video enhancement~\cite{ntire2025ugc}, night photography rendering~\cite{ntire2025night}, image super-resolution (x4)~\cite{ntire2025srx4}, real-world face restoration~\cite{ntire2025face}, efficient super-resolution~\cite{ntire2025esr}, HR depth estimation~\cite{ntire2025hrdepth}, efficient burst HDR and restoration~\cite{ntire2025ebhdr}, cross-domain few-shot object detection~\cite{ntire2025cross}, short-form UGC video quality assessment and enhancement~\cite{ntire2025shortugc,ntire2025shortugc_data}, text to image generation model quality assessment~\cite{ntire2025text}, day and night raindrop removal for dual-focused images~\cite{ntire2025day}, video quality assessment for video conferencing~\cite{ntire2025vqe}, low light image enhancement~\cite{ntire2025lowlight}, light field super-resolution~\cite{ntire2025lightfield}, restore any image model (RAIM) in the wild~\cite{ntire2025raim}, raw restoration and super-resolution~\cite{ntire2025raw} and raw reconstruction from RGB on smartphones~\cite{ntire2025rawrgb}.

\section{Challenge}
\label{sec:challenge}

\noindent\textbf{Track1: Efficient Short-form UGC Video Quality Assessment}

The first track is efficient short-form UGC video quality assessment, and the second track is Diffusion-based Super-resolution for the Short-form UGC Images in the Wild. The first track utilizes the KVQ, i.e., the large-scale Kaleidoscope short Video database for Quality assessment, for training, and evaluation. The KVQ database comprises 600 user-uploaded short videos and 3600 processed videos through diverse practical processing workflows. Moreover, it contains nine primary content scenarios in the practical short-form video platform, including landscape, crowd, person, food, portrait, computer graphic (termed as CG), caption, and stage, covering almost all existing creation modes and scenarios, and the ratio of each category of content satisfies the practical online statistics. The quality score of each short-form video and the partial ranked score are annotated by professional researchers on image processing.

\noindent\textbf{Track2: Diffusion-based Image Super-resolution for Short-form UGC Images in the Wild}

\begin{figure*}[t]
    \centering
    \includegraphics[width=\linewidth]{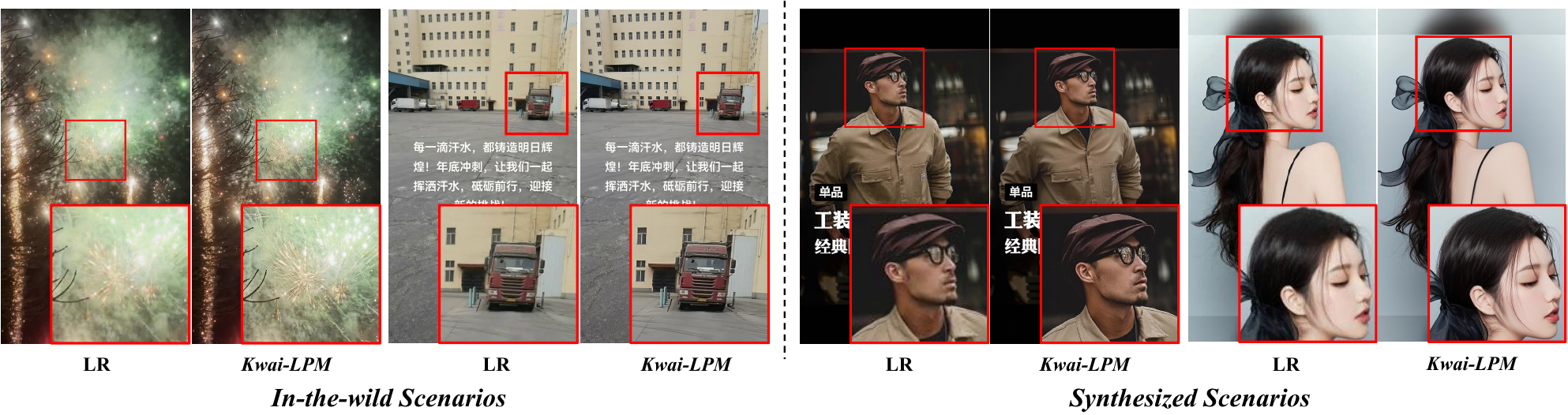}
    \vspace{-5mm}
    \caption{Enhancement results by \textit{Kwai-LPM (Large Processing Model)}, which is a diffusion-based SR method.}
    \label{fig:lpm}
\end{figure*}

The second track collected 1800 synthetic paired images with a simulation strategy from the real-world Kwai Platform and 1900 real-world in-the-wild images with only low-quality images. The contents are from the same source as the KVQ datasets. The purpose is to improve the perceptual quality of images in the wild while maintaining the generalization capability. It is encouraged to utilize the diffusion models for methods. And Fig.~\ref{fig:lpm} shows processing results as a reference. Other methods are also welcome.

\section{Challenge Results}
\label{challenge_results}
The challenge results are presented in Table~\ref{tab:results_1}. We report the performances of teams that submitted their fact sheets. The teams with top performances including SharpMind, ZQE, ZX-AIE-Vector, ECNU-SJTU VQA Team, and TenVQA achieved excellent results in both PLCC and SROCC, with all values exceeding 0.91. Notably, the top three teams also maintained strong Rank1 and Rank2 scores, highlighting their models' robustness under diverse evaluation metrics. The first-place team, SharpMind, attained the highest final score of 0.922, with relatively low computational complexity of 47.39 GFLOPs and 33.01M parameters. ZQE and ZX-AIE-Vector follow closely with scores of 0.916 and 0.912 respectively, while operating at moderate computational budgets. Interestingly, GoldenChef and ZQE exhibit relatively large model sizes (over 150M parameters), whereas DAIQAM achieved competitive accuracy (0.844) with only 66.36 GFLOPs and 27.98M parameters, demonstrating a good balance between performance and efficiency. We also observe that several teams have achieved superior results while maintaining computational constraints below 120 GFLOPs, demonstrating the community's continued advancement in building efficient and accurate VQA models.

\begin{figure*}[htbp]
  \centering
  \begin{subfigure}[b]{0.1865\textwidth}
    \includegraphics[height=5.8cm]{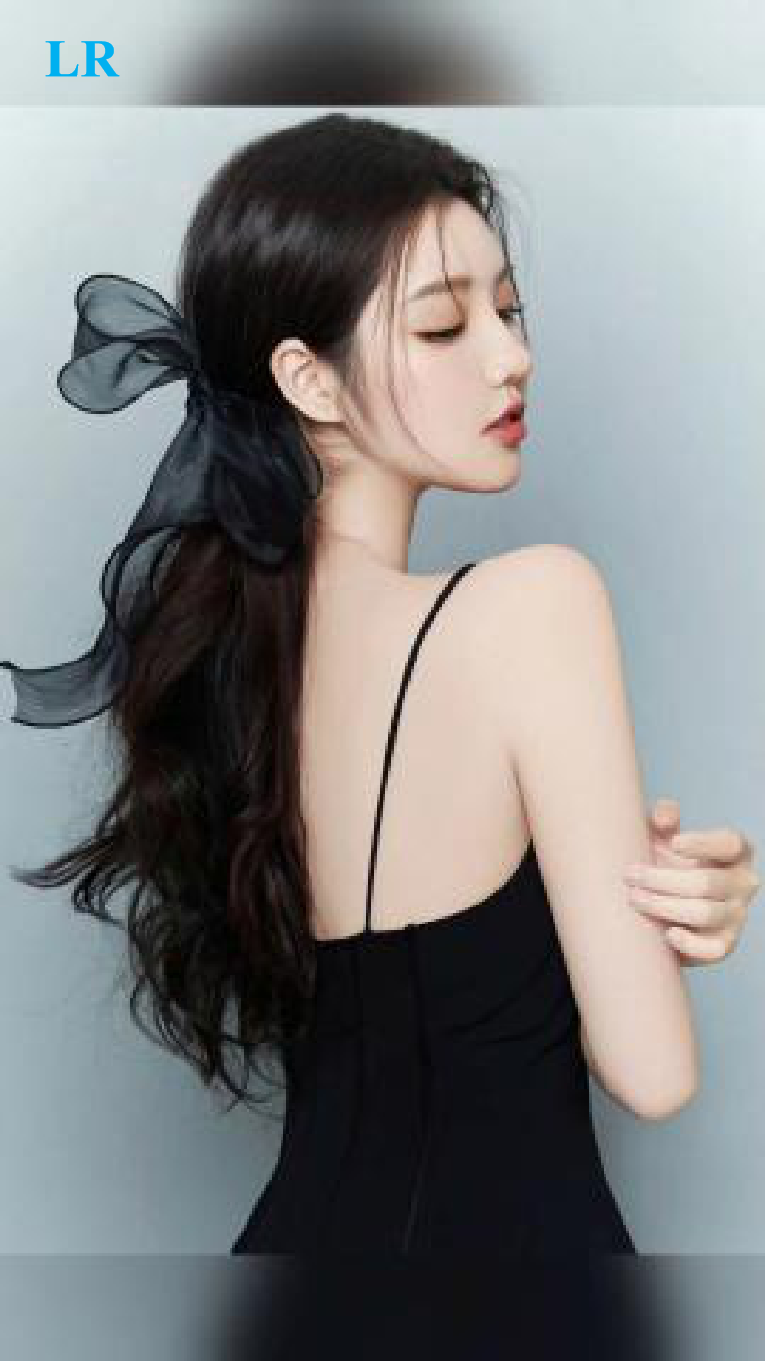}
  \end{subfigure}
  \begin{subfigure}[b]{0.265\textwidth}
    \includegraphics[height=5.8cm]{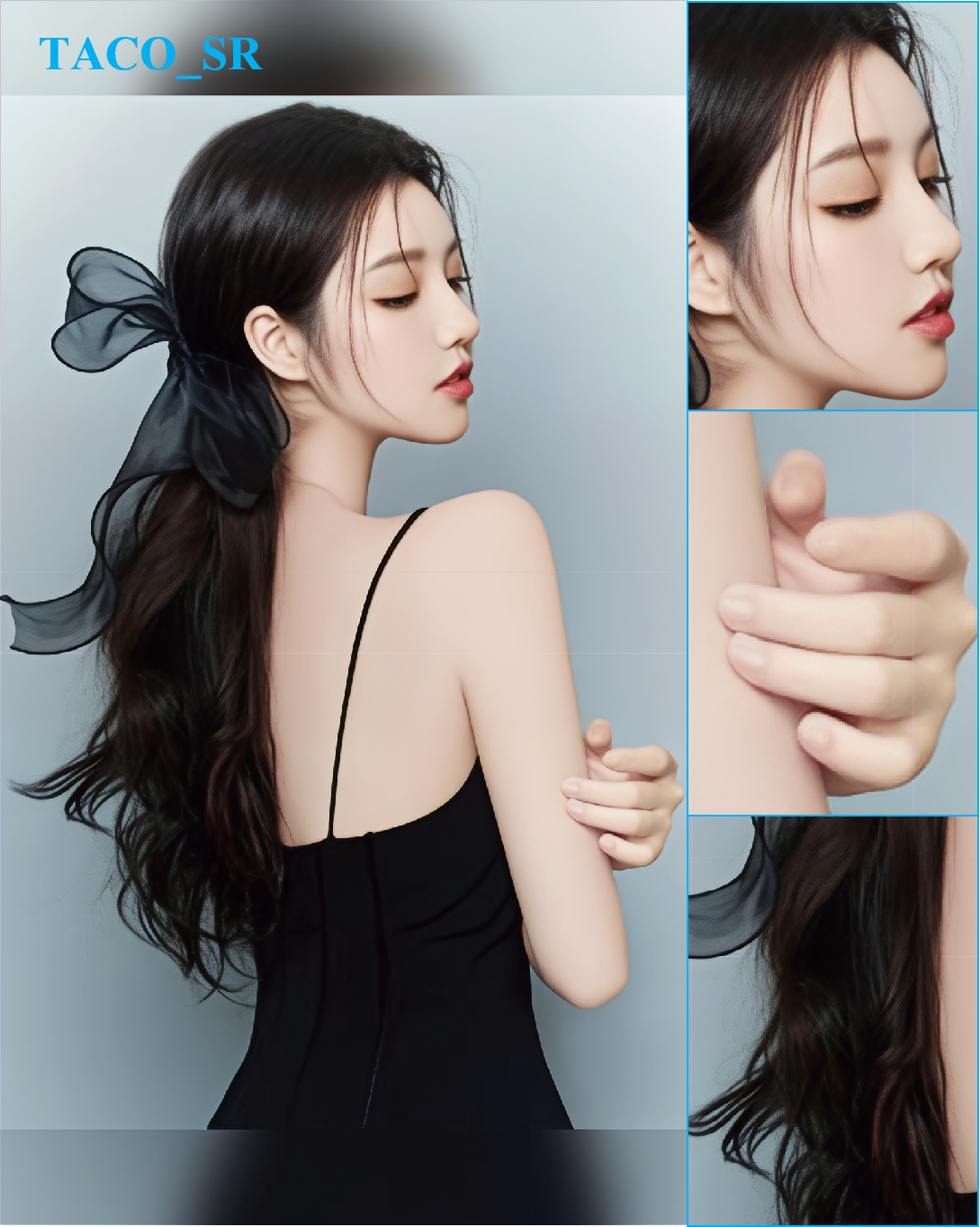}
  \end{subfigure}
  \begin{subfigure}[b]{0.265\textwidth}
    \includegraphics[height=5.8cm]{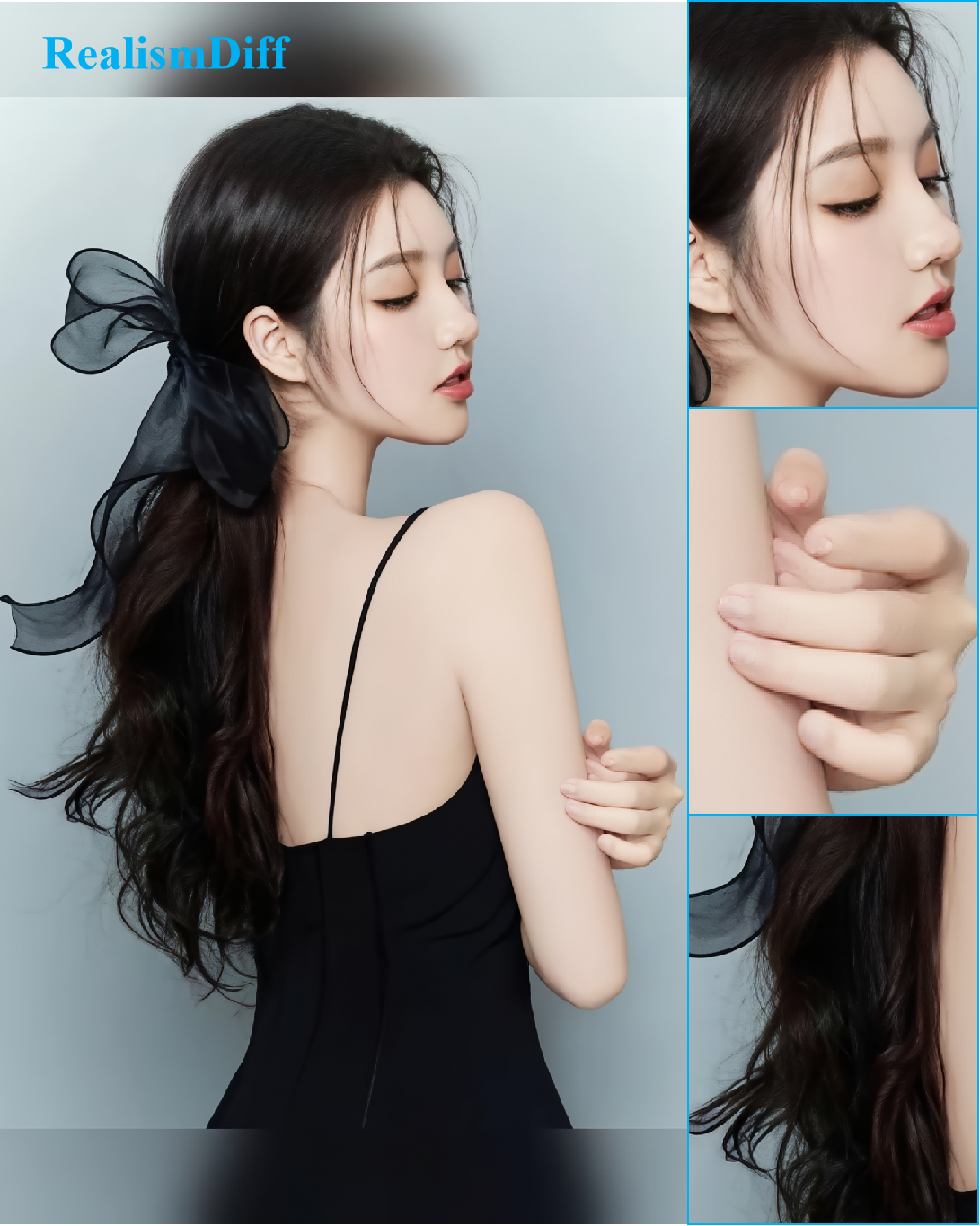}
  \end{subfigure}
  \begin{subfigure}[b]{0.265\textwidth}
    \includegraphics[height=5.8cm]{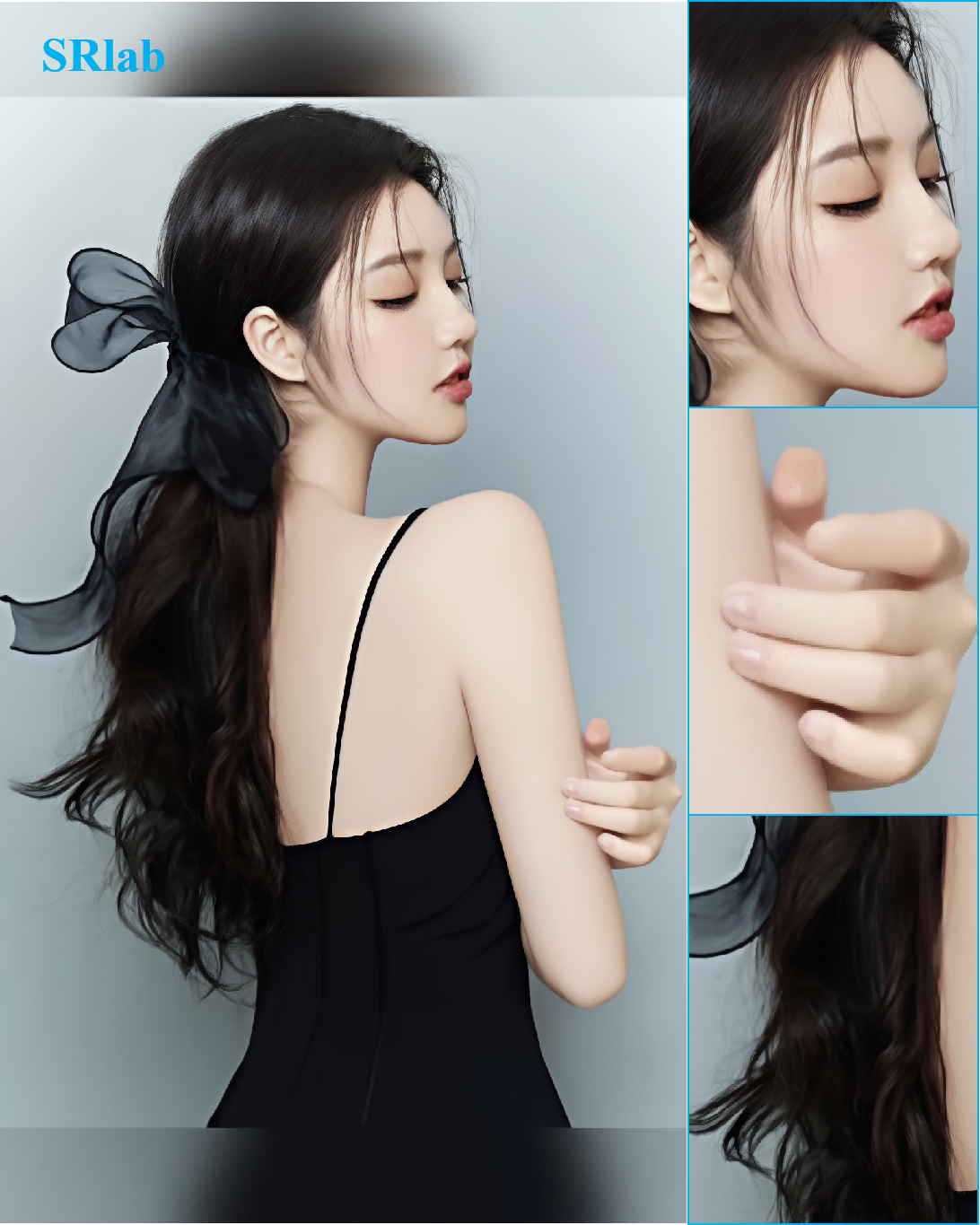}
  \end{subfigure}

  \vspace{0.2em} % 行间距
\begin{subfigure}[b]{0.1865\textwidth}
    \includegraphics[height=5.8cm]{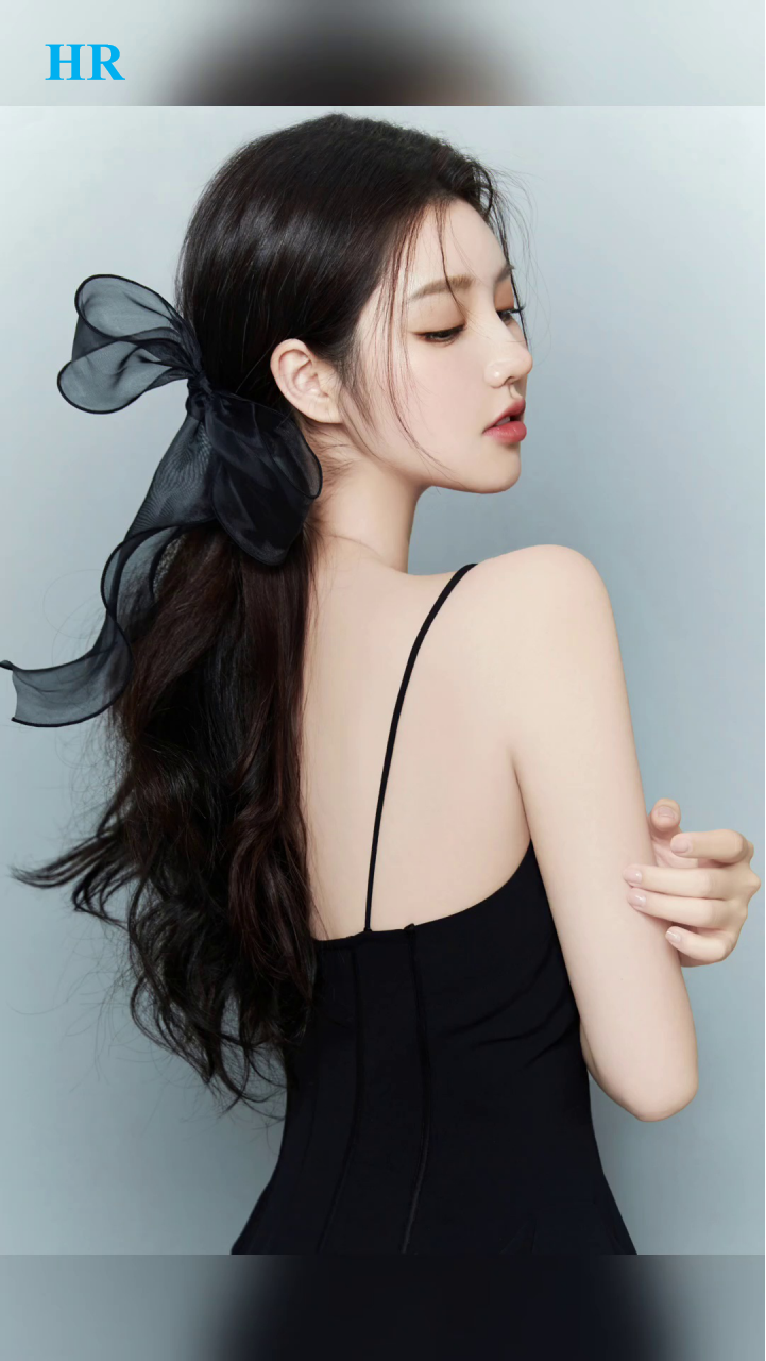}
  \end{subfigure}
  \begin{subfigure}[b]{0.265\textwidth}
    \includegraphics[height=5.8cm]{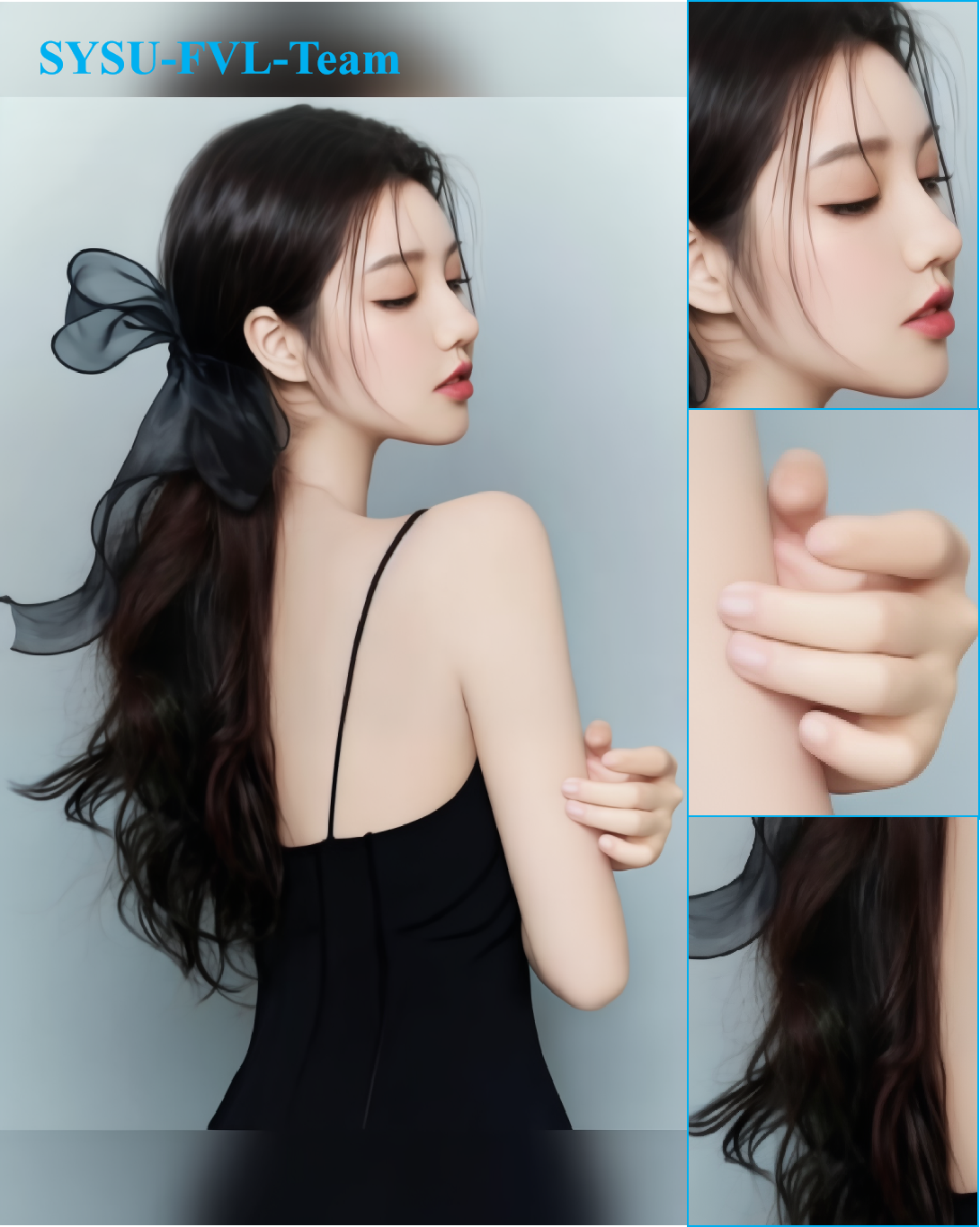}
  \end{subfigure}
  \begin{subfigure}[b]{0.265\textwidth}
    \includegraphics[height=5.8cm]{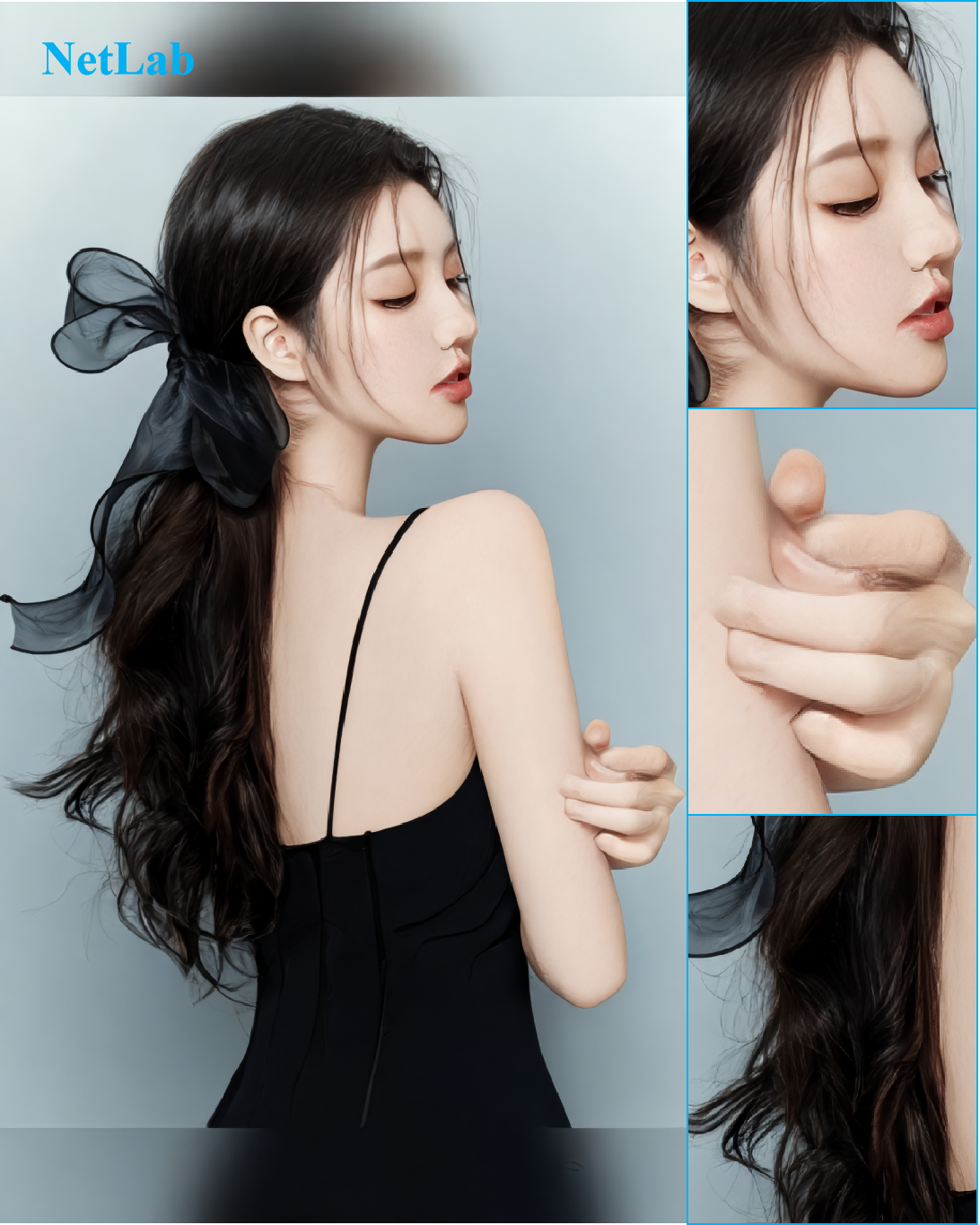}
  \end{subfigure}
  \begin{subfigure}[b]{0.265\textwidth}
    \includegraphics[height=5.8cm]{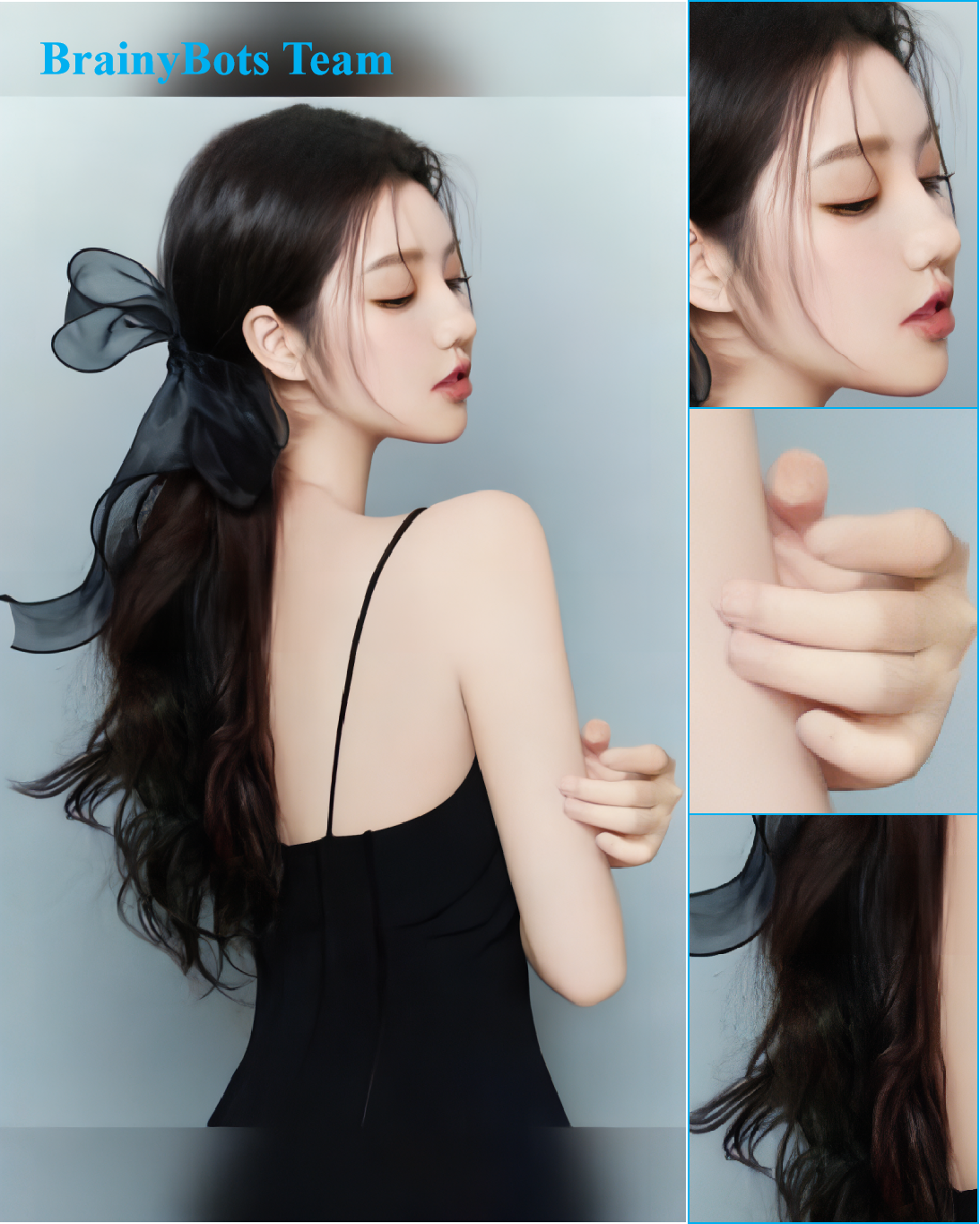}
  \end{subfigure}
  \caption{A comparison of the subjective quality between six teams on the synthetic dataset part: Example 1.}
  \label{label:s1}
\end{figure*}

\begin{figure*}[htbp]
  \centering
  \begin{subfigure}[b]{0.180\textwidth}
    \includegraphics[height=5.6cm]{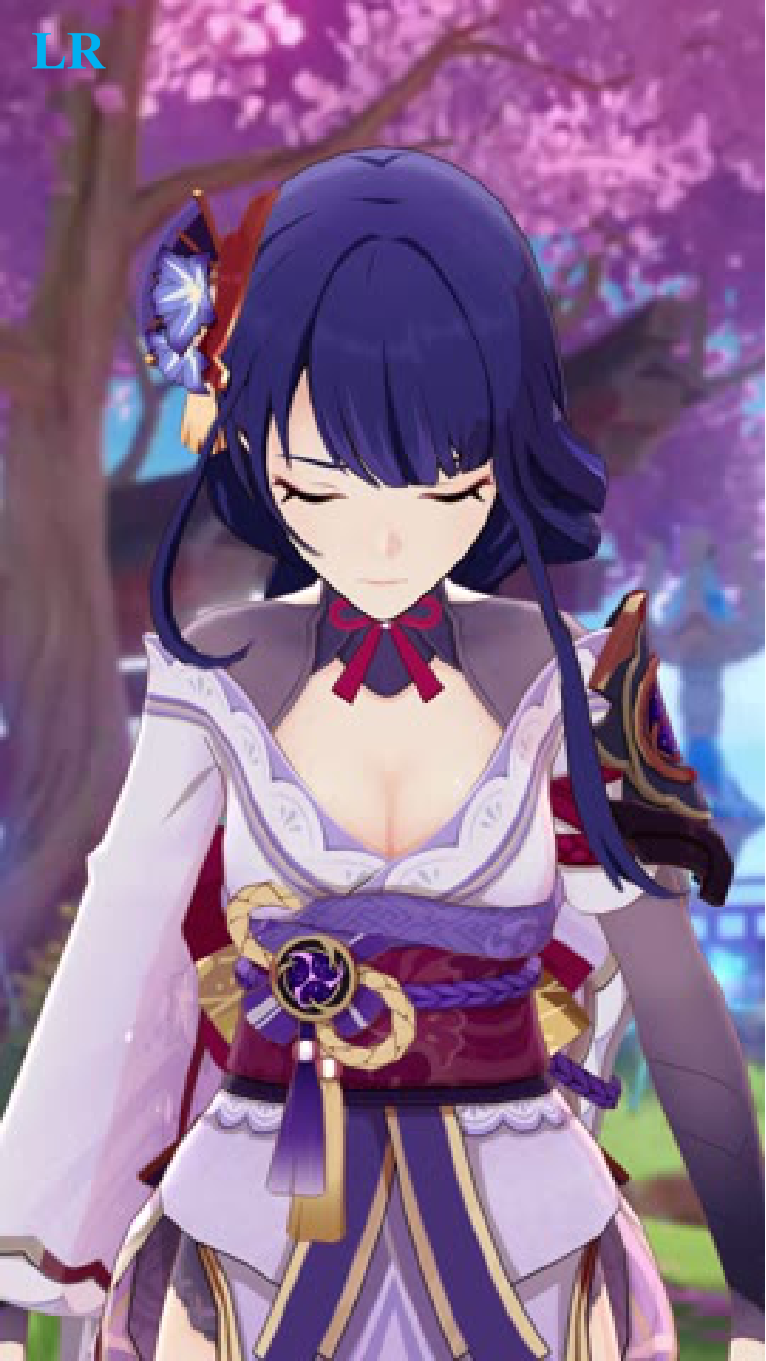}
  \end{subfigure}
  \begin{subfigure}[b]{0.265\textwidth}
    \includegraphics[height=5.6cm]{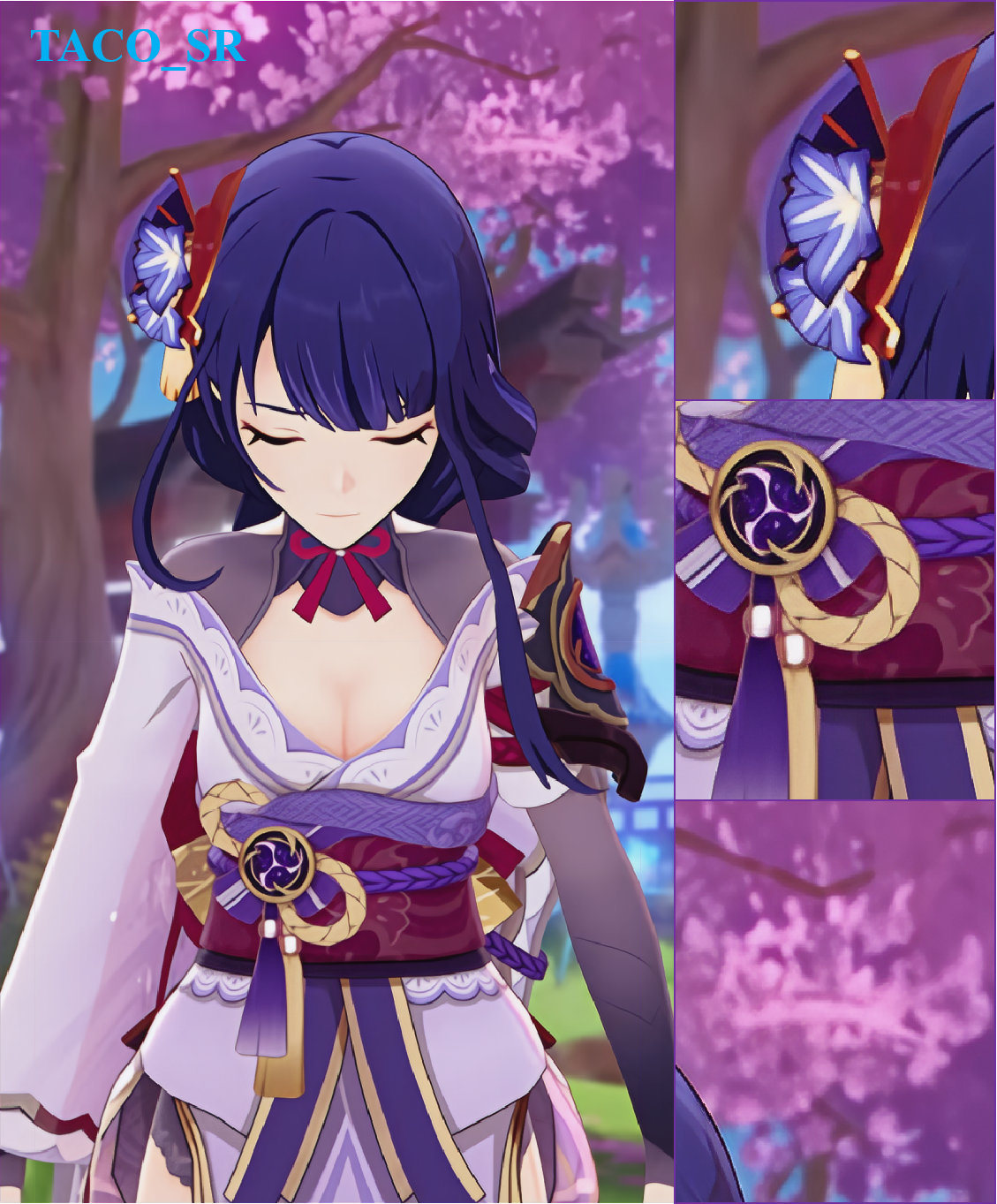}
  \end{subfigure}
  \begin{subfigure}[b]{0.265\textwidth}
    \includegraphics[height=5.6cm]{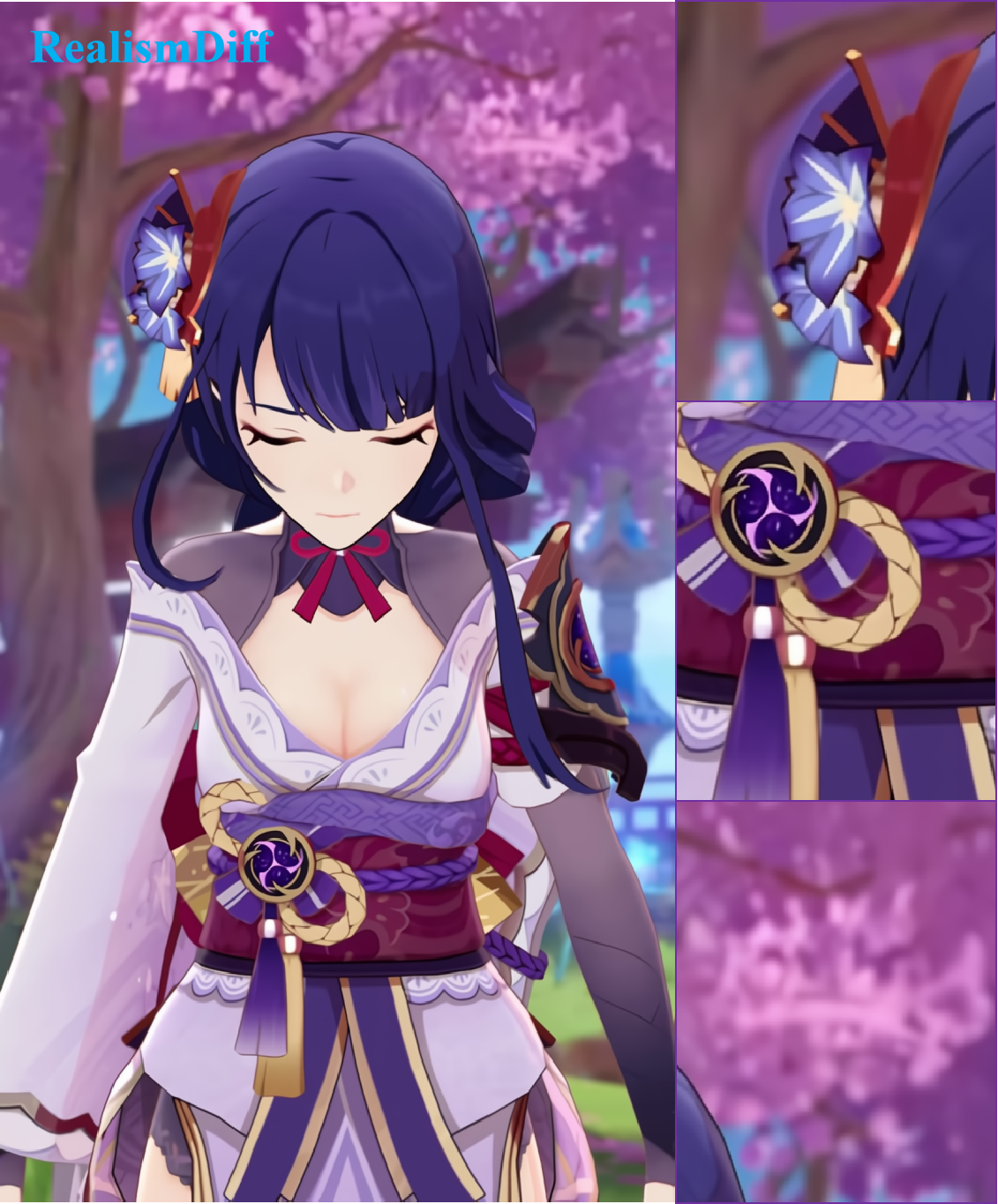}
  \end{subfigure}
  \begin{subfigure}[b]{0.265\textwidth}
    \includegraphics[height=5.6cm]{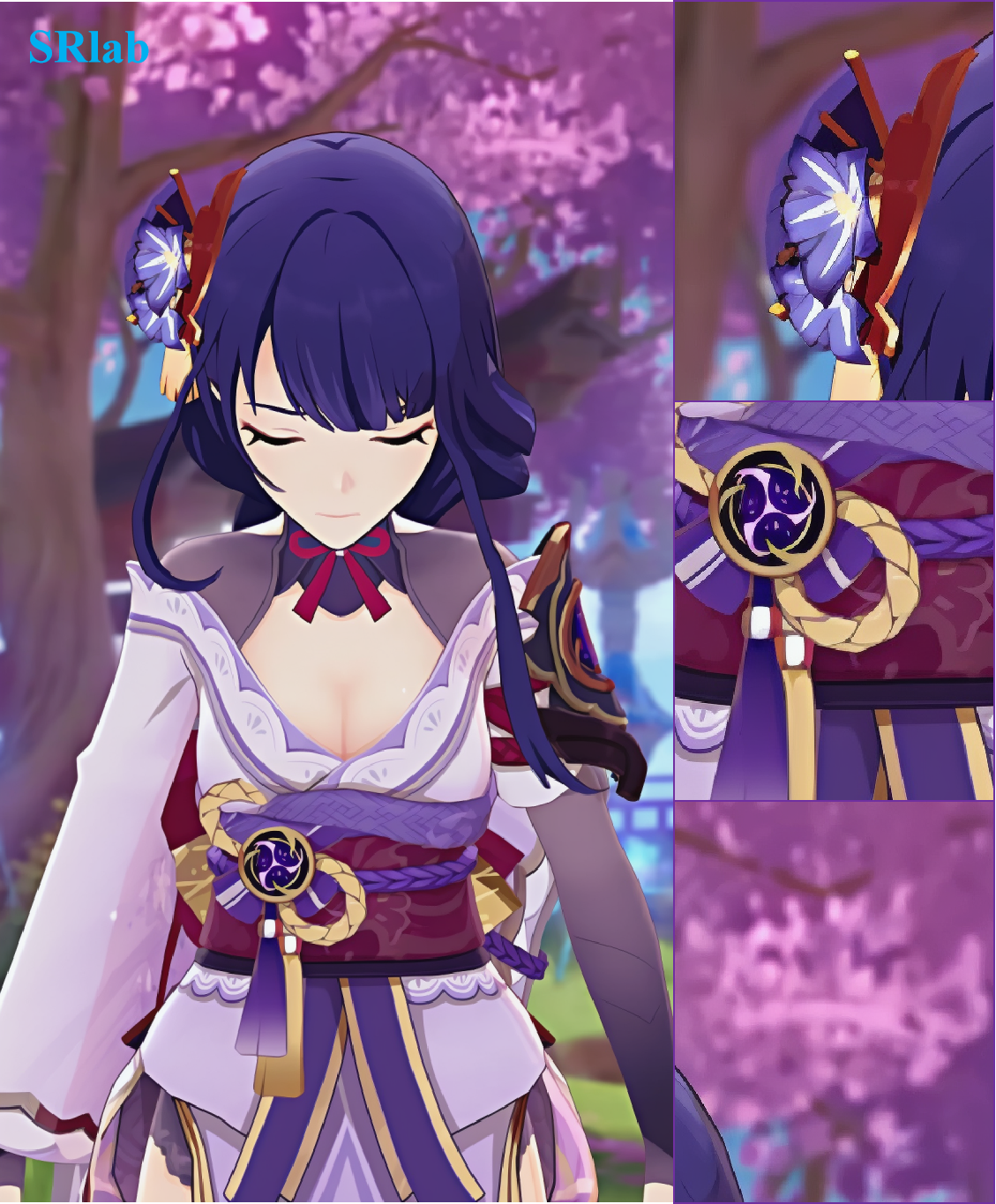}
  \end{subfigure}

  \vspace{0.2em} % 行间距
\begin{subfigure}[b]{0.180\textwidth}
    \includegraphics[height=5.6cm]{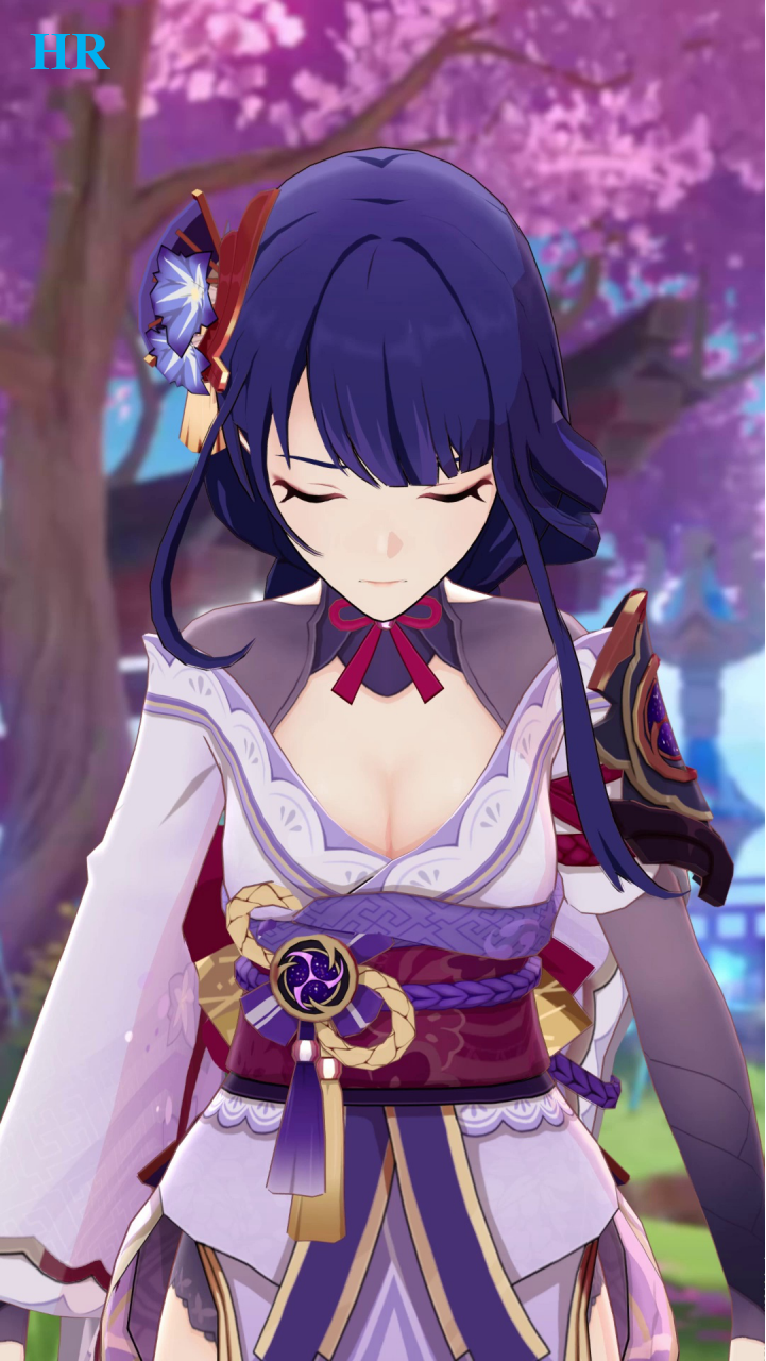}
  \end{subfigure}
  \begin{subfigure}[b]{0.265\textwidth}
    \includegraphics[height=5.6cm]{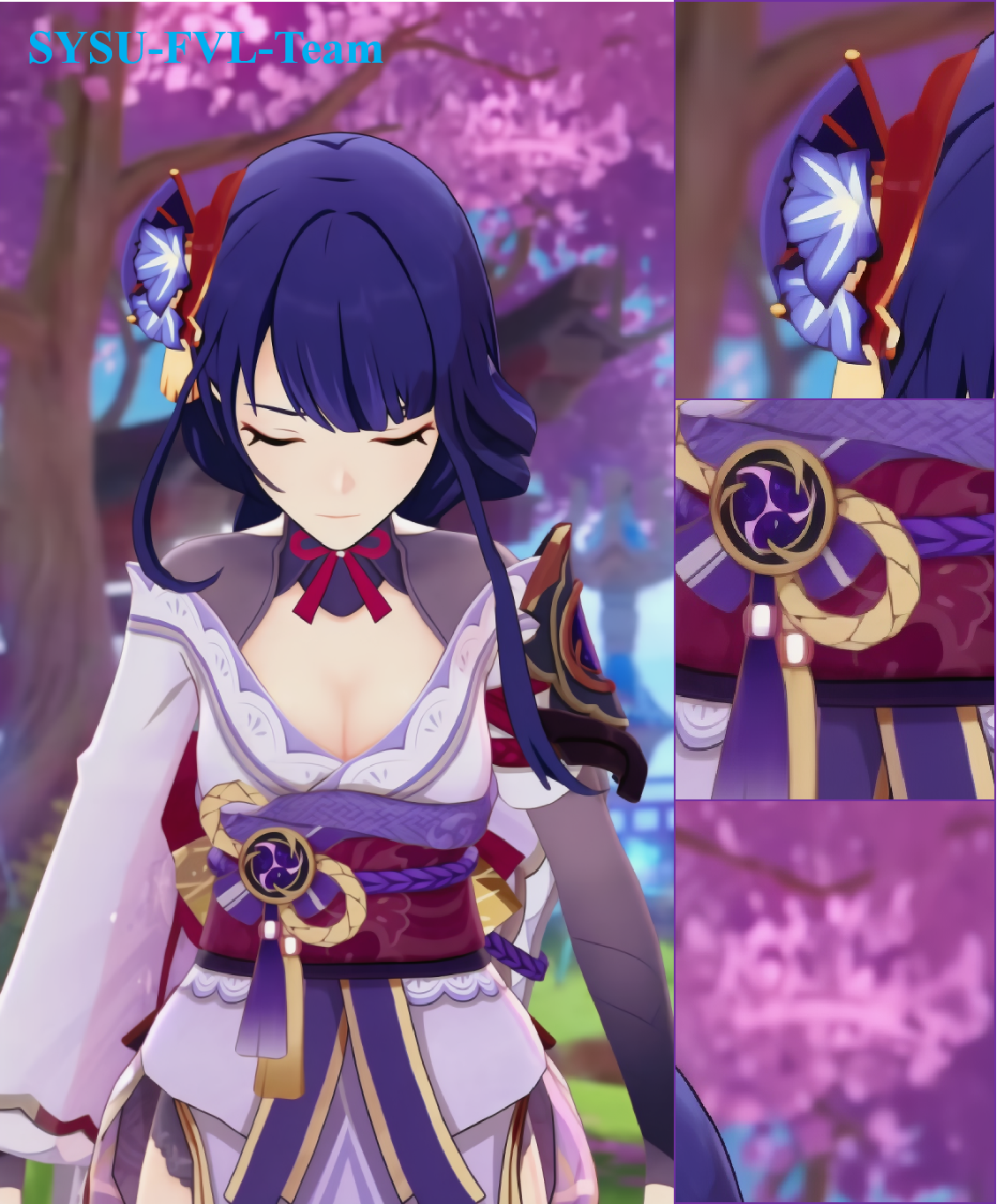}
  \end{subfigure}
  \begin{subfigure}[b]{0.265\textwidth}
    \includegraphics[height=5.6cm]{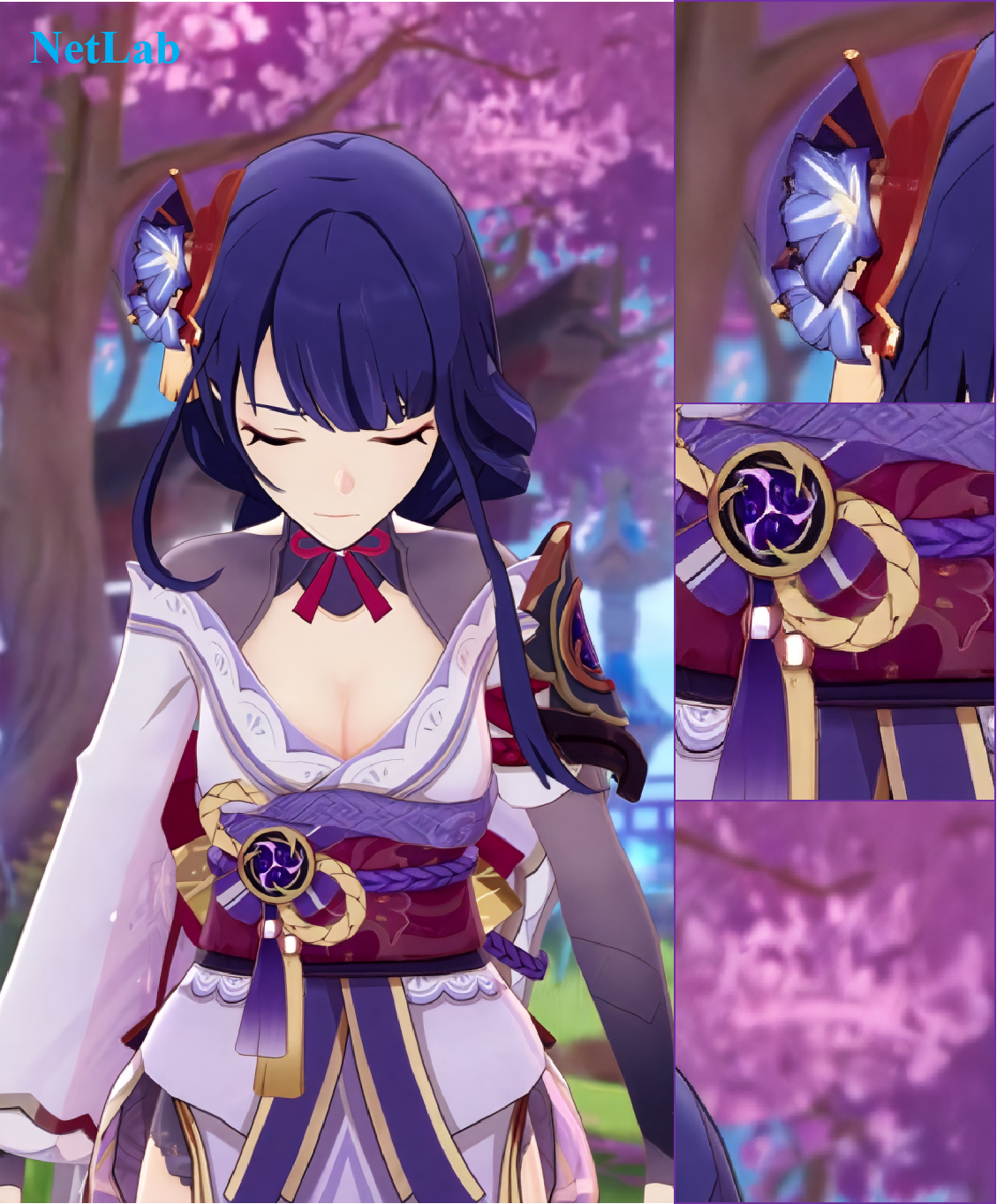}
  \end{subfigure}
  \begin{subfigure}[b]{0.265\textwidth}
    \includegraphics[height=5.6cm]{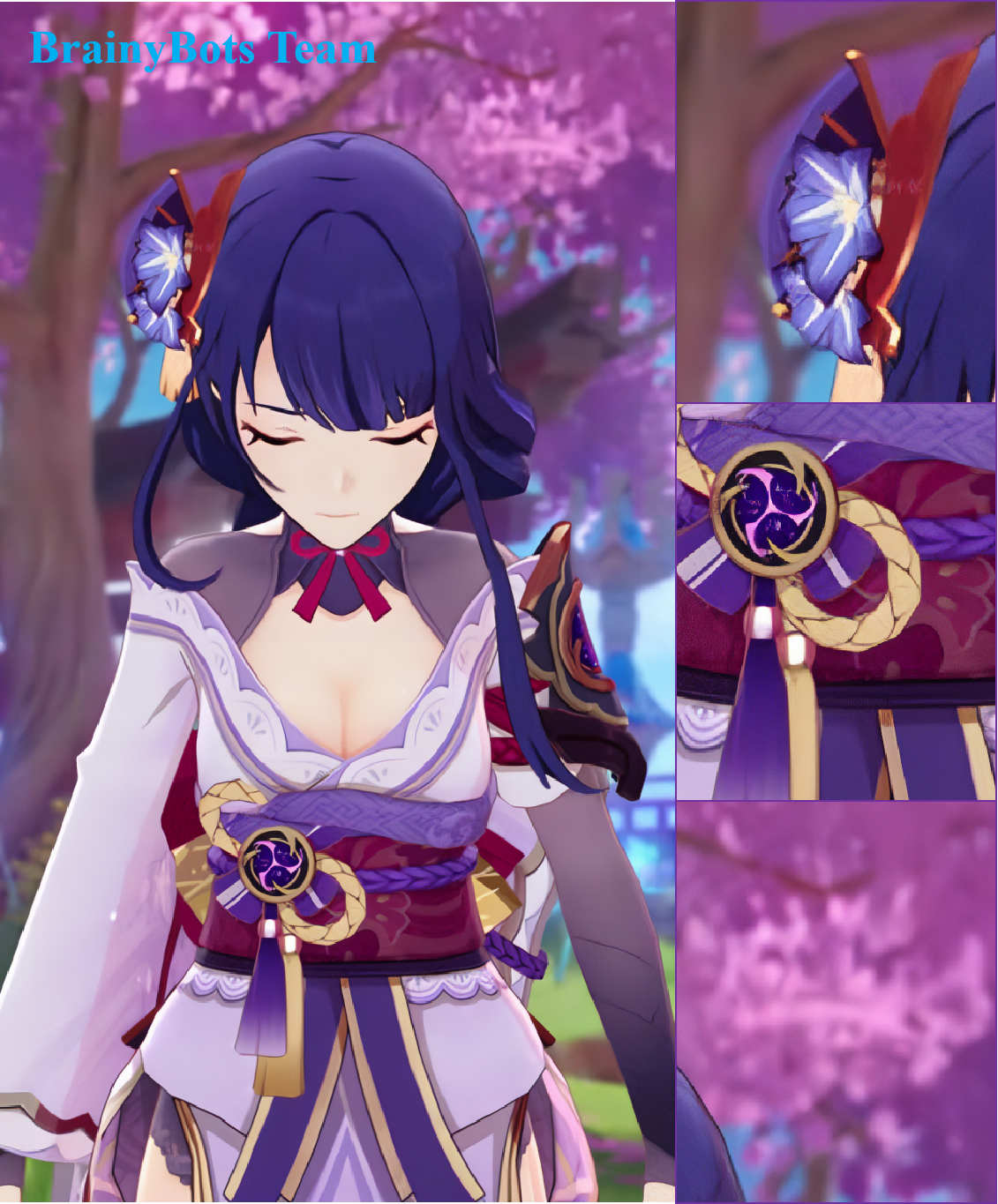}
  \end{subfigure}
  \caption{A comparison of the subjective quality between six teams on the synthetic dataset part: Example 2.}
   \label{label:s2}
\end{figure*}

\begin{figure*}[htbp]
  \centering
  \begin{subfigure}[b]{0.180\textwidth}
    \includegraphics[height=5.7cm]{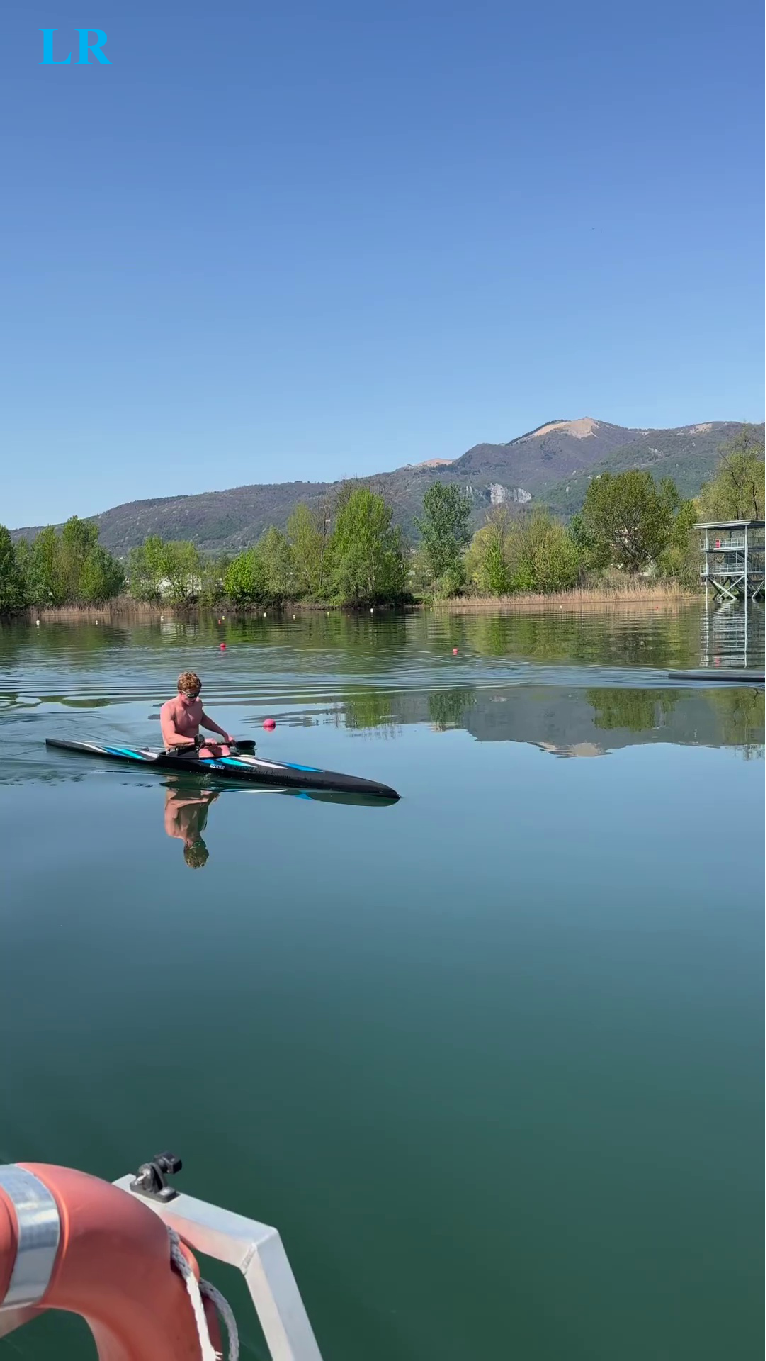}
  \end{subfigure}
  \begin{subfigure}[b]{0.26\textwidth}
    \includegraphics[height=5.7cm]{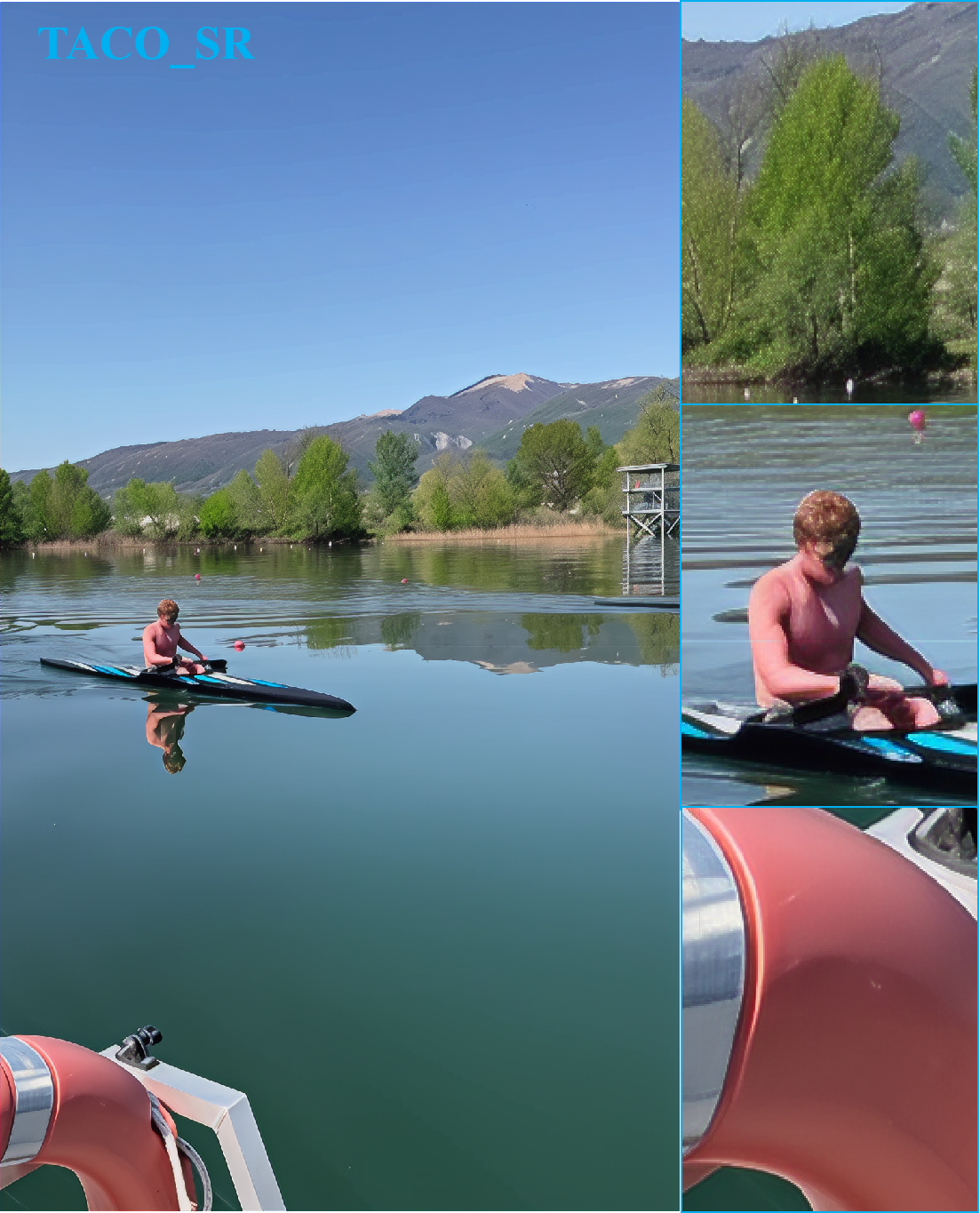}
  \end{subfigure}
  \begin{subfigure}[b]{0.26\textwidth}
    \includegraphics[height=5.7cm]{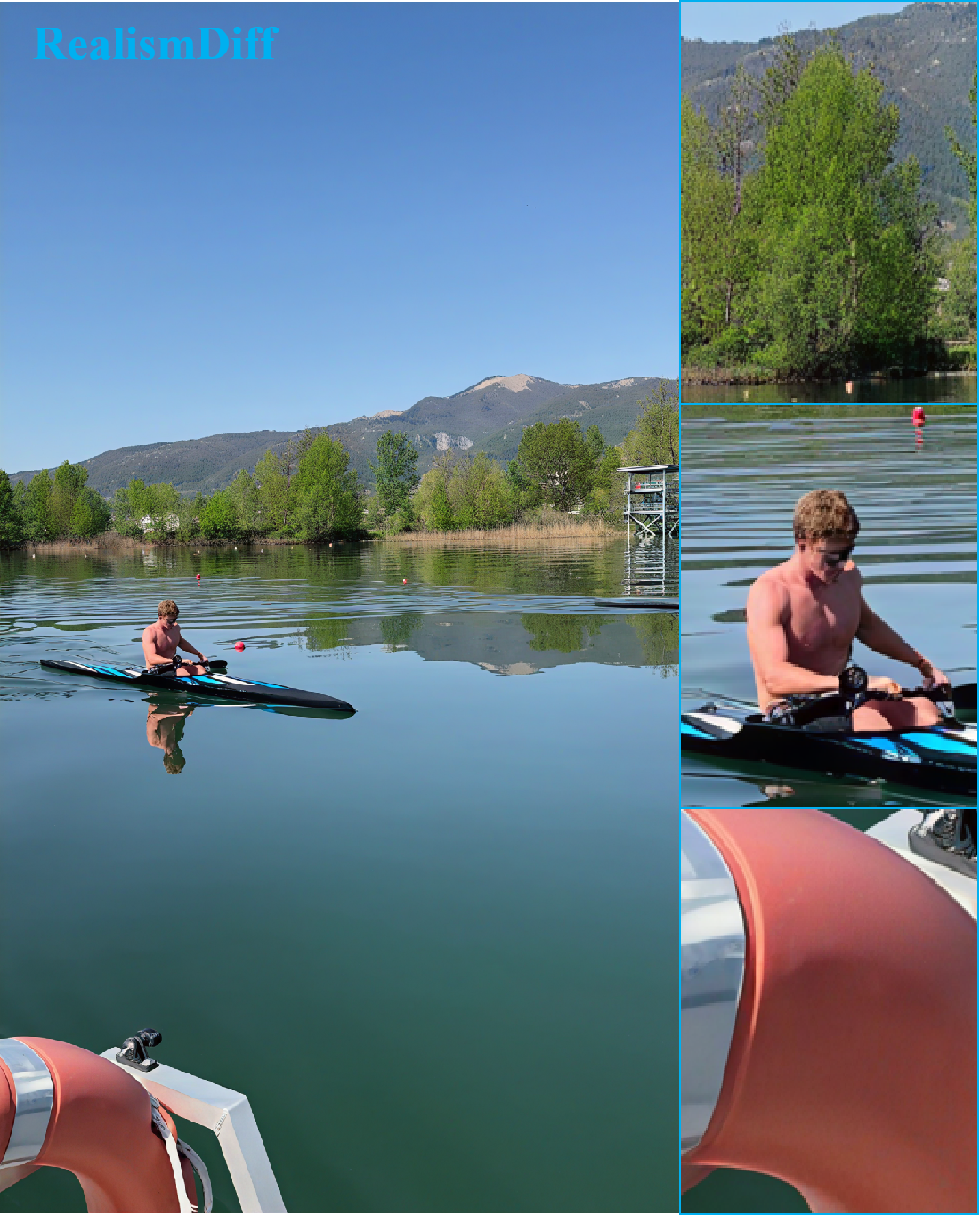}
  \end{subfigure}
  \begin{subfigure}[b]{0.26\textwidth}
    \includegraphics[height=5.7cm]{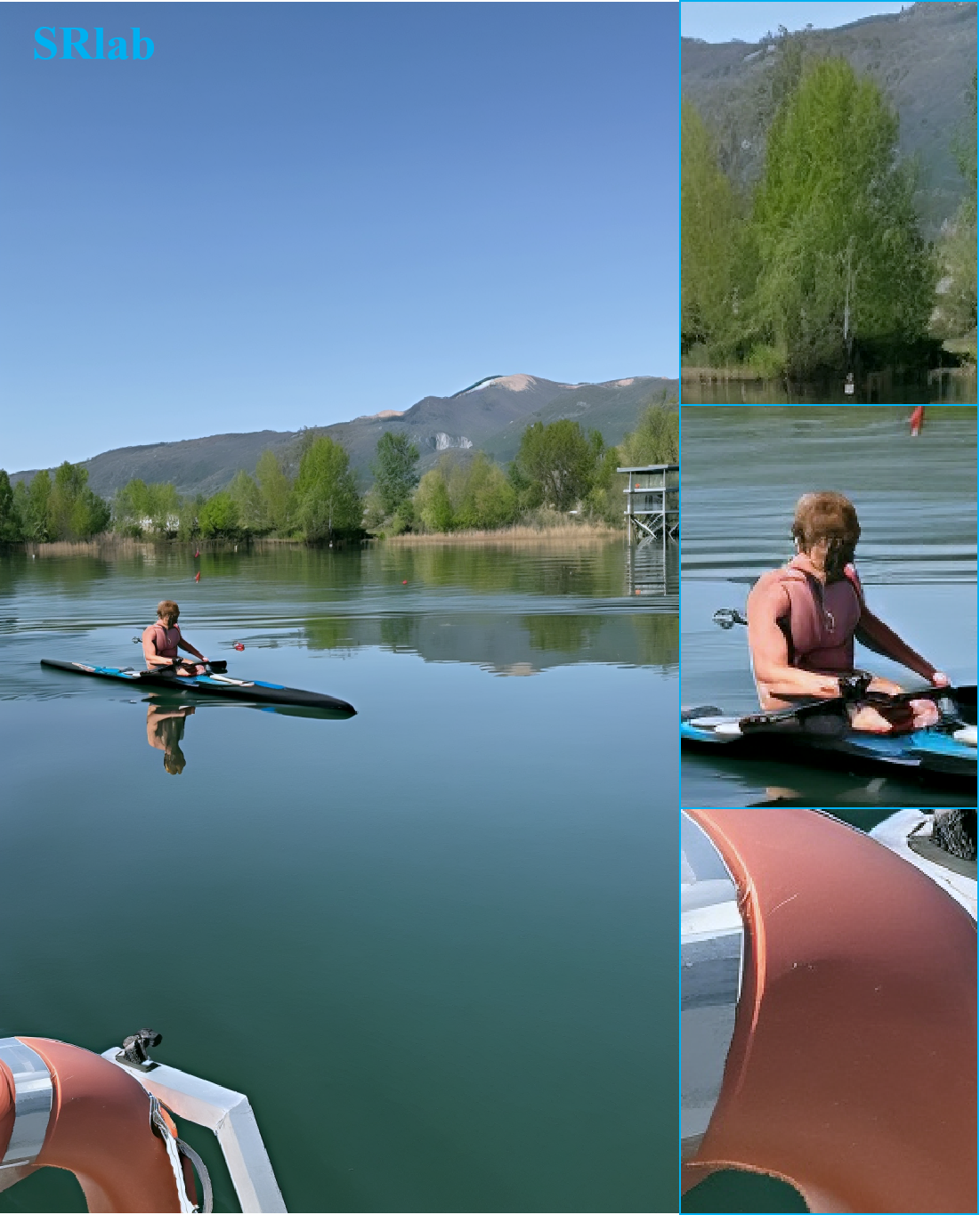}
  \end{subfigure}

  \vspace{0.2em} % 行间距
\begin{subfigure}[b]{0.180\textwidth}
    \includegraphics[height=5.7cm]{compare-figs/3-LR.png}
  \end{subfigure}
  \begin{subfigure}[b]{0.26\textwidth}
    \includegraphics[height=5.7cm]{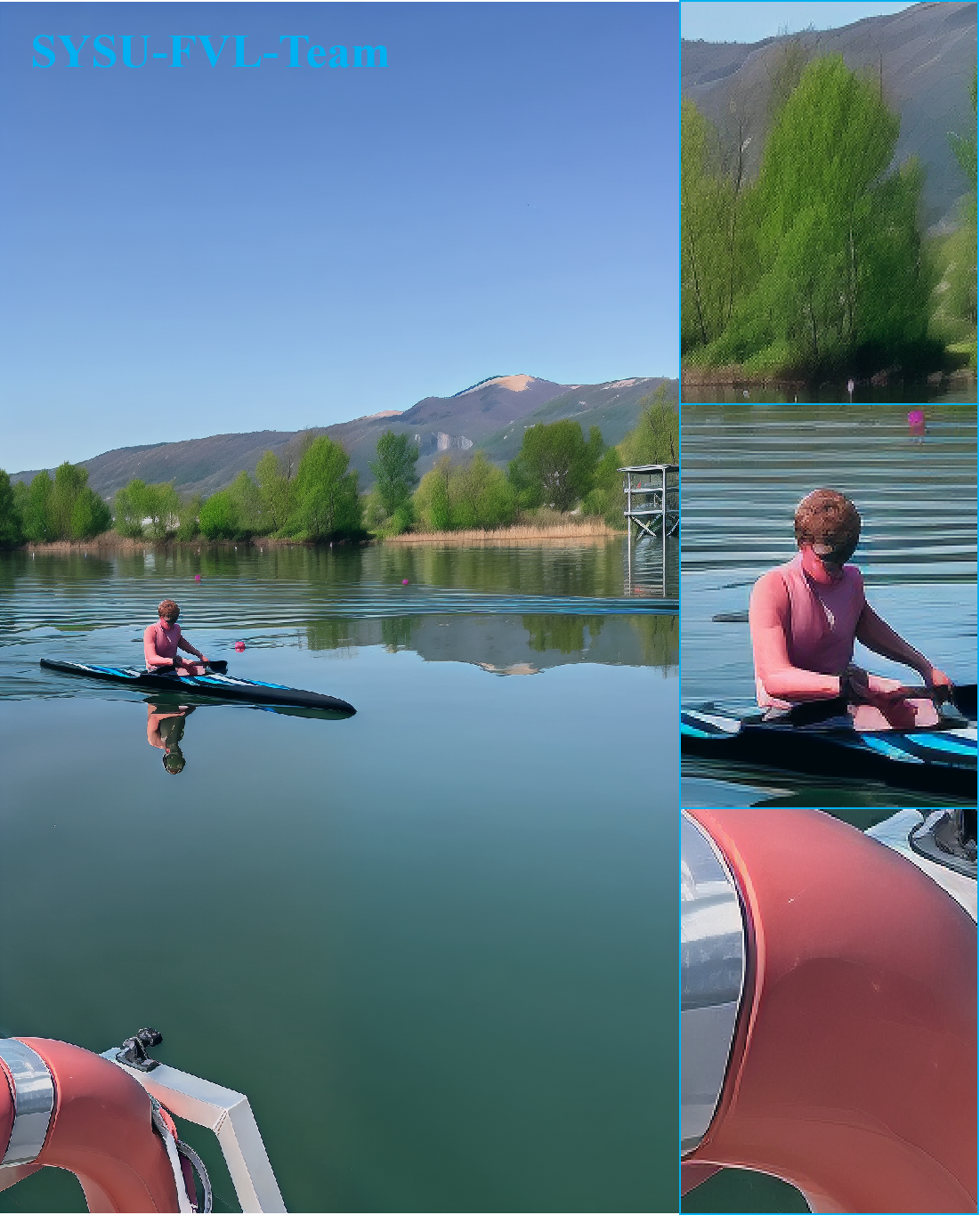}
  \end{subfigure}
  \begin{subfigure}[b]{0.26\textwidth}
    \includegraphics[height=5.7cm]{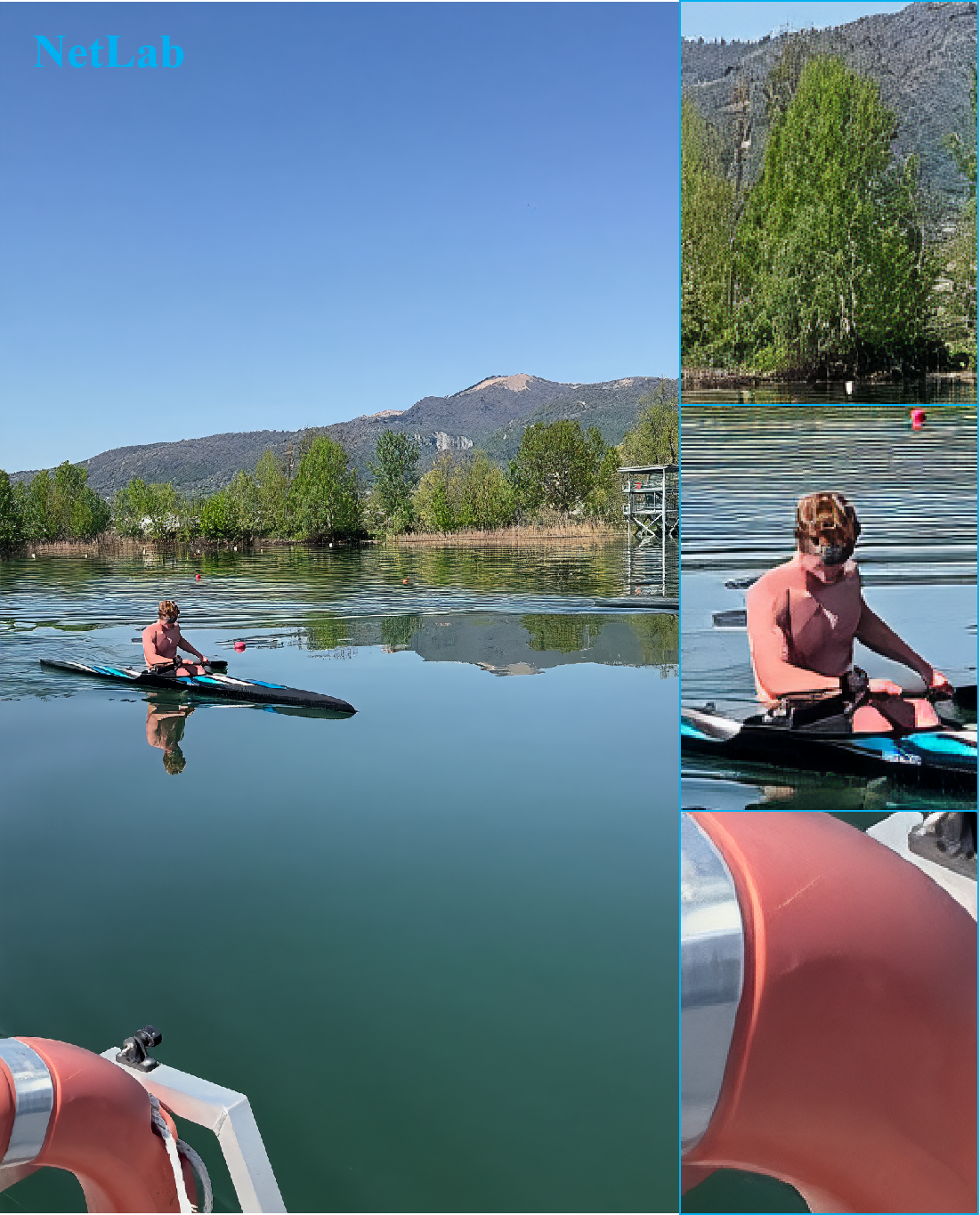}
  \end{subfigure}
  \begin{subfigure}[b]{0.26\textwidth}
    \includegraphics[height=5.7cm]{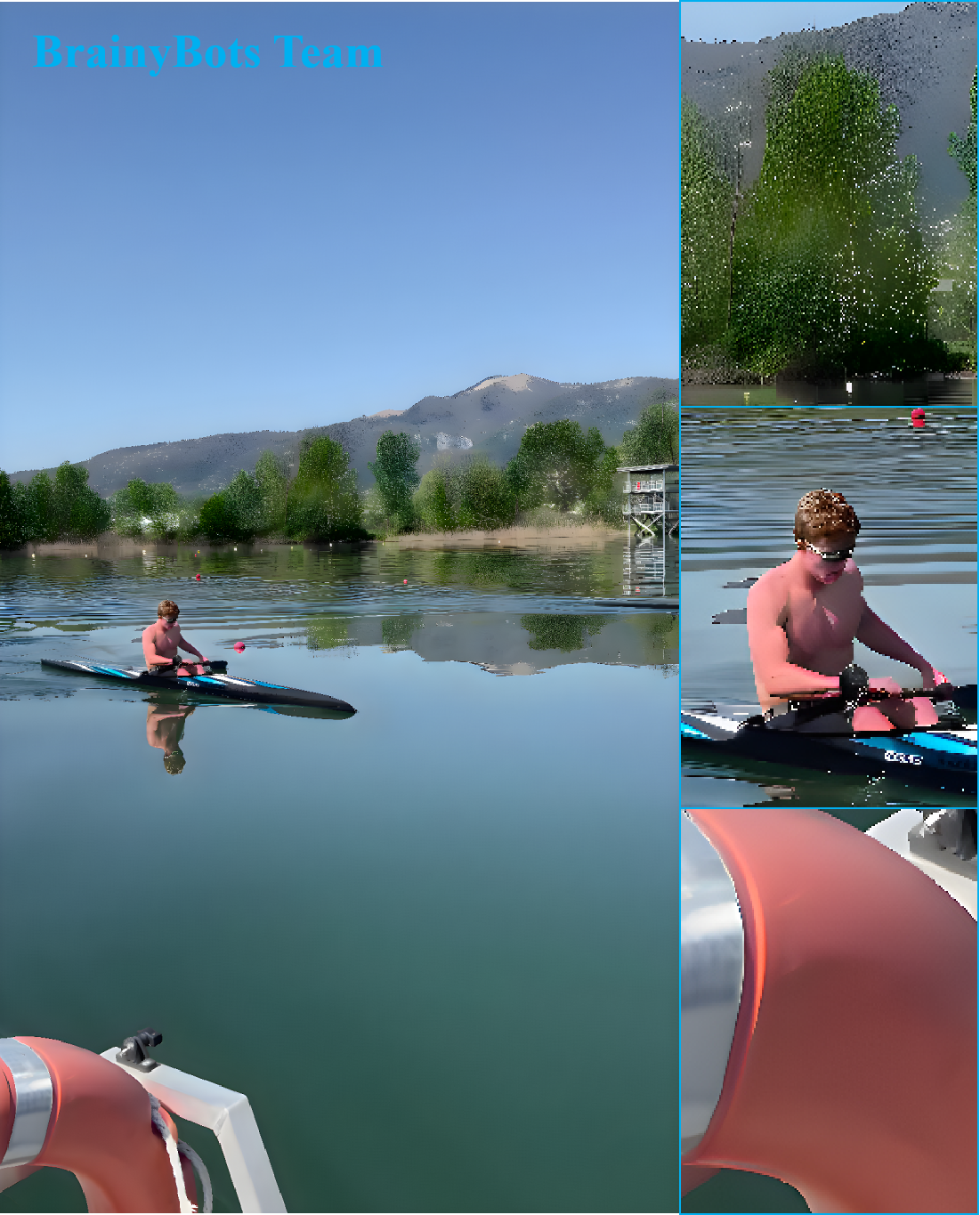}
  \end{subfigure}
  \caption{A comparison of the subjective quality between six teams on the wild dataset part: Example 1.}
   \label{label:w1}
\end{figure*}

\begin{figure*}[htbp]
  \centering
  \begin{subfigure}[b]{0.1765\textwidth}
    \includegraphics[height=5.5cm]{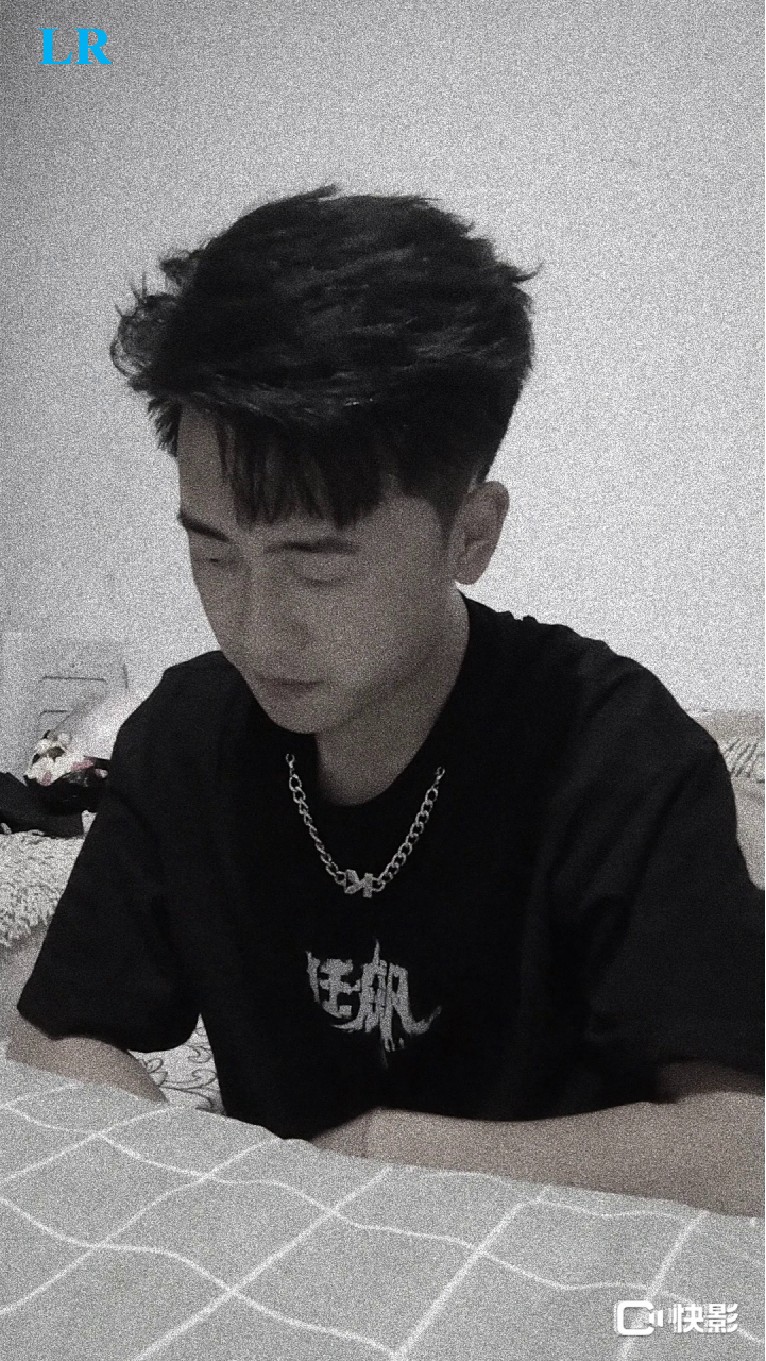}
  \end{subfigure}
  \begin{subfigure}[b]{0.26\textwidth}
    \includegraphics[height=5.5cm]{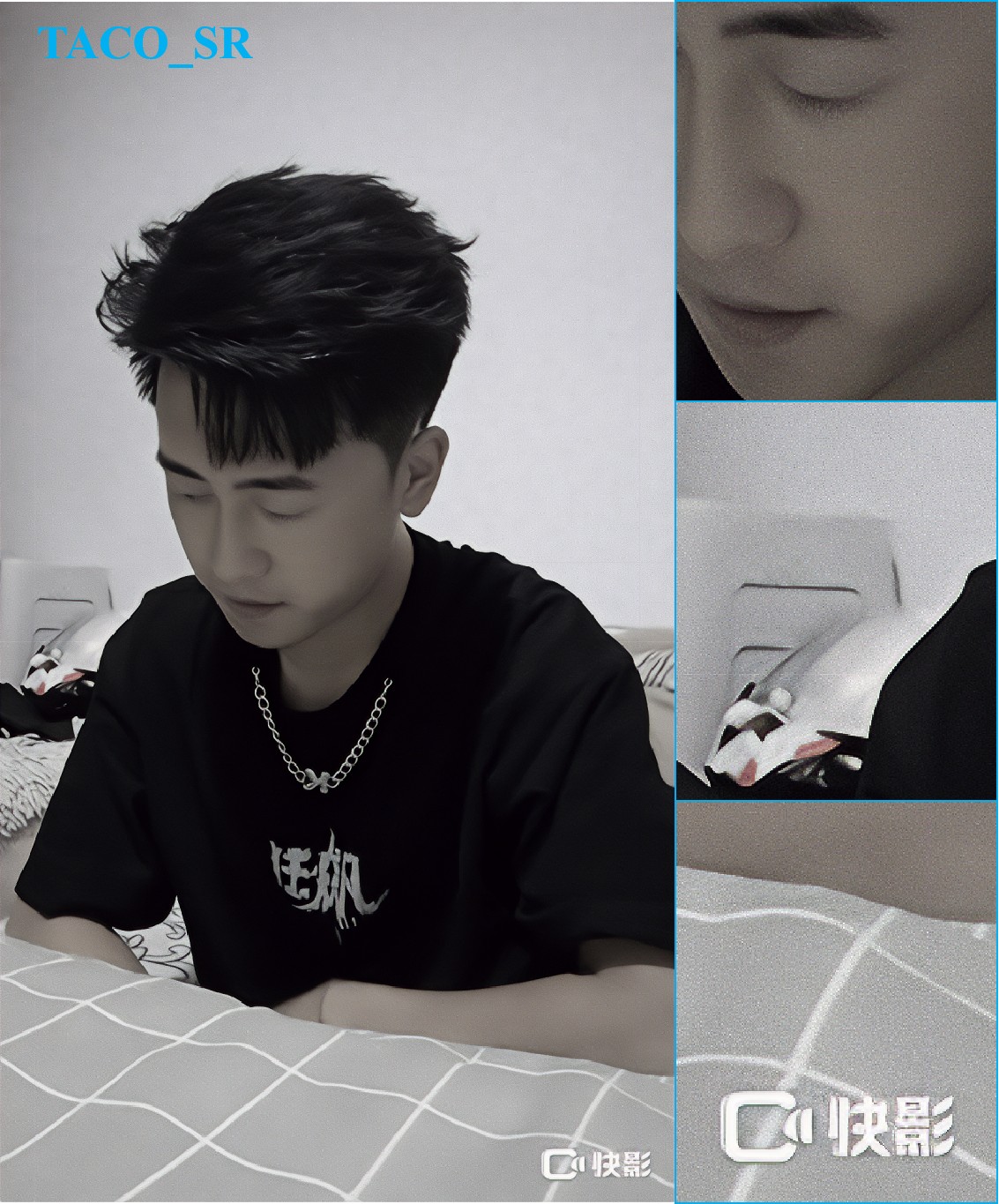}
  \end{subfigure}
  \begin{subfigure}[b]{0.26\textwidth}
    \includegraphics[height=5.5cm]{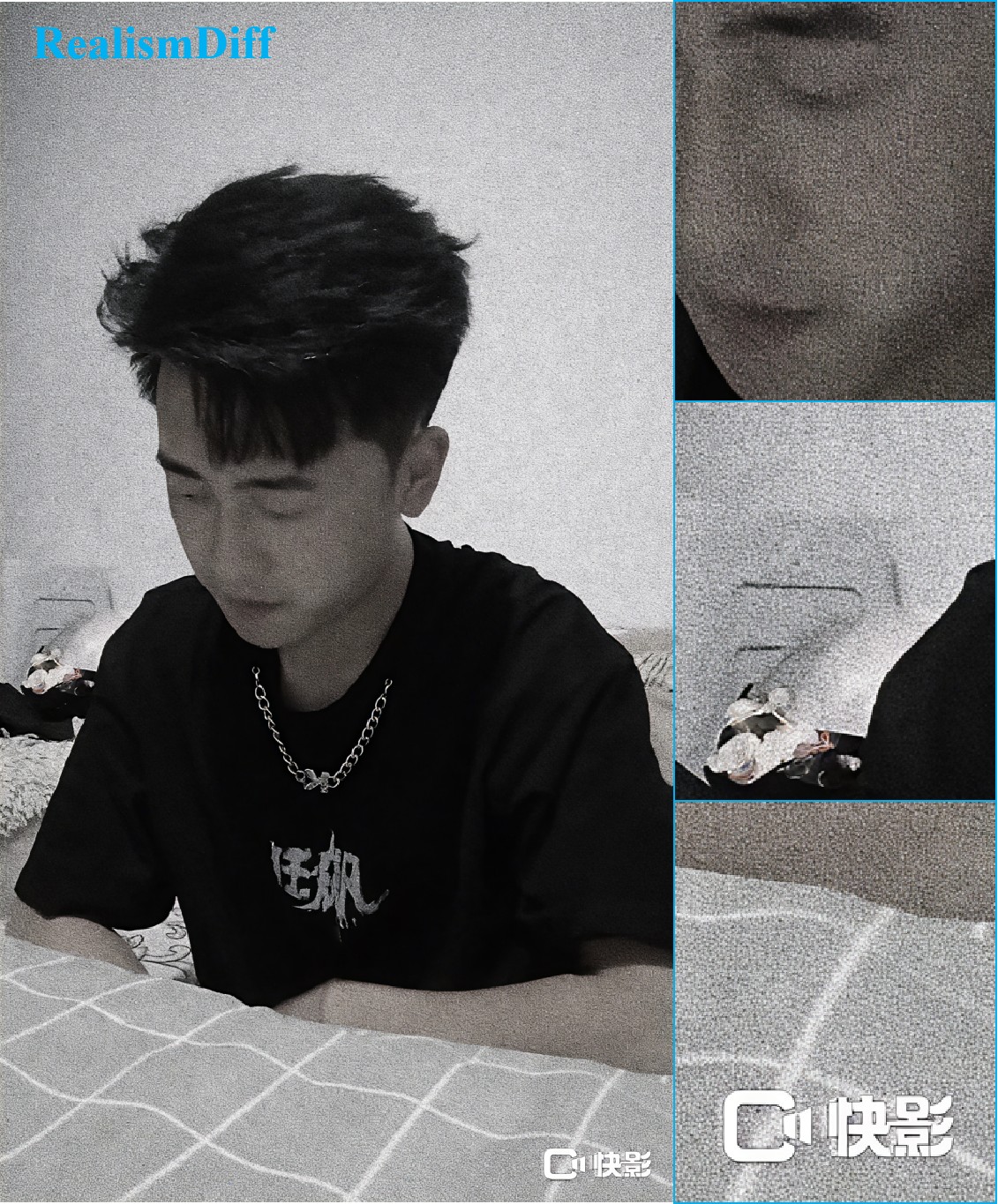}
  \end{subfigure}
  \begin{subfigure}[b]{0.26\textwidth}
    \includegraphics[height=5.5cm]{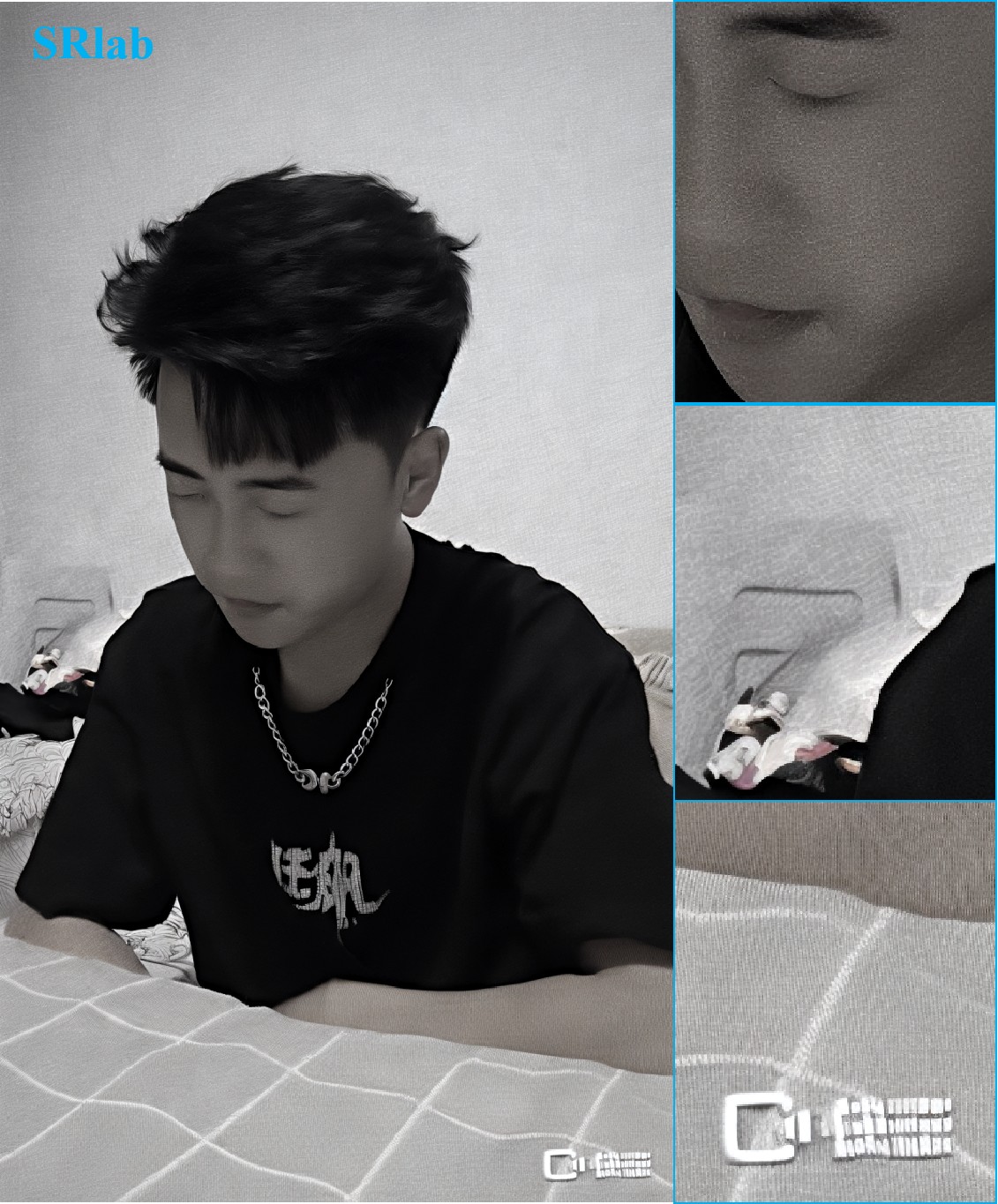}
  \end{subfigure}

  \vspace{0.2em} % 行间距
\begin{subfigure}[b]{0.1765\textwidth}
    \includegraphics[height=5.5cm]{compare-figs/8-LR.jpg}
  \end{subfigure}
  \begin{subfigure}[b]{0.26\textwidth}
    \includegraphics[height=5.5cm]{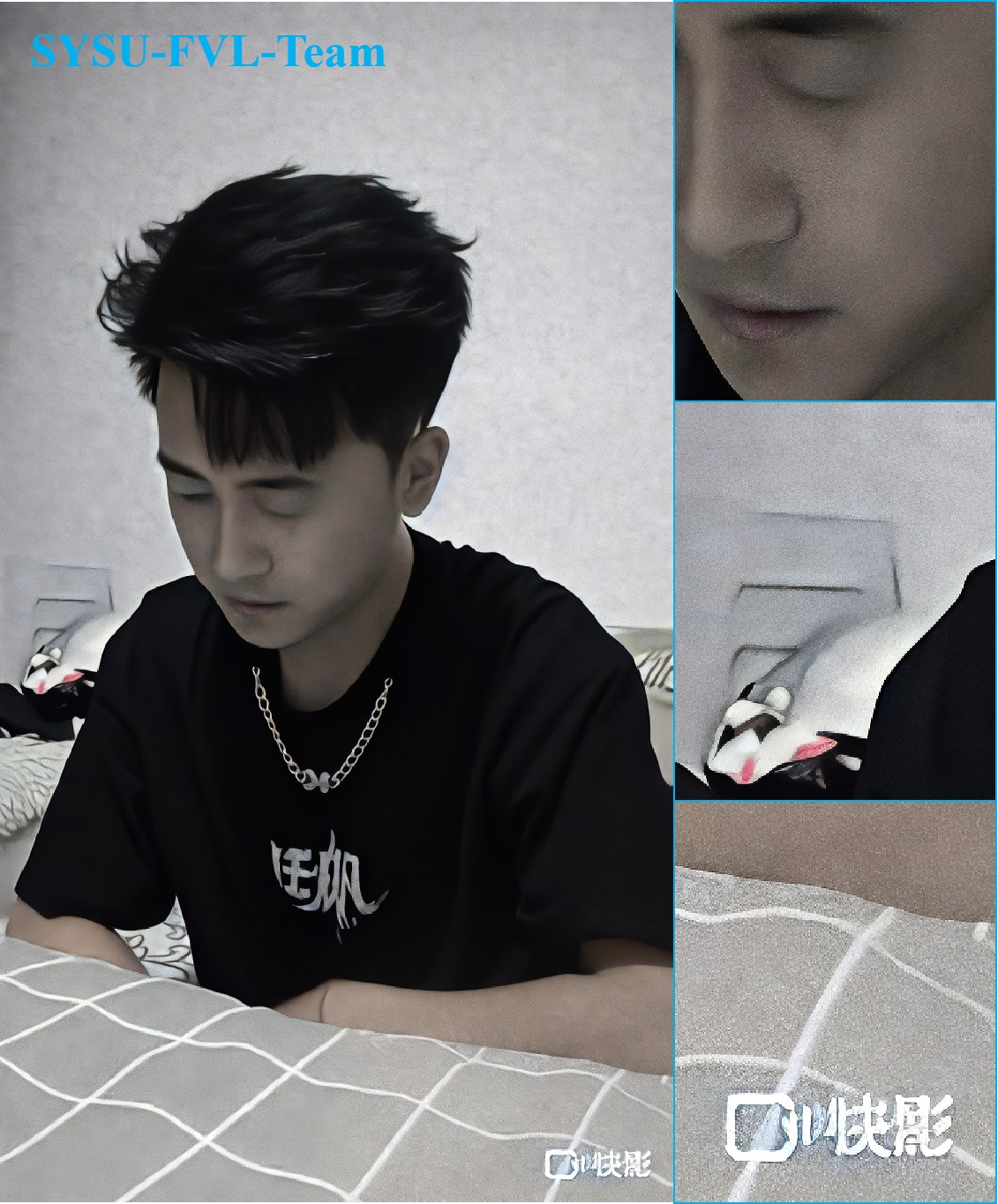}
  \end{subfigure}
  \begin{subfigure}[b]{0.26\textwidth}
    \includegraphics[height=5.5cm]{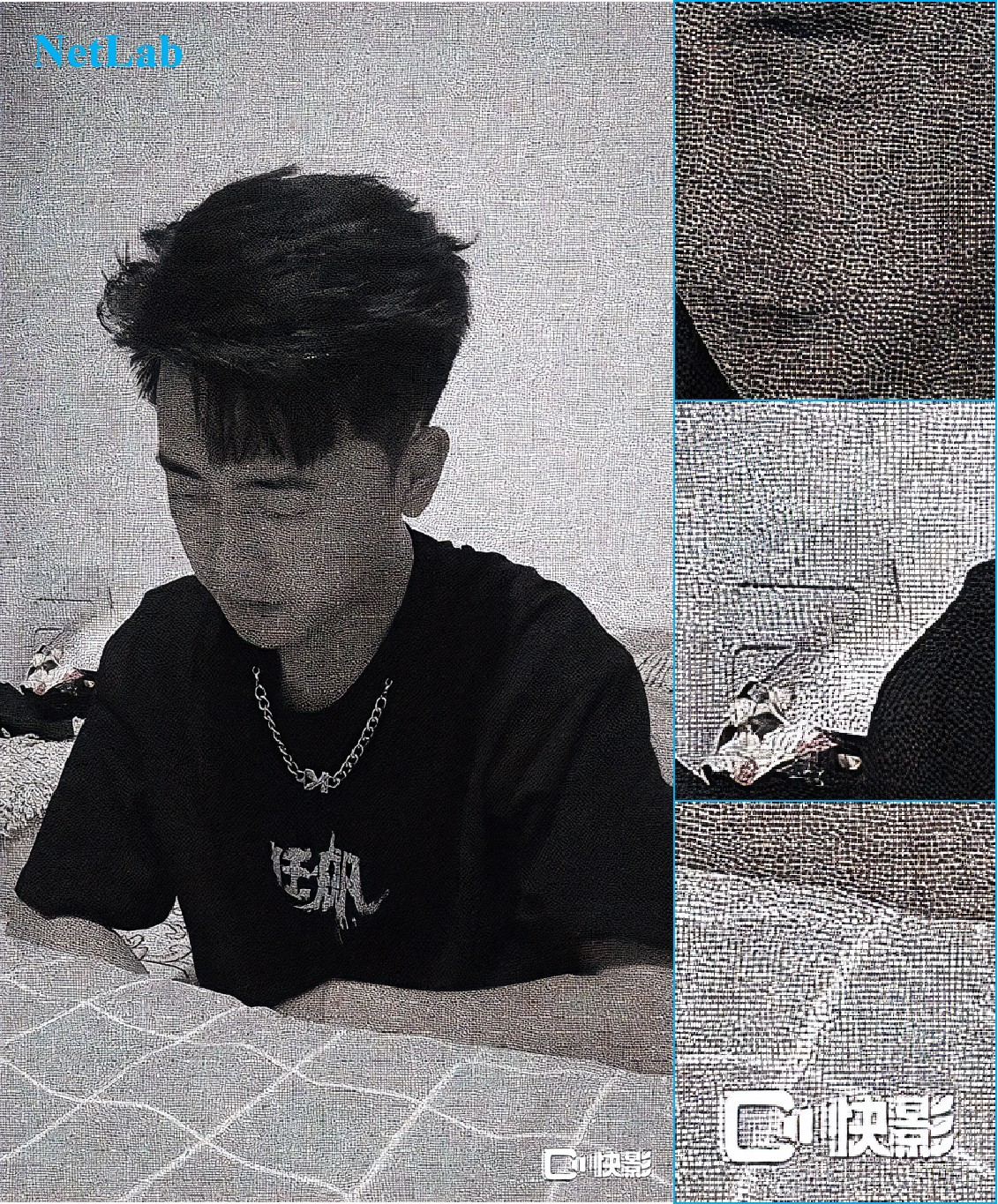}
  \end{subfigure}
  \begin{subfigure}[b]{0.26\textwidth}
    \includegraphics[height=5.5cm]{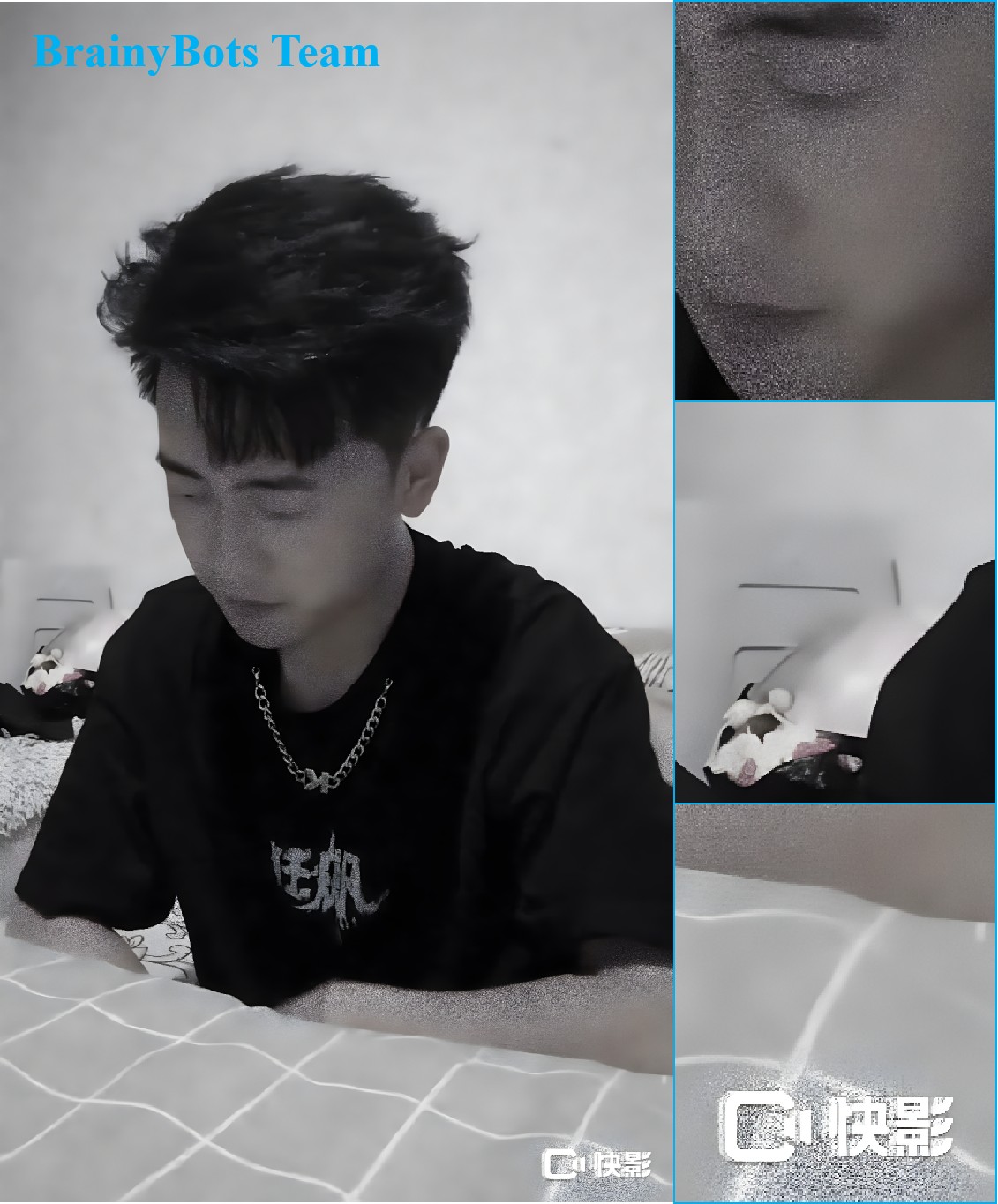}
  \end{subfigure}
  \caption{A comparison of the subjective quality between six teams on the wild dataset part: Example 2.}
   \label{label:w2}
\end{figure*}

For the second track, we focus on the subjective comparison among different teams. After the testing stage, the top six teams were shortlisted based on objective performance metrics as candidates for the user study. We then organize five professional image processing experts to assess the perceptual quality and realism of the super-resolved S-UGC images. Each expert spent approximately eight hours meticulously comparing image details and selecting the most visually convincing and realistic result among the six teams.
As shown in Table~\ref{tab:results_2}, the winning rates of user study for both synthetic and wild datasets are presented in the “User Study” part. Notably, TACO\_SR and RealismDiff demonstrated consistent performance in terms of subjective quality on wild and synthetic datasets. SRLab achieved great performance on synthetic datasets, whereas SYSU-FVL-Team excelled in the wild scenario. Additionally, SYSU-FVL-Team attained the best objective performance, while BrainyBots Team ranked third in objective quality.
Interestingly, the results highlight a noticeable inconsistency between subjective preferences and objective metrics, suggesting that current perceptual metrics may not reliably reflect perceptual quality in generative model-based S-UGC image super-resolution. The subjective comparison between the top six teams can be found in Figs.~\ref{label:s1},~\ref{label:s2},~\ref{label:w1} and \ref{label:w2}.

\begin{table*}[]
\resizebox{\textwidth}{!}{
\begin{tabular}{c|c|c|c|cccc|ccc}
\hline
Rank & Team name          & Team leader            & Final Score & SROCC & PLCC  & Rank1 & Rank2 & GFLOPs & Params (M) \\ \hline
1    & SharpMind          & Yabin Zhang            & 0.922       & 0.934 & 0.933 & 0.788 & 0.846 & 47.39G   & 33.01 \\
2    & ZQE                & Yufan Liu              & 0.916       & 0.930 & 0.933 & 0.732 & 0.817 & 95.75G   & 150.11 \\
3*   & ZX-AIE-Vector      & Yunchen Zhang          & 0.912       & 0.926 & 0.927 & 0.775 & 0.787 & 100.3G   & 98.76 \\
3    & ECNU-SJTU VQA Team & Wei Sun                & 0.910       & 0.926 & 0.924 & 0.736 & 0.817 & 114.83G  & 37.90 \\
5    & TenVQA             & Yuhai Lan              & 0.900       & 0.914 & 0.915 & 0.745 & 0.775 & 118G     & 28.00 \\
6    & GoldenChef         & MingYin Bai            & 0.871       & 0.881 & 0.886 & 0.693 & 0.817 & 119G     & 154.76 \\
7    & DAIQAM             & Ha Thu Nguyen          & 0.844       & 0.856 & 0.855 & 0.667 & 0.811 & 66.36G   & 27.98 \\
8    & 57VQA              & Zhiye Huang            & 0.264       & 0.220 & 0.239 & 0.541 & 0.598 & --    & -- \\
9    & Nourayn            & Nourine Mohammed Nadir & 0.127       & 0.070 & 0.085 & 0.455 & 0.686 & --       & --    \\ \hline
\end{tabular}
}
\caption{Result of Track 1: Efficient KVQ.}
\label{tab:results_1}
\end{table*}

\begin{table*}[]
\resizebox{\textwidth}{!}{\begin{tabular}{c|c|c|c|cccccc|c|c}
\hline
 Team name          & Team leader            & User Study & \makecell{Score\\ (objective)} & PSNR  & SSIM & LPIPS & MUSIQ &   ManIQA &  CLIPIQA & \makecell{Ranking\\(User Study)}  & \makecell{Ranking \\(Objective)}   \\ \hline
 TACO\_SR          &    Yushen Zuo            &   0.2775/0.3529 & 51.1168 &	27.5625&	0.7877 &	0.2232 &	65.7152 &	0.4489	& 0.6849 & 1 & 6 \\
 RealismDiff                & Kexin Zhang             & 0.2640/0.2834  & 51.6289 &	28.3801 &	0.7979&	0.2150&	68.7706&	0.4600 &	0.5943 & 2 & 5 \\
SRlab      & Ying Liang          &  0.2492/0.0932     & 53.4643 &	27.4647 &0.7890 &	0.2121 &	71.1964 &	0.5532&	0.7579 & 3 & 2\\
SYSU-FVL-Team & Zhi Jin                & 0.0733/0.1540        & 54.1403  & 	28.2410  & 	0.8069 & 	0.2366  & 	70.5872  & 	0.5451  & 	0.7687 & 4 & 1\\
NetLab            & Hengyuan Na              & 0.0751/0.0947         & 52.5092 &	27.5176 &	0.7862 &	0.2105 &	68.3565 &	0.4911 &0.74885 & 5 & 4 \\
  BrainyBots Team      & Xinglin Xie            & 0.0609/0.0219     & 53.4161 &	27.9241 &	0.7836 &0.2373 &	64.6089 &	0.6071 &0.7497 & 6 & 3 \\
  BP-SR & Qi Tang & - & 46.4382 &	27.2520 &	0.7744 &	0.2257&	58.9467 &	0.3387 &	0.4418 & - & 7 \\
  NVDTOFCUC & Qingmiao Jiang & - & 44.8025 &	22.3412 &	0.6761 &	0.3589 &	69.7200 & 0.5080 &	0.7238 & - & 8 \\
  BVIVSR & Yuxuan Jiang & - & 44.1724 & 	27.6412 & 	0.7808 & 	0.3857 & 51.8542 & 	0.3146 & 	0.4249&  - & 9 \\\hline
\end{tabular}}
\caption{Result of Track 2: KwaiSR.}
\label{tab:results_2}
\end{table*}

% \begin{table*}[]
% \resizebox{\textwidth}{!}{\begin{tabular}{c|c|cccccccccc}
% \hline
%  Team name          & Team leader            & User Study & Score (objective) & PSNR  & SSIM & LPIPS & MUSIQ & CLIPIQA & ManIQA & \makecell{Ranking\\User Study}  & \makecell{Ranking \\Objective}  \\ \hline
%  TACO\_SR          &    Yushen Zuo            &        & 0.934 & 0.933 & 0.788 & 0.846 \\
%  ZQE                & Yufan Liu              & 0.916       & 0.930  & 0.933 & 0.732 & 0.817 \\
%  ZX-AIE-Vector      & Yunchen Zhang          & 0.912       & 0.926 & 0.927 & 0.775 & 0.787 \\
%  ECNU-SJTU VQA Team & Wei Sun                & 0.910        & 0.926 & 0.924 & 0.736 & 0.817 \\
% TenVQA             & Yuhai Lan              & 0.900         & 0.914 & 0.915 & 0.745 & 0.775 \\
%  GoldenChef         & MingYin Bai            & 0.871       & 0.881 & 0.886 & 0.693 & 0.817 \\
%  DAIQAM             & Ha Thu Nguyen          & 0.844       & 0.856 & 0.855 & 0.667 & 0.811 \\
%  57VQA              & Zhiye Huang            & 0.264       & 0.220  & 0.239 & 0.541 & 0.598 \\
%  Nourayn            & Nourine Mohammed Nadir & 0.127       & 0.070  & 0.085 & 0.455 & 0.686 \\ \hline
% \end{tabular}}
% \caption{Result of Track 1.}
% \label{tab:results_1}
% \end{table*}

\section{Teams and Methods of Track 1}
\label{sec:teams_and_methods_1}

\subsection{SharpMind}
\begin{figure}[t]
  \centering
  \includegraphics[width=1.0\linewidth]{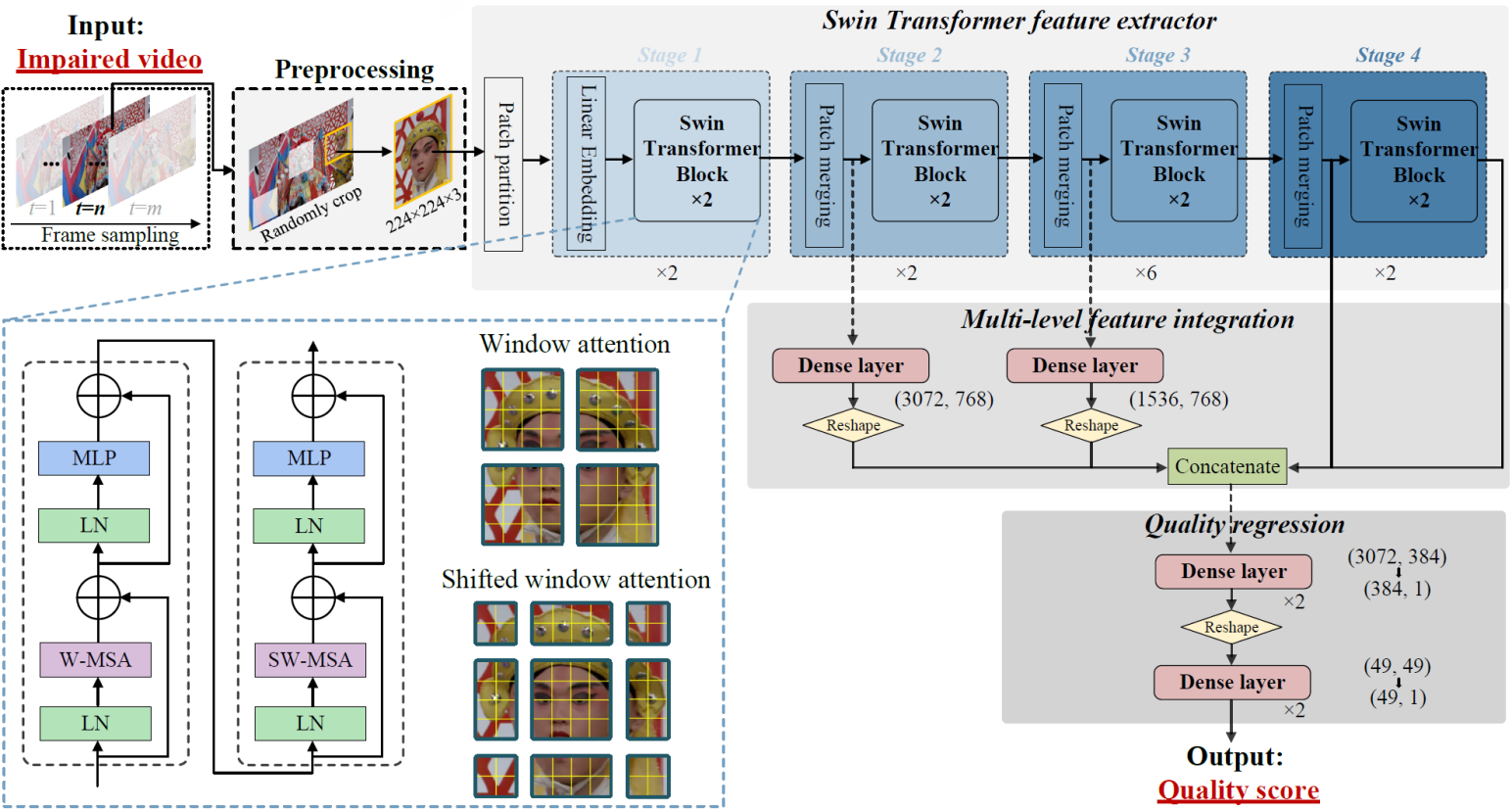} % 替换为你的图像文件名
  % 图注
  % \caption{The student network in the second stage. }
  \caption{The overall framework of Team SharpMind. }
  \label{fig:SharpMind}
\end{figure}
% \subsubsection{Method Description}
This team employs several powerful backbone networks to extract features associated with the human vision system. Subsequently, they train a comprehensive video quality assessment model, which is then utilized as a teacher model to label a series of User-Generated Content (UGC) video datasets. Leveraging these labeled datasets, they train a lightweight student network. As a result, the small-scale model can also effectively discern video quality.

Their method consists of two stages: (i) In the first stage, a comprehensive teacher network is trained by extracting video features through a series of powerful backbone networks. In the second stage, a series of closed-source UGC videos are annotated with pseudo labels by the teacher network and then used to train a lightweight model. They introduce the two stages in detail below.

In the first stage, they extract keyframes of the video following the RQ-VQA \cite{RQ-VQA} strategy. Then, they extract features of these keyframes from three aspects: spatial, temporal, and spatiotemporal. Specifically, they extract the motion features of the video through SlowFast, FAST-VQA features, LIQE features, and DeQA \cite{DeQA} features. In addition, considering that UGC videos are uploaded by ordinary users, these videos often exhibit eye-related characteristics such as edge masking that need to be taken into account. Therefore, they refer to HVS-5M \cite{HVS-5M} to further extract edge and content features of the keyframes. Based on the above features, they fully retain the information of all dimensions of the video. To better preserve the learning of quality-aware features by some learnable parameters, they incorporate an additional Swin-B network. By integrating these features, they train a powerful teacher network. Specifically, they concatenate all of the above features and obtain the final video score through two MLP layers. Due to the integration of a series of features, the teacher model trained in the first stage possesses strong capabilities in identifying the quality of UGC videos. They then use it to annotate a large set of closed-source UGC videos (nearly 30{,}000), providing pseudo labels for training the student model in the second stage.

In the second stage, they utilize a lightweight student network to fit the labels annotated by the teacher network. Their expectation is that the student model can acquire the proficiency to discriminate the quality of UGC videos. Specifically, they first use the keyframe extraction method of RQ-VQA to convert all videos into frame-level representations. Subsequently, the score of each frame is annotated as the score of the corresponding video. After that, they randomly crop a patch with a resolution of $224\times224$ from these frames and feed it into a Swin-T network. Considering the importance of multi-scale features in quality assessment tasks, they obtain the features of each layer of the Swin-T and concatenate them together. Finally, they compute the video quality score through two fully connected layers. 

% \subsubsection{Training details}
\noindent \textbf{Training details} They employ several powerful backbone networks to extract features associated with the human vision system. Subsequently, they train a comprehensive video quality assessment model, which is utilized as a teacher model to label a series of closed-source User-Generated Content (UGC) video datasets. Leveraging these labeled datasets, they train a lightweight network. As a result, the small-scale model can also effectively discern video quality. Differentiable SRCC and PLCC losses are adopted in both training phases.

In the first stage, the keyframes are randomly and centrally cropped to a resolution of $384\times384$ and then fed into the Swin-B network. The Adam optimizer with an initial learning rate of $1 \times 10^{-5}$ and a batch size of 6 is used to train the proposed teacher model on one A100 GPU.

In the second stage, the Adam optimizer with an initial learning rate of $1 \times 10^{-3}$ and a batch size of 64 is used to train the proposed student model on one A100 GPU.

\subsection{ZQE}

\begin{figure}[t]
  \centering
  \includegraphics[width=1.0\linewidth]{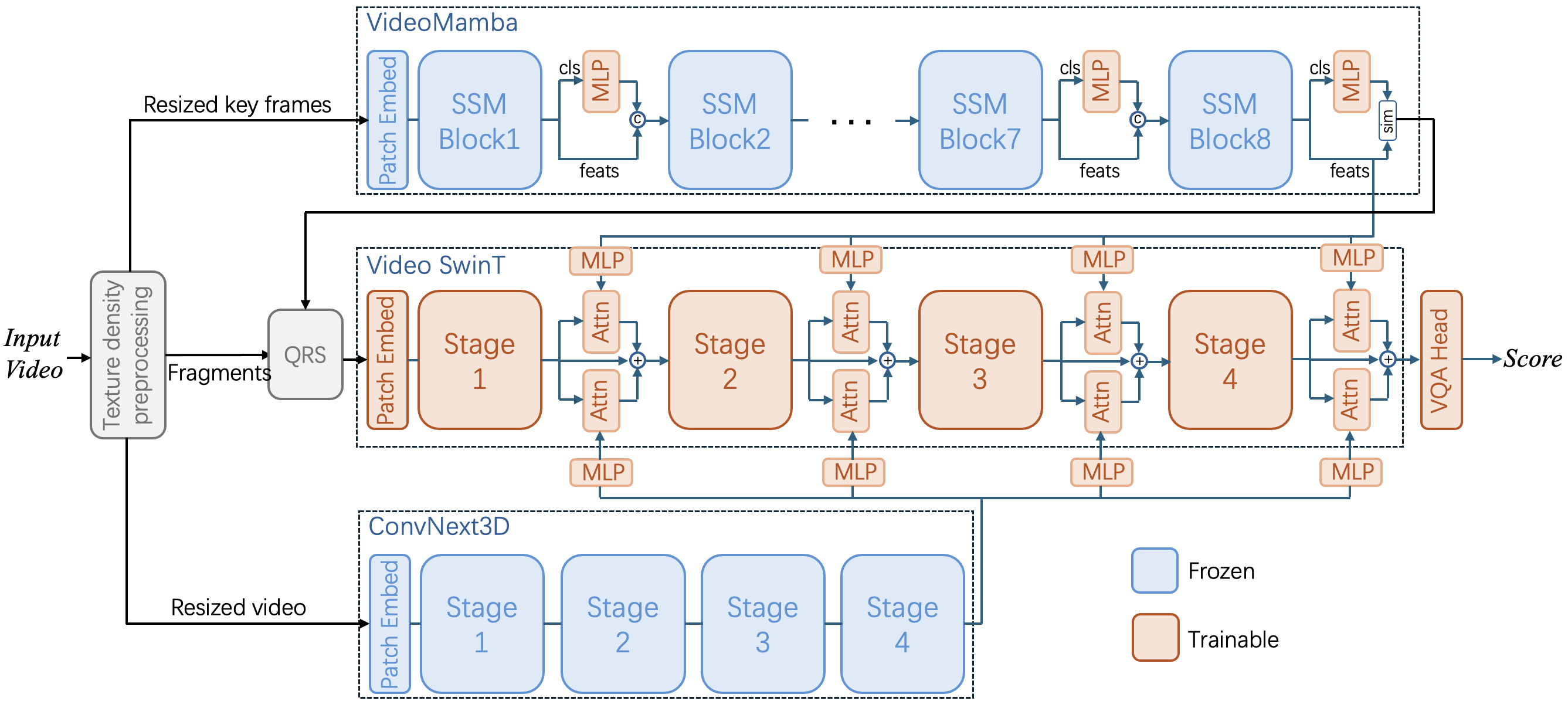} % 替换为你的图像文件名
  % 图注
  % \caption{The architecture of the proposed method.}
  \caption{The overall framework of Team ZQE. }
  \label{fig:ZQE}
\end{figure}

This team extracts semantic, local texture, and global features from videos through three branches, aggregates them via an attention mechanism, and finally maps them into a quality score. Their model is a hybrid model built upon and improved from DOVER \cite{DOVER} and KSVQE \cite{KSVQE}.

First, they pretrain DOVER on the KVQ dataset, freeze the aesthetic branch weights, and discard the original VQA head. They then integrate the attention-based fusion modules proposed in KSVQE between each stage of the technical branch. Next, through extensive ablation experiments, they select VideoMamba-middle \cite{VideoMamba} as their video semantic extractor, as it is an effective and lightweight video/image classification model. Specifically, they freeze VideoMamba's pretrained weights obtained from ImageNet-1k \cite{ImageNet} and insert eight trainable MLP modules into the backbone (one after every four SSM layers) to facilitate fine-tuning for the VQA task. Additionally, to address the issue of large irrelevant regions commonly found in short-form videos, they not only incorporate the quality-aware region selection module from KSVQE but also introduce a novel slicing preprocessing step. Specifically, they horizontally divide each video into multiple slices of equal height, compute the texture density of each slice using the Laplacian operator, and discard slices whose texture density falls below a threshold determined empirically from their experiments on the KVQ dataset. This preprocessing approach significantly improves the performance of their model.

Their model is pretrained end-to-end on a combined dataset consisting of the KVQ training set and their private UGC dataset. Finally, they fine-tune the best-performing pretrained model (selected based on validation performance on the KVQ validation set) using the KVQ training set to obtain the optimal model.

% \subsubsection{Training details}
\noindent \textbf{Training details} They first pretrain DOVER on the KVQ dataset and freeze the weights of the aesthetic branch. The technical branch is initialized using the official weights of KSVQE. The model is pretrained end-to-end on a mixed dataset containing the KVQ training set and their private UGC dataset, and is then fine-tuned on the KVQ training set. All training stages are evaluated on the KVQ validation set.

% \subsubsection{Testing details}
\noindent \textbf{Testing details} The test video is directly fed into the model for inference to obtain the predicted quality score. It is important to note that under the current Torch/CUDA version, they observe that multithreading and cuBLAS may introduce non-determinism in the results. Specifically, the output score may vary from the 7th or 8th decimal place onward with some probability. They ensure that no randomness is artificially introduced at any stage of inference. This does not affect the overall score on the Codalab leaderboard (rounded to 4 decimal places: 0.9158).

\subsection{ZX-AIE-Vector}
\begin{figure}[t]
  \centering
  \includegraphics[width=1.0\linewidth]{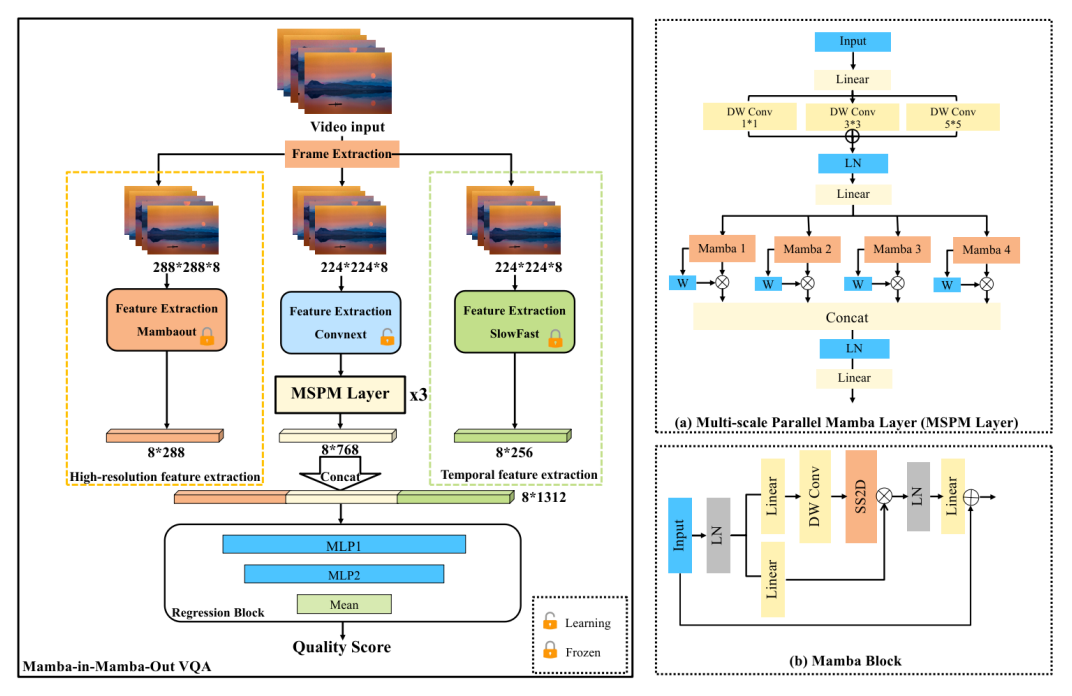} 
  \caption{The overall framework of Team ZX-AIE-Vector. }
  \label{fig:ZX-AIE-Vector}
\end{figure}

This team proposes a lightweight and efficient VQA framework to address the limitations of high computational cost and suboptimal performance in many existing Video Quality Assessment methods. The total computational complexity of their framework is constrained to approximately 100 GFLOPS. Specifically, their approach adopts a widely used dual-stream architecture that separately extracts spatial-temporal and temporal features. For spatial-temporal modeling, they design two lightweight modules: a low-resolution branch to capture coarse global context, and a high-resolution branch to extract fine-grained local details, which are subsequently fused. Meanwhile, a lightweight SlowFast \cite{slowfast} module is employed to extract multi-scale temporal features. 

To further enhance the model’s representation capability, they introduce a Multi-Scale Parallel Mamba (MSPM) layer. This module leverages the global modeling ability and linear complexity of Mamba \cite{mamba} to perform deep feature extraction, improving long-range dependency modeling. Finally, spatial-temporal and temporal features are fused to form a comprehensive video representation. Experiments on the S-UGC competition dataset KVQ \cite{KSVQE} demonstrate that their method achieves competitive performance while maintaining high efficiency, validating the effectiveness of the proposed lightweight dual-stream architecture and the MSPM module.

% \subsubsection{Training details}
\noindent \textbf{Training details} The training pipeline consists of three main stages: pretraining, pseudo-labeling, and two-phase fine-tuning. They first pretrain their lightweight model on the large-scale LSVQ dataset \cite{LSVQ} for 10 epochs to learn generalizable video representations, strictly following the original training-validation split. 

Next, they fine-tune a large-scale model on the KVQ dataset \cite{KSVQE} (merging the official training and validation sets), and use it to generate pseudo-labels for the test set. These pseudo-labels are then merged with the original KVQ training and validation data to construct an augmented dataset.

They conduct the first-phase fine-tuning of their lightweight model using this augmented set via 10-fold cross-validation, training each fold for 30 epochs. The average prediction across all folds is used as a refined pseudo-label for each test sample. In the second-phase fine-tuning, they use the full KVQ training and validation sets along with the refined pseudo-labeled test data to retrain their lightweight model for an additional 30 epochs. This two-stage fine-tuning strategy enables them to fully exploit the unlabeled test data and improve generalization.

They implement their method using PyTorch and train it on 8 NVIDIA V100 GPUs. The Adam optimizer is used with an initial learning rate of $3\times10^{-5}$, combined with a Cosine Annealing learning rate scheduler that gradually decreases the learning rate to a minimum of $1\times10^{-6}$. The entire training process takes approximately 9 hours.

% \subsubsection{Testing details}
\noindent \textbf{Testing details} During testing, they evaluate their final lightweight model on the KVQ test set. For each video, they uniformly sample 8 frames, consistent with the training pipeline. The input resolution during testing is kept the same as in training to ensure consistency in feature extraction. The model directly predicts a quality score for each video. To ensure reproducibility, they fix the random seed to 8 during all testing procedures.

\subsection{ECNU-SJTU VQA Team}
\begin{figure}[t]
  \centering
  \includegraphics[width=1.0\linewidth]{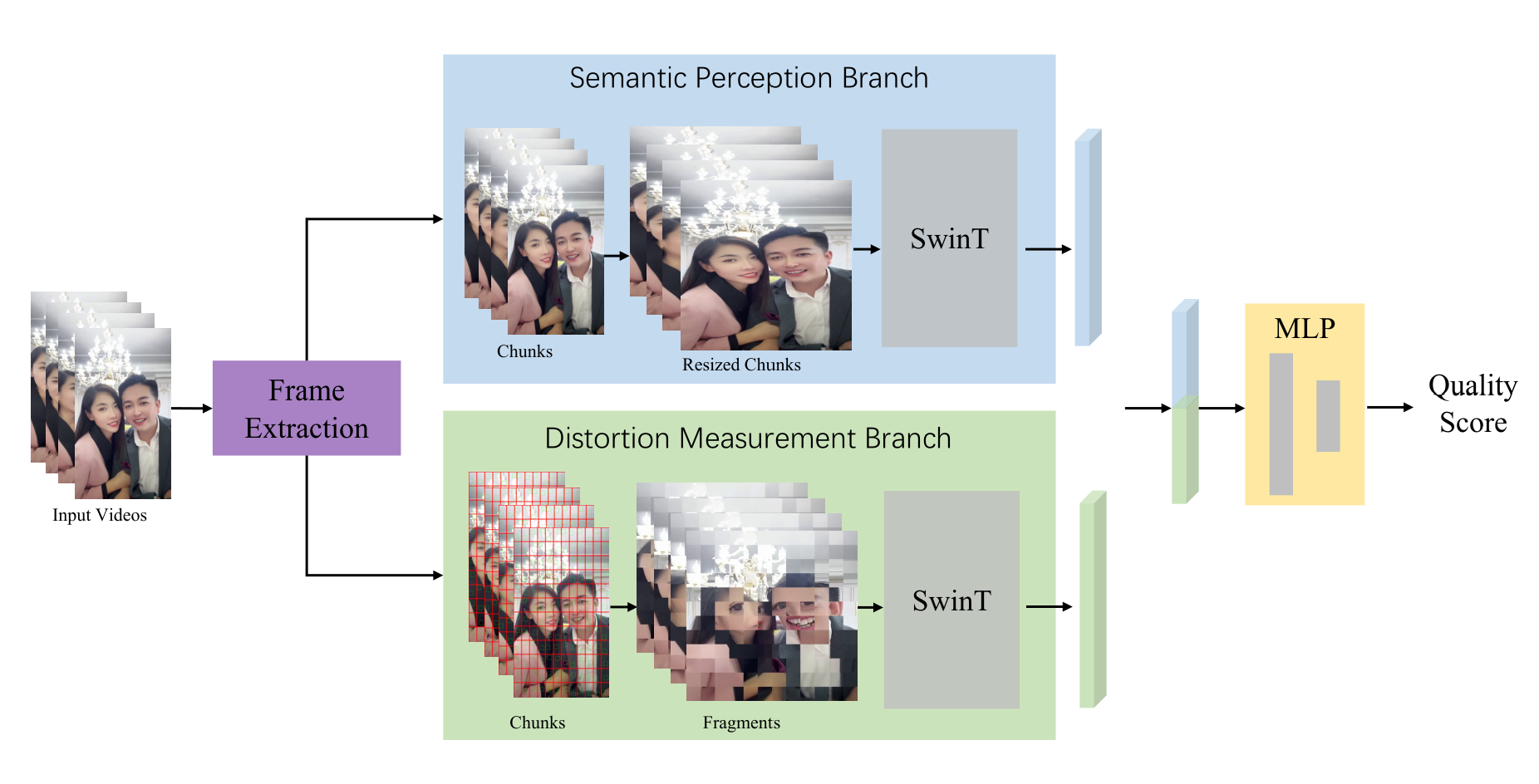} % 替换为你的图像文件名
  % 图注
  % \caption{The framework of the proposed model. }
  \caption{The overall framework of Team ECNU-SJTU. }
  \label{fig:ECNU-SJTU VQA Team}
\end{figure}
% \subsubsection{Method Description}
This team proposes an efficient video quality assessment (VQA) model named E-VQA~\cite{ntire2025EVQA}, designed to achieve high performance while maintaining low computational complexity. Inspired by previous successful efforts in developing efficient VQA models such as SimpleVQA \cite{SimpleVQA}, FAST-VQA \cite{Fast-vqa}, and MinimalisticVQA \cite{MinimalisticVQA}, they empirically explore a combination of best practices from these models along with techniques from other efficient deep neural networks (DNNs) to develop E-VQA. By combining task-specific and general optimizations, their method balances accuracy and efficiency for practical deployment.

They begin with the lightweight MinimalisticVQA VIII model \cite{MinimalisticVQA}, replacing its backbone from Swin Transformer-Base to Swin Transformer-Tiny (Swin-T) \cite{swin}, and evaluate the optimal frame count and resolution under the competition’s computational constraint of 120 GFLOPs. Next, they incrementally integrate motion features (extracted via X3D \cite{X3d}) and fragment frame features (extracted using 2D/3D FAST-VQA variants \cite{Fast-vqa}) into the baseline to assess performance gains.

Their experiments reveal that a dual Swin-T architecture with shared weights achieves optimal results. One branch aligns with MinimalisticVQA VIII, processing frames resized to $384\times384$, while the other employs a 2D FAST-VQA variant operating on fragment frames composed of $12 \times 12$ image patches with a resolution of $32 \times 32$. For feature extraction, they uniformly sample 4 frames from 8-second video clips.

To enhance generalization, they adopt an offline knowledge distillation strategy using RQ-VQA \cite{RQ-VQA}—the winning solution of the NTIRE 2024 VQA Challenge—as the teacher model. They curate a pretraining dataset of 52,000 videos, consisting of 40,000 web-sourced videos and 12,000 synthetically compressed samples. Specifically, they generate 2,000 originals from the web-sourced videos and compress them using H.264 at 6 levels \cite{KSVQE}. RQ-VQA is used to generate pseudo-labels for all videos. Finally, they pretrain E-VQA on this dataset using fidelity loss \cite{fidelity} and RMSE loss, and fine-tune it on the KVQ training set using PLCC loss. The source code and dataset will be publicly released to ensure reproducibility. The final test score achieved is 0.91083.

% \subsubsection{Training details}
\noindent \textbf{Training details} They initialize the Swin-T backbone using weights from MinimalisticVQA \cite{MinimalisticVQA}. For training, they sample one keyframe every two seconds (i.e., 0.5 fps). In the semantic perception branch, these keyframes are further resized to a resolution of $384\times384$. In the distortion measurement branch, fragments are extracted from $12\times12$ partitions, with each fragment having a resolution of $32\times32$. They train the proposed model using 4 NVIDIA RTX 3090 GPUs. The training consists of 3 epochs on the collected dataset and 30 epochs on the KVQ dataset \cite{KSVQE}, with a batch size of 6. The learning rate is set to $1 \times 10^{-5}$, and the optimizer used is Adam. The total training time is approximately 3 hours for the pretraining stage and 4 hours for fine-tuning on the KVQ dataset. No additional training strategies or specific efficiency optimizations are applied.

% \subsubsection{Testing details}
\noindent \textbf{Testing details} During testing, the procedure for obtaining resized frames and fragments is kept exactly the same as in the training stage. The model uses the same input structure to ensure consistency in evaluation.

\subsection{TenVQA}
% \subsubsection{Method Description}
This team proposes a robust baseline for video quality assessment (VQA) using a single convolutional neural network (CNN). The core of their approach is based on the ConvNeXtV2-Tiny architecture, which is pre-trained on ImageNet and selected for its favorable balance between computational efficiency and classification performance.

They implement a two-stage training strategy with a differentiable global rank loss, namely RaMBO loss, to improve rank-aware quality prediction. For data processing, videos are resized to 720p while maintaining aspect ratio. From each video, four equidistant frames are sampled and $576 \times 576$ patches are cropped. During training, they apply data augmentations including random cropping, horizontal flipping, and brightness/contrast jittering to improve model robustness.

% \subsubsection{Training details}
\noindent \textbf{Training details} The training consists of two stages. In Stage 1, the model is jointly trained on the KVQ dataset and a private dataset using a combination of L2 loss and PLCC loss. In Stage 2, they fine-tune the model on the KVQ dataset using PLCC loss and a differentiable SRCC loss, specifically the RaMBO loss. During training, four equidistant frames are randomly sampled from each video, and $576 \times 576$ random crops are applied. The training is implemented in PyTorch using Distributed Data Parallel (DDP) on 4 NVIDIA A100 GPUs. The AdamW optimizer is used, with a learning rate of $3 \times 10^{-5}$ for Stage 1 and $1 \times 10^{-5}$ for Stage 2. The total training time is approximately 4 hours for Stage 1 and 0.5 hours for Stage 2. Automatic Mixed Precision (AMP) is used to improve training efficiency, and Exponential Moving Average (EMA) is enabled for more stable optimization.

% \subsubsection{Testing details}
\noindent \textbf{Testing details} During testing, the model samples four equidistant frames from each video and applies a center crop of resolution $576 \times 576$. The inference process follows the same preprocessing strategy as in training to ensure consistency.

\subsection{GoldenChef}
\begin{figure}[t]
  \centering
  \includegraphics[width=1.0\linewidth]{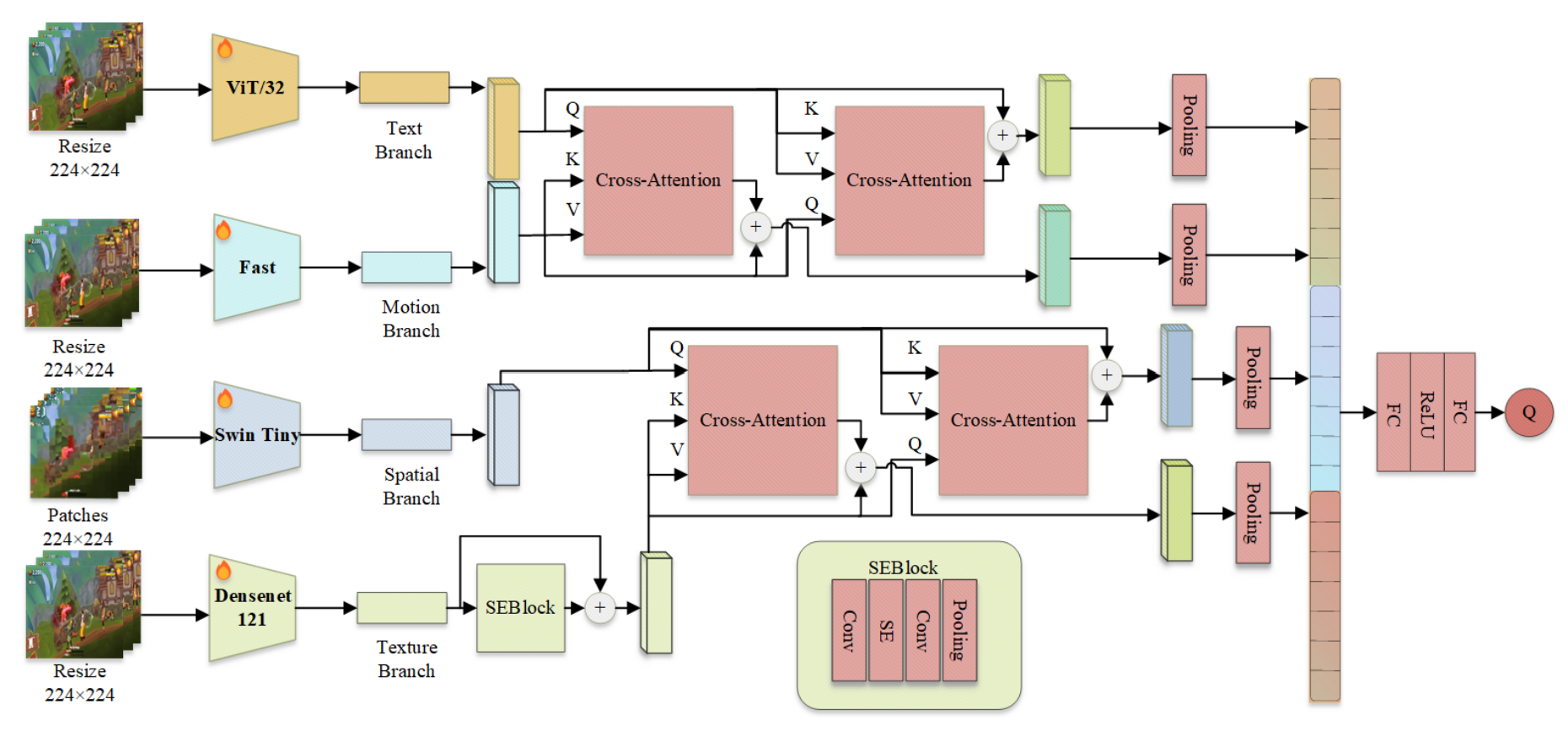} % 替换为你的图像文件名
  % 图注
  % \caption{The LMFVQA Framework. }
  \caption{The overall framework of Team GoldenChef. }
  \label{fig:GoldenChef}
\end{figure}
% \subsubsection{Method Description}
This team proposes a lightweight multi-feature fusion model for short-form UGC video quality assessment (LMFVQA), which integrates several advanced components to capture diverse quality-related features of videos. The model incorporates the visual component of the CLIP model with a ViT-32 backbone to extract semantic features, a ResNet50-3D-based network to extract temporal motion features (similar to the Fast pathway of SlowFast), a Swin Transformer-Tiny (Swin-T) to capture spatial global features, and a DenseNet121 to extract local texture features from video frames.

To achieve comprehensive feature integration, they employ a cross-attention fusion module to model the interaction between semantic and motion features, as well as between global and local spatial structures. This enhances the robustness and expressiveness of the representation for effective and efficient VQA. The fused features are subsequently processed through a multilayer perceptron (MLP) with two fully connected layers to derive the final video quality score. Notably, within the DenseNet121 branch dedicated to texture feature extraction, they introduce a channel attention mechanism (SE) to dynamically recalibrate channel weights and refine feature representations before the cross-attention fusion.

% \subsubsection{Training details}
\noindent \textbf{Training details} They extract two keyframes per second from each video (i.e., 2 fps) as input to the model. These keyframes are resized to a resolution of $224 \times 224$ and fed into the ViT-32-based semantic branch, the DenseNet121-based texture branch, and the ResNet50-3D-based motion branch to extract semantic, texture, and motion features, respectively. Furthermore, the frame difference between keyframes is utilized to guide a grid-based sampling process.

In detail, each keyframe is divided into a uniform $7 \times 7$ grid. Frame difference information is used to sample one $32 \times 32$ patch at the original resolution within each grid cell. These sampled patches are then stitched together according to their original spatial locations to construct a $224 \times 224$ frame-difference sampling map. This map is subsequently used as input to the Swin Transformer-T for spatial feature extraction.

The model is trained on a single NVIDIA RTX 3090 GPU with a batch size of 4 over 30 epochs. The learning rate is set to $1 \times 10^{-5}$, and the optimizer used is Adam. The total training time is approximately 2.51 hours. They apply multiscale feature fusion and exponential decay of the learning rate as efficiency optimization strategies.

% \subsubsection{Testing details}
\noindent \textbf{Testing details} They train the model on the KVQ training set, select the best-performing checkpoint based on the KVQ validation set, and evaluate the final performance on the KVQ test set. The same preprocessing and input sampling strategies used during training are applied during inference to ensure consistency.

\subsection{DAIQAM}
\begin{figure}[t]
  \centering
  \includegraphics[width=1.0\linewidth]{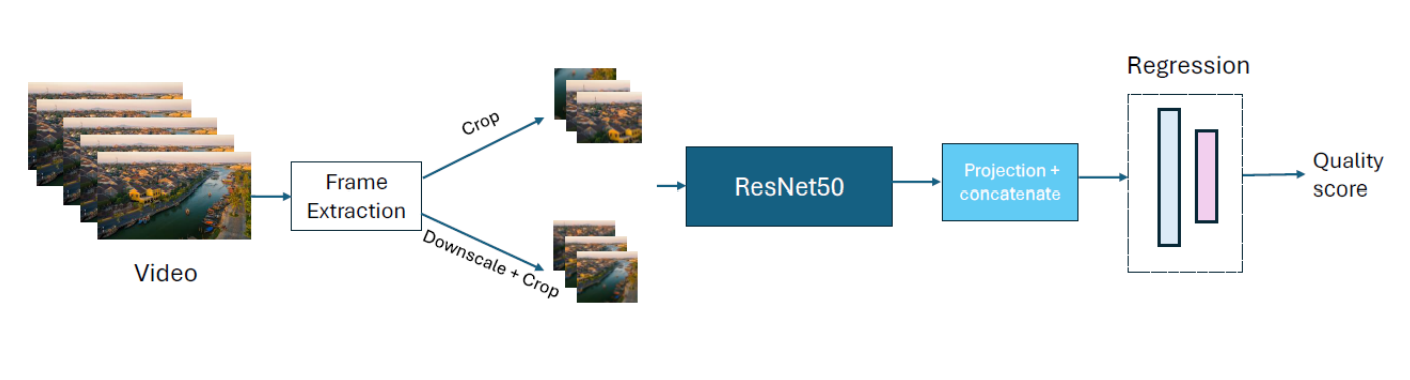} % 替换为你的图像文件名
  % 图注
  % \caption{The network architecture of their solution. }
  \caption{The overall framework of Team DAIQAM. }
  \label{fig:DAIQAM}
\end{figure}
% \subsubsection{Method Description}
This team uses ResNet-50 \cite{resnet} to extract spatial features from patches of two versions of keyframes: original and downsampled. To leverage the effectiveness of image distortion manifold learning introduced in ARNIQA, they adopt the ResNet-50 encoder from that model as the initialization for their feature extractor. After extracting features from each frame, a projector module is applied to reduce the feature dimensionality. The features from all frames are then concatenated and passed through a multilayer perceptron (MLP) for quality regression.

% \subsubsection{Training details}
\noindent \textbf{Training details} During training, they sample one keyframe for each one-second segment of the video. Since most videos in the training set are 8 seconds long, the total number of keyframes is typically 8. In cases where a video is shorter than 8 seconds, replicate padding is applied, duplicating the last available keyframe to maintain a consistent frame count.

They find that this sampling technique outperforms uniform frame sampling, improving the validation accuracy by 3.2\%. A downsampled version of each keyframe is created by resizing the frame while maintaining its aspect ratio, setting the shorter side to 224 pixels. Both the original and downsampled keyframes are then randomly cropped to a resolution of $224 \times 224$ during training.

The model is trained using the Adam optimizer with an initial learning rate of $1 \times 10^{-5}$ and a batch size of 8 on an NVIDIA RTX 4090 GPU. The learning rate is decayed by a factor of 10 after every 10 epochs, and the total number of training epochs is 20.

% \subsubsection{Testing details}
\noindent \textbf{Testing details} During testing, videos are processed in the same way as in the training procedure. However, instead of random cropping, a center crop of $224 \times 224$ is applied to the keyframes to ensure consistent evaluation.

\subsection{57VQA}
% The submitted project is just an adaption for the StarVQA to the specified dataset so that it work.\\
% \textbf{Technical details}
% Framework is the Slowfast, Optimezer is the Adam, GPU is RTX4090, lr is
% provided in code, training time is about 3 hours, training strategie is trainingdirectly.\\
% \subsubsection{Method Description}
This team submits a project adapted from the StarVQA model, making the necessary modifications for it to work on the specified competition dataset. The core architecture and methodology are inherited from the original StarVQA implementation without major structural changes.

% \subsubsection{Training details}
\noindent \textbf{Training details} They use the SlowFast framework for model training, with the Adam optimizer and an NVIDIA RTX 4090 GPU. The learning rate is provided in the code and not explicitly stated in the report. The total training time is approximately 3 hours. Their training strategy is direct training on the competition dataset without additional pretraining or fine-tuning stages.

\subsection{Nourayn}
\begin{figure}[t]
  \centering
  \includegraphics[width=1.0\linewidth]{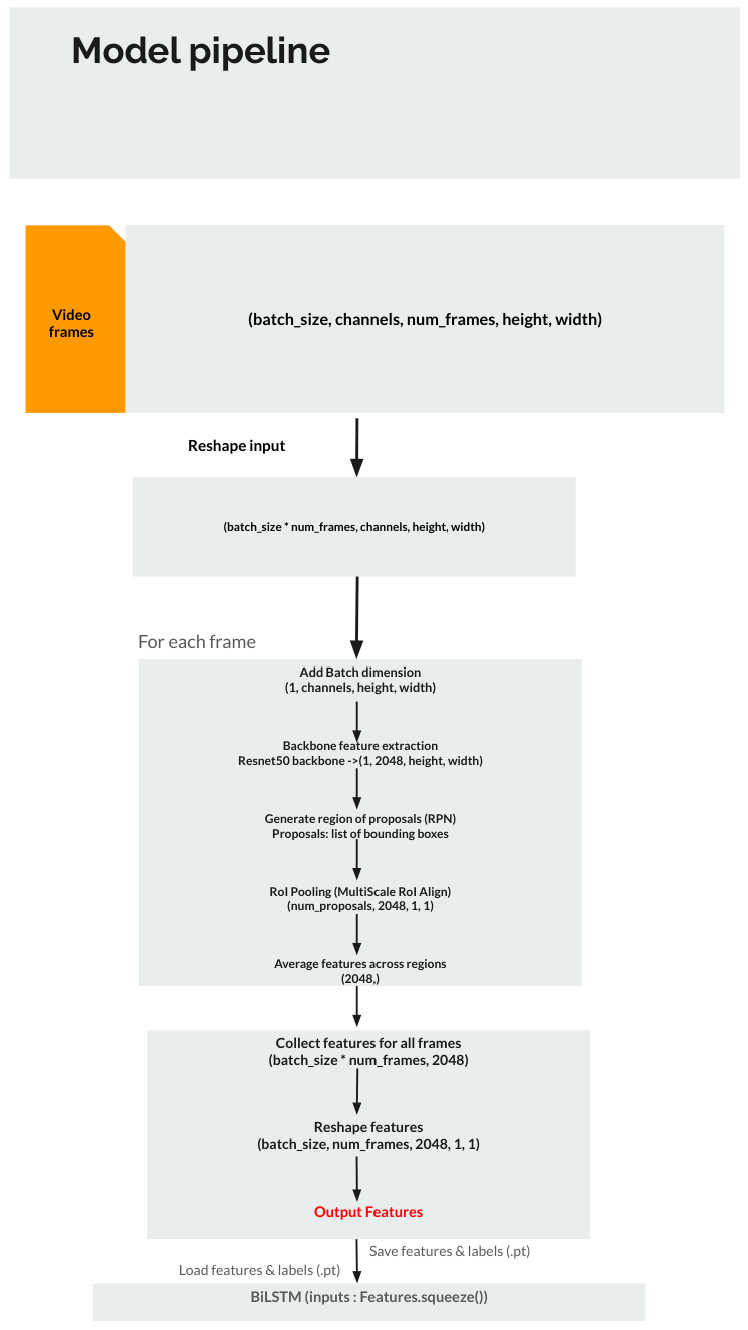} % 替换为你的图像文件名
  % 图注
  % \caption{The pipeline of the proposed model. }
  \caption{The overall framework of Team Nourayn. }
  \label{fig:Nourayn}
\end{figure}

This team presents a two-stage deep learning solution for the NTIRE2025 Video Quality Assessment Challenge. Their model combines spatial feature extraction with a pre-trained ResNet-50 backbone and Faster-RNN, alongside temporal modeling using a Bidirectional LSTM (BiLSTM). The overall design aims to predict the Mean Opinion Score (MOS) for video quality by capturing both spatial and temporal information from video sequences.

% \subsubsection{Training details}
\noindent \textbf{Training details} They train the BiLSTM component using features extracted from the spatial encoder. A composite loss function is employed to guide the training process. To improve generalization, the model is trained on a combination of the official training and validation datasets.

% \subsubsection{Testing details}
\noindent \textbf{Testing details} The trained model is evaluated on the test dataset. They compute standard performance metrics including SROCC, PLCC, KROCC, and RMSE to assess the accuracy and robustness of the predicted video quality scores.

\section{Teams and Methods of Track 2}
\label{sec:teams_and_methods_2}
\subsection{TACO\_SR}
\begin{figure}
    \centering
    \includegraphics[width=1\linewidth]{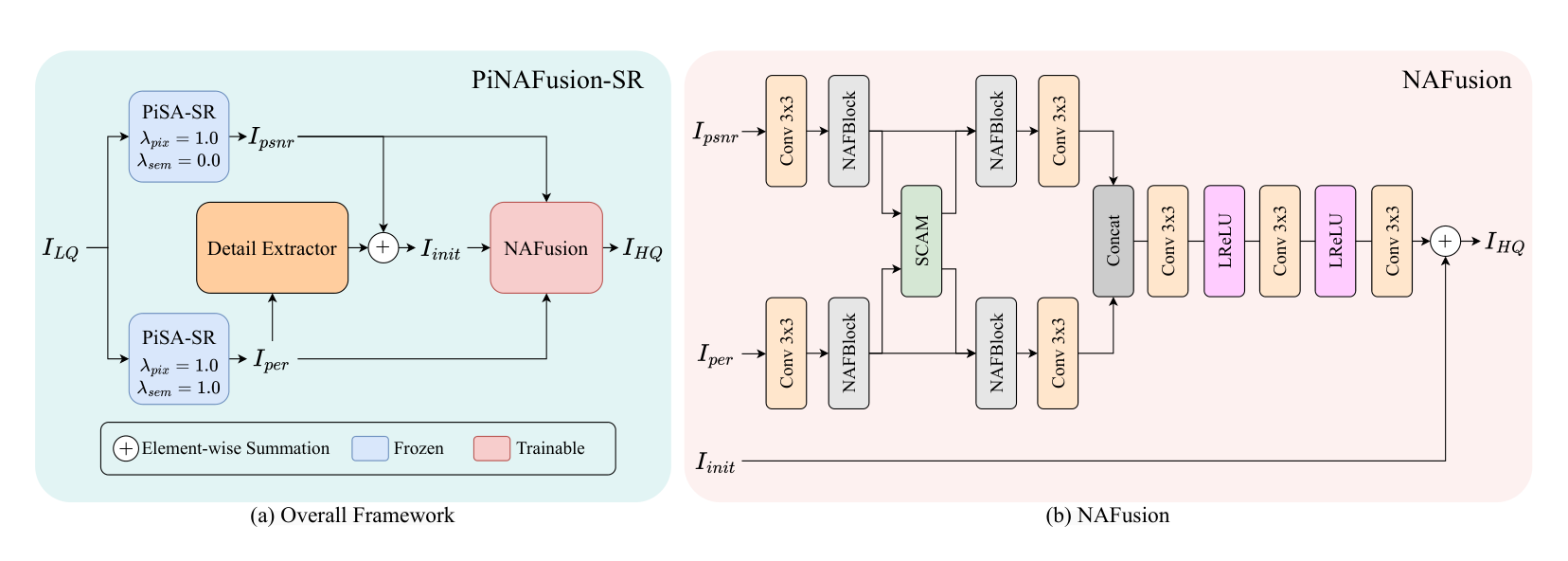}
    \caption{The overall framework of Team PiNAFusion-SR.}
    \label{fig:TACO_SR}
\end{figure}

% Done!
This team develops a two-stage super-resolution method, including an image super-resolution phase and an image fusion phase, as shown in Figure~\ref{fig:TACO_SR}. Their method consists of three key components designed to generate a high-quality image: (1) PiSASR~\cite{sun2024pixelPiSASR}, (2) Detail Extractor, and (3) NAFusion (Inspired by NAFSSR~\cite{chu2022nafssr}). They optimize the NAFusion module based on KwaiSR dataset, while keeping pre-trained PiSASR fixed.

In the first phase, they develop their method based on PiSASR~\cite{sun2024pixelPiSASR}. They generate candidate images uder two different settings: (1) $\lambda_{\text{pix}} = 1.0, \lambda_{\text{sem}} = 0.0$: This configuration prioritizes high fidelity, producing an image denoted as $I_{\text{psnr}}$; and (2) $\lambda_{\text{pix}} = 1.0, \lambda_{\text{sem}} = 1.0$: This setting enhances perceptual quality, resulting in an image referred to as $I_{\text{per}}$. They observe that $I_{\text{psnr}}$ achieves superior fidelity (PSNR, SSIM) but lacks perceptual quality (NIQE, MANIQA, CLIPIQA). Conversely, $I_{\text{per}}$ excels in perceptual quality but exhibits lower fidelity. They take advantage of $I_{\text{psnr}}$ and $I_{\text{per}}$ by an image fusion model. Specifically, they first combine $I_{\text{psnr}}$ and $I_{\text{per}}$ to $I_{\text{init}}$ with Detail Extractor: use high pass filter to extract high frequency component from $I_{\text{per}}$ and add to $I_{\text{psnr}}$. 

In the second phase, they construct NAFusion using $I_{\text{psnr}}$, $I_{\text{per}}$, and $I_{\text{init}}$. Inspired by NAFSSR~\cite{chu2022nafssr}, they design the NAFusion module with NAFBlock and SCAM. Finally, they add $I_{\text{init}}$ to the output of the main branch of NAFusion to generate $I_{\text{HQ}}$.

\noindent \textbf{Training Details}
First, they generate $I_{\text{psnr}}$ and $I_{\text{per}}$ from low-quality images in the provided training dataset using PiSASR under different settings. ($I_{\text{psnr}}$: $\lambda_{\text{pix}} = 1.0, \lambda_{\text{sem}} = 0.0$, $I_{\text{per}}$: $\lambda_{\text{pix}} = 1.0, \lambda_{\text{sem}} = 1.0$). Then, they crop image patches from $I_{\text{psnr}}$, $I_{\text{per}}$ and $I_{\text{GT}}$ with $512 \times 512$ resolution (Overlap is set to 128). They train NAFusion $f$ using data triplet $(I_{\text{psnr}}, I_{\text{per}}, I_{\text{GT}})$ with a loss function $L_{\text{total}}$:
\begin{equation}
L_{\text{total}} = \lambda_1 L_{\text{pix}} + \lambda_2 L_{\text{ssim}} + \lambda_3 L_{\text{lpips}},
\end{equation}
where
\begin{align*}
L_{\text{pix}}\  &= L_1(f(I_{\text{psnr}}, I_{\text{per}}), I_{\text{GT}}) \text{ is L1 loss}, \\
L_{\text{ssim}} &= L_{\text{SSIM}}(f(I_{\text{psnr}}, I_{\text{per}}), I_{\text{GT}}) \text{ is SSIM loss}, \\
L_{\text{lpips}} &= L_{\text{LPIPS}}(f(I_{\text{psnr}}, I_{\text{per}}), I_{\text{GT}}) \text{ is LPIPS loss}.
\end{align*}

They set the batch size and the learning rate to 8 and $1e^{- 5}$. They train NAFusion for 5 epochs, with $\lambda_1 = 10.0, \lambda_2 = 0.5, \lambda_3 = 1.0$.

\noindent \textbf{Testing Details}
They set the scale factor to 4 and 1 when running inference with PiNAFusion-SR on images from the ``synthetic'' and ``wild'' subfolders in the validation and test datasets. The scaling strategy is determined based on the resolution of the input image. Specifically, if the larger dimension (height or width) of the input image is below 500 pixels, they apply a scale factor of 4. Otherwise, they use a scale factor of 1. They perform inference on both the validation and test datasets using the same settings.

\subsection{RealismDiff}
\begin{figure}
    \centering
    \includegraphics[width=1\linewidth]{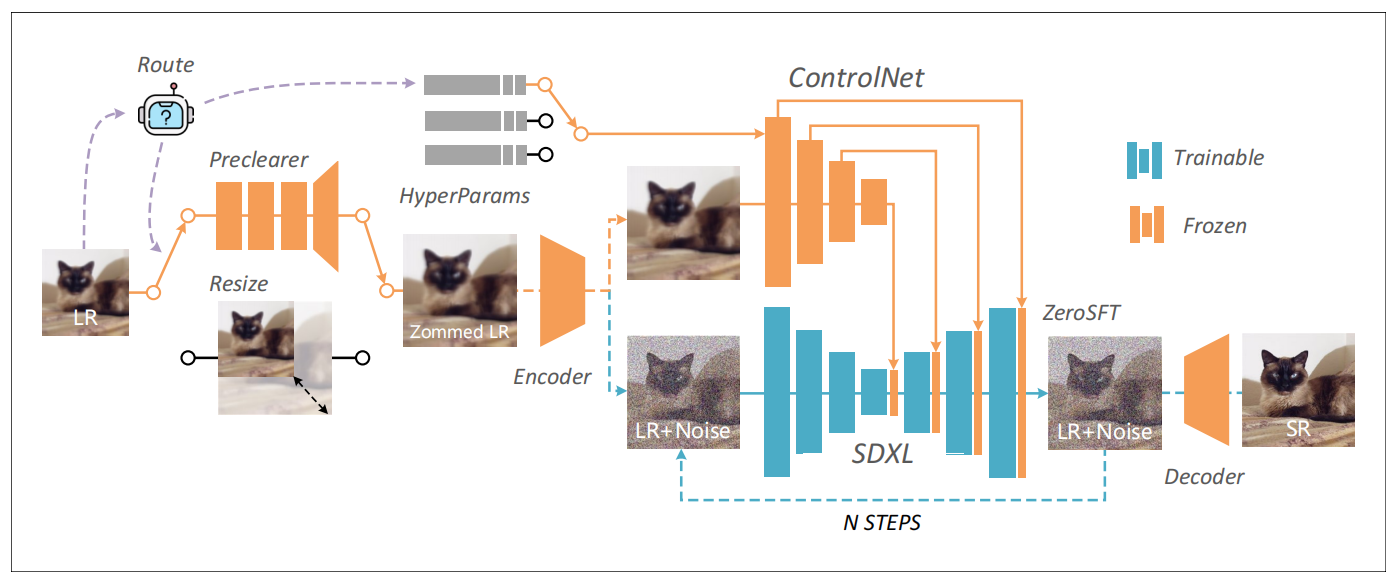}
    \caption{The framework proposed by Team RealismDiff.}
    \label{fig:realismdiff}
\end{figure}

% Done!
This team proposes a two-stage diffusion-based image super-resolution method consisting of the PreCleaner (a lightweight CNN) and SUPIR~\cite{yu2024scalingsupir}. Specifically, as shown in Figure.~\ref{fig:realismdiff}, they construct a dynamic pipeline to assign different processing paths according to the image quality assessment. For images with weak degradation, they use a light restoration module (or resize function) to maintain the high-frequency information and fewer step diffusion, while using more extensive settings (larger CNN and diffusion model with more steps and higher CFG) for heavily distorted images. Besides, they collect a training dataset of high-resolution images from DIV2K~\cite{agustsson2017ntireDIV2K}, Flicker2K~\cite{lim2017enhancedFlik2K}, FFHQ~\cite{kazemi2014oneFFHQ}, and Laion5B~\cite{schuhmann2022Laion5B}.

\noindent \textbf{Training Details}
They train the proposed method using 8 A100 GPUs and adopt the AdamW optimizer with default parameters. During training, they crop images into patches of 512$\times$512 pixels and set the batch size as 32. They first pre-train the PreCleaner for 1000 iterations with an initial learning rate of $5\times e^{-5}$. They then fine-tune the entire pipeline in an end-to-end manner for 40,000 iterations. During the fine-tuning stage, the learning rate for the ControlNet is initialized as $5\times e^{-6}$. The learning rates are updated using the Cosine
Annealing scheme.

\subsection{SRlab}
\begin{figure}
    \centering
    \includegraphics[width=1.0\linewidth]{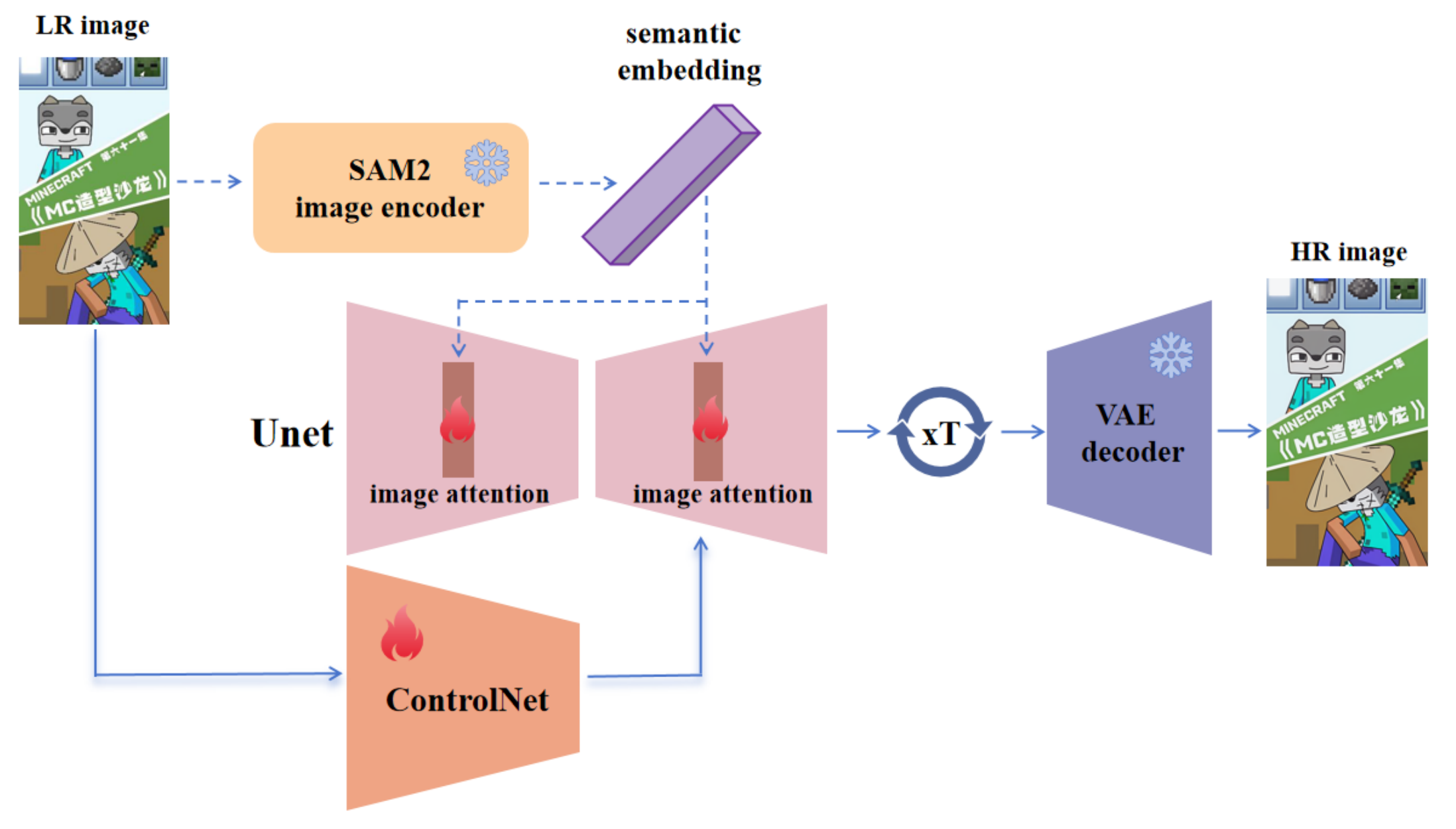}
    \caption{Overall Pipeline of the solution of Team SRlab.}
    \label{fig:SRlab}
\end{figure}

% Done!
As shown in Figure~\ref{fig:SRlab}, the method of this team~\cite{wang2025enhancedtopthree} is based on the diffusion framework. They employ a VAE to encode the input low-quality images and obtain their corresponding latent representations. These latents are then processed by a Denoising U-Net, which iteratively refines them through multiple denoising steps. To ensure the generated images maintain high fidelity with the low-resolution (LR) inputs, they incorporate the ControlNet architecture, which allows for precise control over the generation process. Furthermore, recognizing that the official synthetic and wild test sets exhibit varying degrees of degradation and require different super-resolution scaling factors, they enhance their approach by utilizing the Segment Anything Model 2~\cite{ravi2024sam} (SAM2). SAM2 is employed to extract rich semantic embeddings from these images, providing additional contextual information that aids in the denoising process. The extracted latents, enriched with semantic embeddings, are subsequently fed into the Denoising U-Net for T steps of iterative refinement. During training, they optimize their model by minimizing the denoising objective:
\begin{equation}
    \mathcal{L} = \mathbb{E}_{X_0,X_{lr},t,c,c_{\text{sem}},\epsilon} \left\|\epsilon - \epsilon_{\theta}(X_t, X_{lr}, t, c, c_{\text{sem}})\right\|^2,
\end{equation}
where $X_{lr}$ represents the low-resolution (LR) latent, $c$ denotes the tag prompt, and $c_{\text{sem}}$ is the semantic embedding. The noise estimation network $\epsilon_{\theta}$ is responsible for predicting the noise $\epsilon \sim \mathcal{N}(0, I)$.

\noindent \textbf{Training Details}
To enhance the model’s performance in short-form UGC scenarios, they constructed a new training dataset by combining the synthetic training set provided by the competition with the LSDIR dataset~\cite{li2023lsdir}. First, they processed the high-resolution images from the LSDIR training set by applying a $4\times$ downsampling and degradation with a 50\% probability for each of two degradation modes to simulate the image degradation scenarios in the synthetic and wild datasets, respectively. Second, they refined the synthetic training set (1440 pairs of gt and lr images) by cropping and overlapping its high-resolution (1080×1920) images to generate $512\times 512$ sub-images, which served as ground truth (GT) images. These GT images were then 4× downsampled and degraded to create the corresponding low-resolution images. Finally, they merged the processed LSDIR training set with the synthetic training set to form the final dataset for training.

They trained the model on a synthesized 512$\times$512 dataset for 90,000 steps using an Nvidia RTX 3090 GPU and the Adam optimizer with a learning rate of $5\times e^{-5}$. The original Stable Diffusion parameters were frozen, and only the ControlNet component and semantic embedding transformer module were trained. The implementation was built on PyTorch, with mixed precision training (FP16) and gradient accumulation to optimize efficiency.

\noindent \textbf{Testing Details}
During the testing phase, they analyzed the impact of three key parameters: ``start point'', ``guidance scale(gs)'', and positive/negative prompts on the experimental results.

First, they evaluated the model’s performance on both the synthetic and wild datasets by setting ``start point'' to either ``noise'' or ``lr'' while keeping ``gs = 5.5'' without adding additional prompts. Next, with ``gs = 5.5'', they introduced positive prompts (``ultra-detailed'', ``ultra-realistic'') and negative prompts (``distorted'', ``deformed'') separately, comparing the model’s overall performance in each case. Finally, they examined the model’s final scores across different ``gs'' values. Based on the results, they selected the best-performing parameter combination for the final test settings.

\subsection{SYSU-FVL-Team}
% Done!
This team develops the framework based on diffusion. They leverage the diffusion model to predict the residual between LQ latent feature and HQ latent feature rather than directly predict HQ latent feature itself. Such a residual learning formulation helps the model focus on learning the desired high-frequency information from the HQ latent features, and it can also accelerate the convergence of the model training process~\cite{he2016deep}. The predicted HQ latent feature will be decoded into high-quality images by the decoder. In order to adjust the preference between pixel and visual semantics, they add a dual LoRA module controlling the pixel-wise and semantic-wise quality based on the pretrained diffusion model SD21, similar to PiSASR~\cite{sun2024pixelPiSASR}.

Following PiSASR, they introduce a pair of pixel and semantic guidance factors, denoted by $\lambda_{\text{pix}}$ and $\lambda_{\text{sem}}$, to control the SR results as follows:
\begin{equation}
\epsilon_\theta(z_L) = \lambda_{\text{pix}}\epsilon_{\theta_{\text{pix}}}(z_L) + \lambda_{\text{sem}}(\epsilon_{\theta_{\text{PiS A}}}(z_L) - \epsilon_{\theta_{\text{pix}}}(z_L)),
\end{equation}
where $\epsilon_{\theta_{\text{pix}}}(z_L)$ is the output with only pixel-level LoRA, and $\epsilon_{\theta_{\text{PiS A}}}(z_L)$ is the output with both pixel and semantic level enhancement. When processing synthetic images, $\lambda_{\text{pix}}$ and $\lambda_{\text{sem}}$ are set as 1.0 and 0.5 to process wild images, separately.

\noindent \textbf{Training Details}
This team utilized an RTX 3090 GPU for model training. The model was optimized using the Adam optimizer with a learning rate of $5\times e^{-5}$, and trained on the released KwaiSR dataset. The training process spanned 96 hours, during which they fine-tuned the model for 100,000 iterations. Both synthetic and wild datasets were processed with a consistent batch size of 4.

\subsection{NetLab}
\begin{figure}
    \centering
    \includegraphics[width=1.0\linewidth]{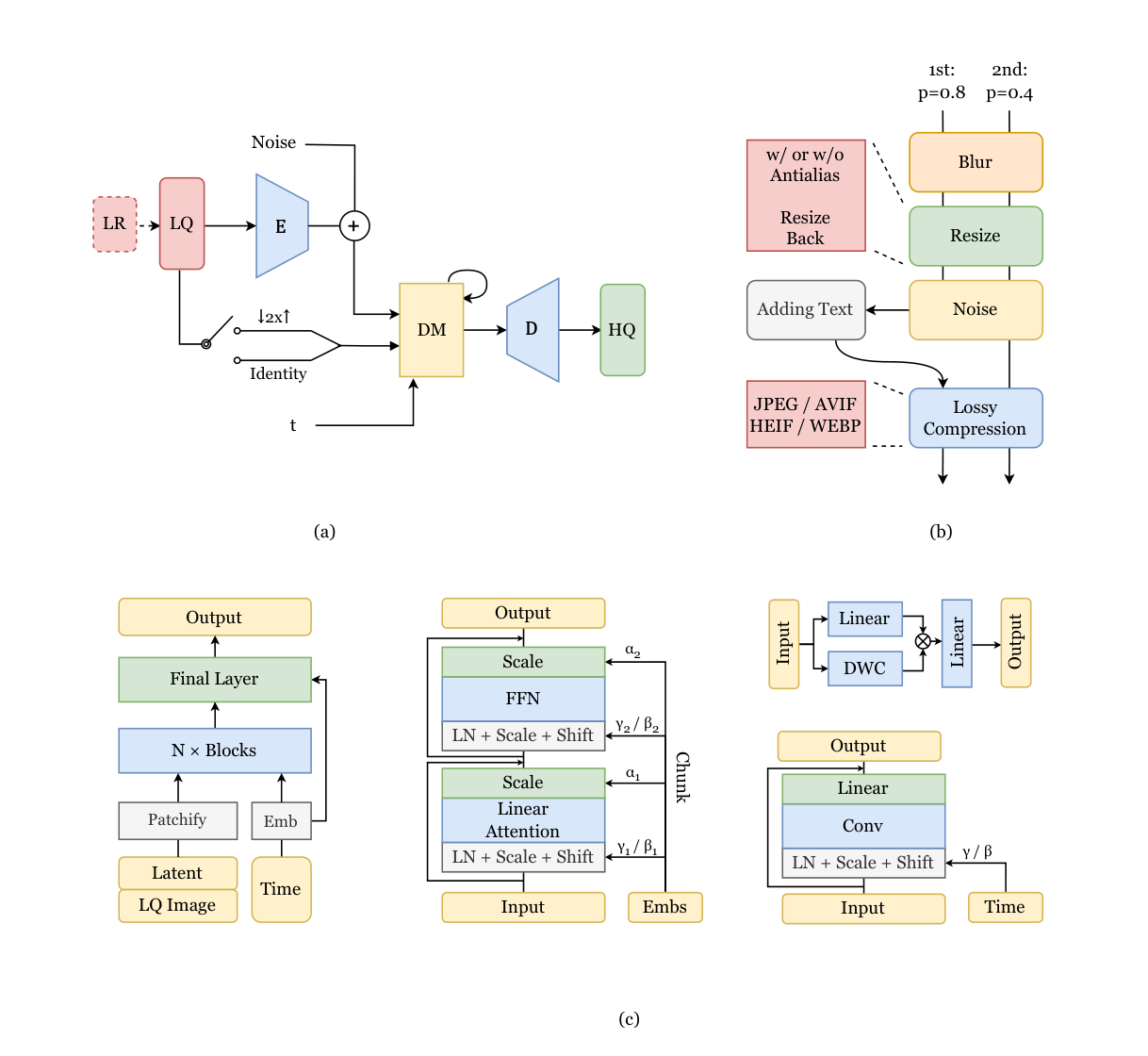}
    \caption{The overall pipeline of the method proposed by Team NetLab. (a) General pipeline for inferencing an image. (b) Proposed degrading process to match the reality production of short-form UGC video. (c) Model structure details.}
    \label{fig:NetLab}
\end{figure}

% Done!
This team addresses two key challenges in diffusion-based super-resolution, \textit{i.e.}, inference efficiency and generalization ability. The pipeline of their method is shown in Figure~\ref{fig:NetLab}. For the first challenge, they design a compact architecture inspired by the Diffusion Transformer (DiT)~\cite{peebles2023scalableDiT} but optimized for high-resolution inputs. To handle mismatched resolutions between input images and latent, they introduce a lightweight $8\times$ downscaling encoder using convolutions. To mitigate DiT's quadratic complexity, they simplify the transformer structure with ReLU activations and a Q(KV) operation~\cite{wang2025lit} while enhancing local detail handling through a hybrid convolutional-linear feed-forward network. For the second challenge, they refine the original RealESRGAN~\cite{wang2021realesrgan} degradation pipeline by adjusting downscaling probabilities (0.4 for the final stage and 0.8 for the initial stage) and incorporating more recent compression formats (WEBP, HEIF, AVIF) to avoid unrealistic artifacts demonstrated in KwaiSR dataset. They also simulate real-world text overlay degradation by injecting synthetic text components into the degradation pipeline before applying compression. They adopt ResShift's efficient Markov chain framework \cite{yue2024efficientResShift} as the diffusion scheduler, achieving significant quality improvements in both synthetic and real-world scenarios.

\noindent \textbf{Training Details}
The training process of their method is conducted on PyTorch using NVIDIA RTX 4090 GPUs. It consists of two distinct phases, each with specific configurations.

In the first phase, they train their model on a composite dataset including WIDER~\cite{yang2016wider}, LSDIR~\cite{li2023lsdir}, DF2K~\cite{agustsson2017ntireDIV2K}, OST~\cite{wang2018ost} and the first 10k images in FFHQ~\cite{karras2019FFHQ}. As for WIDER, they selected it by limiting CLIPIQA $>$ 0.6, MANIQA $>$ 0.4 and MUSIQ $>$ 60. As a result, 1768 images are selected for training the model. For loss functions in this phase, MSE loss was used in the latent space. They employed the AdamW optimizer and a cosine annealing learning rate schedule. This schedule included a warm-up period of 20,000 iterations, during which the learning rate ramped up to a peak value of $1\times e^{-4}$, then ended with $1\times e^{-5}$. With a global batch size of 64, the model are optimized for 400k iterations. This whole phase costs about 3 days on 8 GPUs.

In the second phase, they extend the training dataset to include the high-resolution (HR) images from the KwaiSR Dataset (synthetic portion, while discarding the LR portion and the wild part for ease of augmentation), in addition to the datasets used in the first phase. These images were repeated six times per epoch to make the model better fit the distribution of UGC contents. Additionally, about 6k high-quality images selected by IQA metrics in WIDER dataset were also used. Similar to the first phase, they used the AdamW optimizer and a learning rate schedule with a 2,000-iteration warm-up. The learning rate in this phase annealed from a peak of $3\times e^{-5}$ down to a minimum of $1\times e^{-5}$. In the latent space, an L2 loss was applied. Simultaneously, in the image space, they incorporated L1, L2, SSIM, and LPIPS losses with respective weights of 1, 0.3, 0.3, 0.1, and 0.05. The batchsize was set to 1 on each GPU and 8$\times$ gradient accumulations in order to save VRAM. This phase lasted for 16,000 iterations. It costs 0.75 days working on 7 GPUs. They adopted mixed-precision training with fp16 for efficiency optimization. They randomly cropped each image to 512 in shape and utilized a modified version of the RealESRGAN degradation process for generating low-resolution inputs.

\noindent \textbf{Testing Details}
During testing, they first interpolate images to 1080p and encode these images by VAE to the latent space. The noise and original image (as condition) are applied to the model for the 15-step sampling, which has the same process as ResShift ~\cite{yue2024efficientResShift}. For inferring the Wild content, before being conditioned, images are downscaled then upscaled to the original shape with a factor of 2$\times$.

\subsection{BrainyBots Team}
% Done!
This team develops the method after a comprehensive analysis of existing super-resolution approaches. They propose a hybrid approach combining RealESRGAN~\cite{wang2021realesrgan} and SinSR~\cite{wang2024sinsr} by leveraging their complementary strengths. RealESRGAN excels at removing real-world distortions, whereas SinSR specializes in achieving super-resolution in a single step. Consequently, their method performs super-resolution by sequentially processing distorted images with SinSR followed by RealESRGAN.

\subsection{NVDTOFCUC}
\begin{figure}
    \centering
    \includegraphics[width=1.0\linewidth]{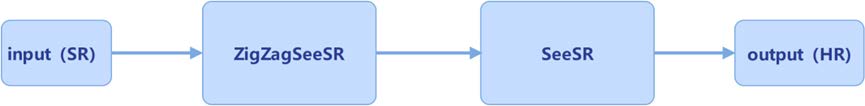}
    \caption{The framework of Team ZigZagSeeSR.}
    \label{fig:NVDTOFCUC}
\end{figure}

% Done!
This team proposes a two-stage training free diffusion-based super-resolution method based on pre-trained SeeSR~\cite{wu2024seesr}, as shown in Figure~\ref{fig:NVDTOFCUC}. In the first stage, they adopt a zigzag sampling method~\cite{bai2024zigzag} to accelerate the denoising process of SeeSR. The denoising trajectory is alternated between deterministic
forward steps and stochastic backward jumps. They dynamically skip denoising steps based on gradient magnitude thresholds, preventing restored images from oversmoothing in high-frequency regions. In the second stage, they adopt the standard DDPM sampling strategy (with 30 denoising steps) to refine the super-resolved image with multi-scale feature fusion. The denoising is accomplished with a single A100-SXM4-80GB GPU.

\subsection{BVIVSR}
\begin{figure}
    \centering
    \includegraphics[width=1.0\linewidth]{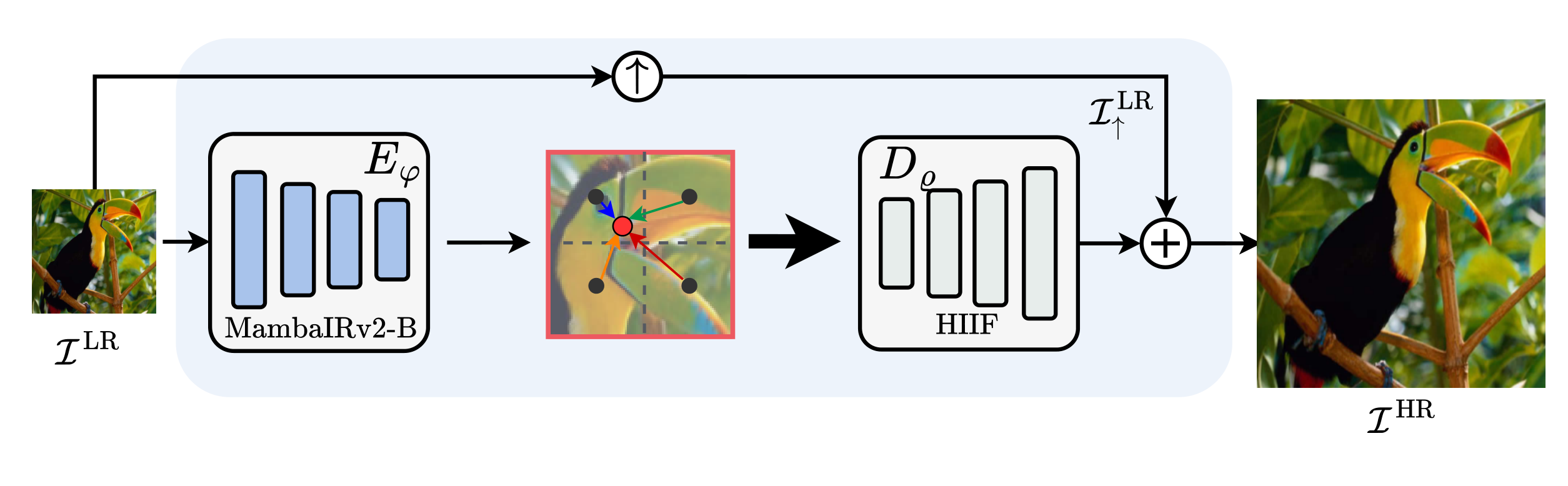}
    \caption{The framework proposed by Team BVIVSR.}
    \label{fig:BVIVSR}
\end{figure}

% Done!
This team proposes to build the method based on the state-of-the-art super-resolution model MambaIRv2~\cite{guo2024mambairv2} and the continuous super-resolution approach HIIF~\cite{jiang2024hiif}, denoted as HIMambaSR. As depicted in Figure~\ref{fig:BVIVSR}, they adopt the MambaIRv2-B model as the latent encoder $E_{\varphi}$ without its upsampling modules. These latent features are subsequently processed by HIIF~\cite{jiang2024hiif} as the latent decoder $D_{\rho}$ to generate restored images. Specifically, $E_{\varphi}$ consists of a sequence of Attentive State Space Groups (ASSG), with each ASSG incorporating multiple Attentive State Space Blocks (ASSBs). Within each ASSB, a progressive local-to-global modeling strategy is employed. Notably, Window Multi-Head Self-Attention (MHSA) is used to capture local interactions, while the Attentive State Space Model (ASSM) models global dependencies. Each block follows a ``Norm $\rightarrow$ Token Mixer $\rightarrow$ Norm $\rightarrow$ FFN'' structure, and incorporates two residual connections with learnable scaling. This encoder $E_{\varphi}$ is responsible for extracting deep latent features from the input low-resolution image. $D_{\rho}$ consists of a multi-scale hierarchical encoding module, multiple multi-head linear attention blocks, and MLPs. Its hierarchical positional encoding captures the local implicit image function across different scales. By progressively injecting these encodings into the network, features at each scale are effectively propagated and shared among neighboring sampling points. This enhances the ability of the network to exploit spatial correlations and reconstruct high-frequency details. 

\noindent \textbf{Training Details}
They adopt the original configurations of MambaIRv2-B and HIIF as the model settings. They utilize a combination of DIV2K~\cite{agustsson2017ntireDIV2K}, 1000 high-resolution images from BVI-AOM~\cite{nawala2024bvi}, Flickr2K~\cite{lim2017enhancedFlik2K} and 5000 images from LSDIR~\cite{li2023lsdir} as the training dataset. For evaluation, they follow common practice in continuous super-resolution task~\cite{jiang2024hiif,jiang2025c2disr} and employ the DIV2K validation set (containing 100 images)~\cite{agustsson2017ntireDIV2K}. The maximum learning rate is set to $4\times e^{-4}$. The learning rate follows a cosine annealing schedule, gradually decreasing after an initial warm-up phase of 50 epochs. L1 loss and the Adam~\cite{kingma2014adam} optimization are adopted to optimize their model during training. Training and testing are implemented based on Pytorch on the 4 NVIDIA A100 GPUs. Their model was trained for 1000 epochs with a batch size of 48.

\section*{Acknowledgments}
This work was partially supported by NSFC under Grant 623B2098 and the China Postdoctoral Science Foundation-Anhui Joint Support Program under Grant Number 2024T017AH. We thank Kuaishou for sponsoring this challenge. This work was also partially supported by the Humboldt Foundation. We thank the NTIRE 2025 sponsors: Kuaishou, ByteDance, Meituan, and University of Wurzburg (Computer Vision Lab).
\appendix
\section{Teams and Affiliations of Track 1}
\subsection*{NTIRE2024 Organizers}

\noindent  \textit{\textbf{Title:}} NTIRE 2024 Challenge on Short-form UGC Video Quality Assessment and Enhancement-Track 1

\noindent  \textit{\textbf{Members:}}
Xin Li\textsuperscript{1} (\textcolor{magenta}{xin.li@ustc.edu.cn}), Kun Yuan\textsuperscript{2} (\textcolor{magenta}{yuankun03@kuaishou.com}), Fengbin Guan\textsuperscript{1}, Zihao Yu\textsuperscript{1}, Yiting Lu\textsuperscript{1}, Wei Luo\textsuperscript{1}, Ming Sun\textsuperscript{2}, Chao Zhou\textsuperscript{2}, Zhibo Chen\textsuperscript{1}, and Radu Timofte\textsuperscript{3} 

\noindent  \textit{\textbf{Affiliations:}}

\noindent  \textsuperscript{1} University of Science and Technology of China

\noindent  \textsuperscript{2} KuaiShou Technology 

\noindent  \textsuperscript{3} Computer Vision Lab, University of Wurzburg, Germany

\subsection*{SharpMind}

\noindent  \textit{\textbf{Title:}} Distillation-based Video Quality Assessment: Aligning with Human
Eye Characteristics for Enhanced Precision

\noindent  \textit{\textbf{Members:}}
\noindent   Yabin Zhang (\textcolor{magenta}{zhangtao.ceb@bytedance.com}), Ao-Xiang Zhang, Tianwu Zhi, Jianzhao Liu, Yang Li, Jingwen Xu, and Yiting Liao

\noindent  \textit{\textbf{Affiliations:}}

\noindent Bytedance Inc.

\subsection*{ZQE}
\noindent  \textit{\textbf{Title:}} ZQE (Z-Tech Video Quality Evaluator)

\noindent  \textit{\textbf{Members:}}
\noindent   Yufan Liu (\textcolor{magenta}{tycholiu@tencent.com}), Xiangguang Chen, Zuowei Cao, Minhao Tang, and Shan Liu   

\noindent  \textit{\textbf{Affiliations:}}

\noindent   Tencent Online Video

\subsection*{ZX-AIE-Vector}

\noindent  \textit{\textbf{Title:}} Mamba-in-Mamba-Out: A Lightweight Video Quality Assessment Network
with Hybrid Mamba-Attention Design

\noindent  \textit{\textbf{Members:}}
\noindent   Yunchen Zhang (\textcolor{magenta}{zhang.yunchen@zte.com.cn}), Xiangkai Xu, Hong Gao, Ji Shi, Yiming Bao, Xiugang Dong, Xiangsheng Zhou, Yaofeng Tu

\noindent  \textit{\textbf{Affiliations:}}

\noindent   ZTE Corporation

\subsection*{ECNU-SJTU VQA Team}

\noindent  \textit{\textbf{Title:}} Towards Good Practices for Efficient Video Quality Assessment

\noindent  \textit{\textbf{Members:}}
\noindent   Wei Sun (\textcolor{magenta}{sunguwei@gmail.com}), Kang Fu, Linhan Cao, Dandan Zhu, Kaiwei Zhang, Yucheng Zhu, Zicheng Zhang, Menghan Hu, Xiongkuo Min and Guangtao Zhai

\noindent  \textit{\textbf{Affiliations:}}

\noindent   East China Normal University, Shanghai Jiao Tong University

\subsection*{TenVQA}

\noindent  \textit{\textbf{Title:}} Strong Baseline Strategies for Video Quality Assessment Tasks

\noindent  \textit{\textbf{Members:}}
\noindent   Yuhai Lan (\textcolor{magenta}{lanyuhai.hit@gmail.com}), Gaoxiong Yi

\noindent  \textit{\textbf{Affiliations:}}

\noindent   Tencent

\subsection*{GoldenChef}

\noindent  \textit{\textbf{Title:}} Lightweight Multi-Feature Cross Attention Fusion Model for Short-form
UGC Video Quality Assessment

\noindent  \textit{\textbf{Members:}}
\noindent   MingYin Bai (\textcolor{magenta}{1984048425@qq.com}), Jiawang Du, Zilong Lu, Zhenyu Jiang, Hui Zeng, Ziguan Cui,
Zongliang Gan, Guijin Tang

\noindent  \textit{\textbf{Affiliations:}}

\noindent   College of Telecommunications and Information Engineering, Nanjing University
of Posts and Telecommunications

\subsection*{DAIQAM}

\noindent  \textit{\textbf{Title:}} Short-form Video Quality Assessment: a simple approach

\noindent  \textit{\textbf{Members:}}
\noindent   Ha Thu Nguyen (\textcolor{magenta}{ha.t.nguyen@ntnu.no}), Katrien De Moor, Seyed Ali Amirshahi,
Mohamed-Chaker Larabi

\noindent  \textit{\textbf{Affiliations:}}

\noindent   Norwegian University of Science and Technology; Universit´e
de Poitiers, CNRS, XLIM, France

\subsection*{57VQA}

\noindent  \textit{\textbf{Title:}} 

\noindent  \textit{\textbf{Members:}}
\noindent   Zhiye Huang (\textcolor{magenta}{huang.zhiye@bupt.edu.cn}), Yi Deng

\noindent  \textit{\textbf{Affiliations:}}

\noindent   Beijing University of Posts and Telecommunications

\subsection*{Nourayn}

\noindent  \textit{\textbf{Title:}} No Title

\noindent  \textit{\textbf{Members:}}
\noindent   Nourine Mohammed Nadir (\textcolor{magenta}{nounadir@gmail.com})

\noindent  \textit{\textbf{Affiliations:}}

\noindent   None

\section{Teams and Affiliations of Track 2}

\subsection*{NTIRE2024 Organizers}

\noindent  \textit{\textbf{Title:}} NTIRE 2024 Challenge on Short-form UGC Video Quality Assessment and Enhancement-Track 2

\noindent  \textit{\textbf{Members:}}
Xin Li\textsuperscript{1} (\textcolor{magenta}{xin.li@ustc.edu.cn}), Kun Yuan\textsuperscript{2} (\textcolor{magenta}{yuankun03@kuaishou.com}), Bingchen Li\textsuperscript{1}, YiZhen Shao\textsuperscript{2}, Xijun Wang\textsuperscript{1}, Suhang Yao\textsuperscript{1}, Ming Sun\textsuperscript{2}, Chao Zhou\textsuperscript{2}, Zhibo Chen\textsuperscript{1}, and Radu Timofte\textsuperscript{3} 

\noindent  \textit{\textbf{Affiliations:}}

\noindent  \textsuperscript{1} University of Science and Technology of China

\noindent  \textsuperscript{2} KuaiShou Technology 

\noindent  \textsuperscript{3} Computer Vision Lab, University of Wurzburg, Germany

\subsection*{TACO\_SR}

\noindent  \textit{\textbf{Title:}} PiNAFusion-SR

\noindent  \textit{\textbf{Members:}}
\noindent   Yushen Zuo\textsuperscript{1} (\textcolor{magenta}{zuoyushen12@gmail.com}), Mingyang Wu\textsuperscript{2}, Renjie Li\textsuperscript{2}, Shengyun Zhong\textsuperscript{3}, Zhengzhong Tu\textsuperscript{2}

\noindent  \textit{\textbf{Affiliations:}}

\noindent   \textsuperscript{1} The Hong Kong Polytechnic University

\noindent   \textsuperscript{2} Texas A\&M University

\noindent   \textsuperscript{3} Northeastern University

\subsection*{RealismDif}

\noindent  \textit{\textbf{Title:}} No Title

\noindent  \textit{\textbf{Members:}}
\noindent   Kexin Zhang (\textcolor{magenta}{cvkexincv@gmail.com}), Jingfen Xie, Yan Wang, Kai Chen, Shijie Zhao

\noindent  \textit{\textbf{Affiliations:}}
Bytedance Inc.

\subsection*{SRlab}

\noindent  \textit{\textbf{Title:}} No Title

\noindent  \textit{\textbf{Members:}}
\noindent   Ying Liang\textsuperscript{1} (\textcolor{magenta}{forest726@sjtu.edu.cn}), Yiwen Wang\textsuperscript{1}, Xinning Chai\textsuperscript{1}, Yuxuan Zhang\textsuperscript{1}, Zhengxue Cheng\textsuperscript{1}, Yingsheng Qin\textsuperscript{2}, Yucai Yang\textsuperscript{2}, Rong Xie\textsuperscript{1}, Li Song\textsuperscript{1}

\noindent  \textit{\textbf{Affiliations:}}

\noindent \textsuperscript{1}Shanghai Jiao Tong University

\noindent \textsuperscript{2}Transsion, China

\subsection*{SYSU-FVL-Team}

\noindent  \textit{\textbf{Title:}} Pixel-level and Semantic-level Adjustable Super-resolution.

\noindent  \textit{\textbf{Members:}}
\noindent   Zhi Jin (\textcolor{magenta}{jinzh26@mail.sysu.edu.cn}), Jiawei Wu, Wei Wang, Wenjian Zhang

\noindent  \textit{\textbf{Affiliations:}}

\noindent  Shenzhen Campus of Sun Yat-sen University

\subsection*{NetLab}

\noindent  \textit{\textbf{Title:}} Make Small Model Diffuse Well to Higher Resolution

\noindent  \textit{\textbf{Members:}}
\noindent   Hengyuan Na (\textcolor{magenta}{neihy@mail2.sysu.edu.cn}), Wang Luo, Di Wu

\noindent  \textit{\textbf{Affiliations:}}

\noindent   Sun Yat-sen University

\subsection*{BrainyBots Team}

\noindent   \textit{\textbf{Title:}} No Title

\noindent  \textit{\textbf{Members:}}  
\noindent Xinglin Xie (\textcolor{magenta}{xinglin\_xie@163.com}), Kehuan Song (skh2272@163.com), Xiaoqiang Lu (xqlu@stu.xidian.edu.cn), Licheng Jiao (lchjiao@mail.xidi\\an.edu.cn), Fang Liu (f63liu@163.com), Xu Liu  (xuliu@xi\\dian.edu.cn), Puhua Chen(phchen@xidian.edu.cn)

\noindent  \textit{\textbf{Affiliations:}}

\noindent   XiDian University

\subsection*{BP-SR}

\noindent  \textit{\textbf{Title:}} 
No title

\noindent  \textit{\textbf{Members:}}
Qi Tang (\textcolor{magenta}{qitang@bjtu.edu.cn}), Linfeng He, Zhiyong Gao, Zixuan Gao, Guohua Zhang, Meiqin Liu, Chao Yao, Yao Zhao

\noindent  \textit{\textbf{Affiliations:}} Bejing Jiaotong University

\subsection*{NVDTOFCUC}

\noindent  \textit{\textbf{Title:}} 
ZigZagSeeSR: Semantic-Driven Super-Resolution
Diffusion Sampling

\noindent  \textit{\textbf{Members:}}
\noindent   Qingmiao Jiang (\textcolor{magenta}{jqmiao@cuc.edu.cn}), Lu Chen, Yi Yang, Xi Liao

\noindent  \textit{\textbf{Affiliations:}}

\noindent   School of Information and Communication Engineering,
Communication University of China

\subsection*{BVIVSR}

\noindent  \textit{\textbf{Title:}} No Title

\noindent  \textit{\textbf{Members:}}
\noindent   Yuxuan Jiang\textsuperscript{1} (\textcolor{magenta}{dd22654@bristol.ac.uk}), Qiang Zhu\textsuperscript{2,1}, Siyue Teng\textsuperscript{1}, Fan Zhang\textsuperscript{1}, Shuyuan Zhu\textsuperscript{2}, Bing Zeng\textsuperscript{2}, and David Bull\textsuperscript{1}

\noindent  \textit{\textbf{Affiliations:}}

\noindent   \textsuperscript{1} University of Bristol

\noindent   \textsuperscript{2} University of Electronic Science and Technology of China
{
    \small
    \bibliographystyle{ieeenat_fullname}
    \bibliography{main}
}

\end{document}